\documentclass[%
superscriptaddress,
preprint,
 amsmath,amssymb,
 aip,
jap,
floatfix,
longbibliography
]{revtex4-2}

\usepackage{booktabs}
\usepackage{microtype}
\usepackage{footmisc}
\usepackage[]{natbib}
\usepackage{graphicx}
\usepackage{placeins}
\usepackage{xcolor}
\usepackage[breaklinks,colorlinks,
citecolor=blue,
urlcolor=blue,
pdfsubject={LaTeX}]{hyperref}
\usepackage{amssymb}
\usepackage{subcaption}

\begin{document}
\title{Radiation-induced segregation in dilute Fe-Cr: A rate-theory framework for the Cr enrichment-depletion transition at the grain boundary}

\author{Russell Oplinger}
\affiliation{Idaho National Laboratory, Idaho Falls, ID 83415, USA}
\affiliation{Texas A\&M University, College Station, TX 77843, USA}

\author{Mukesh Bachhav}
\affiliation{Idaho National Laboratory, Idaho Falls, ID 83415, USA}

\author{Karim Ahmed}
\affiliation{Texas A\&M University, College Station, TX 77843, USA}

\author{Sourabh Bhagwan Kadambi}
\email[]{sourabhbhagwan.kadambi@inl.gov}
\thanks{R. Oplinger and S. B. Kadambi contributed equally to this work.}
\affiliation{Idaho National Laboratory, Idaho Falls, ID 83415, USA}

\date{\today}
\preprint{Revised version submitted to \textit{Journal of Applied Physics}}
\begin{abstract}
Radiation-induced segregation (RIS) poses a significant challenge for ferritic Fe-Cr alloys under irradiation, as it can compromise mechanical integrity and increase susceptibility to intergranular corrosion. Yet, the mechanisms governing Cr segregation remain incompletely understood. 
In this study, We present a physics-based rate-theory model parameterized using self-consistent mean field theory-based Onsager transport coefficients to investigate RIS at the grain boundary (GB) in dilute Fe-(0.1 at.\%) Cr. Under equal production rates of vacancies and self-interstitial atoms (SIA), and their equal absorption rates by bulk dislocations, the model simulates the experimentally observed transition from Cr enrichment at low temperatures to depletion at higher temperatures. Under these unbiased conditions, systematic investigation reveals that while temperature-dependent transport properties dictate the segregation direction, dose rate, grain size, and dislocation density only influence the magnitude and spatial extent of Cr segregation. However, under more realistic conditions of preferential vacancy production within damage cascade and/or preferential SIA absorption by bulk dislocations, the enrichment-to-depletion transition shifts to lower temperatures. Our findings demonstrate that RIS predictions based solely on transport coefficients are valid only under symmetric point defect flux conditions, and that biases in defect production and absorption must be considered for accurate predictions. This work provides a mechanistic framework for understanding RIS in ferritic alloys and informs alloy design for advanced nuclear systems.

\end{abstract}

\keywords{radiation-induced segregation, rate-theory model, Fe-Cr alloys, production bias, dislocation absorption bias}

\maketitle
\section{Introduction}

Ferritic Fe–Cr steels are widely employed in nuclear reactor systems, particularly in reactor pressure vessels (RPVs) and other structural components that support the reactor core. These materials are favored for their excellent thermal conductivity, low swelling rates, and good resistance to radiation-induced embrittlement \cite{klueh2007ferritic}. However, during prolonged service, they are subjected to intense neutron irradiation, which induces a range of microstructural and compositional changes that can degrade mechanical performance. Among the various radiation-induced phenomena, radiation-induced segregation (RIS) is a critical non-equilibrium process that alters the local chemical composition at defect sinks such as grain boundaries (GBs), dislocations, and void surfaces \cite{ardell2016radiation,nastar20121}. In ferritic steels, RIS at GBs can significantly impact material integrity and performance. For instance, Cr enrichment may promote the formation of embrittling Cr-rich precipitates, while Cr depletion can increase susceptibility to intergranular corrosion, both of which compromise the long-term reliability of structural components \cite{bruemmer1999radiation,lu_RIS_microchem}. Despite extensive experimental and modeling efforts, the mechanistic understanding of RIS in Fe–Cr alloys remains incomplete, particularly regarding the transition between Cr enrichment and depletion as a function of irradiation conditions. A predictive understanding of RIS is essential for the development of radiation-tolerant ferritic steels for Generation IV reactors \cite{henry2017irradiation} and for extending the operational lifetimes of current light water reactor (LWR) components \cite{was2011assessment}.

Experimental characterization of RIS in ferritic alloys has revealed complex and often contradictory behavior regarding both the direction and magnitude of Cr segregation \cite{lu_RIS_microchem,was2011assessment}. This contrasts with the more consistent Cr depletion observed in Fe–Cr–Ni austenitic stainless steels \cite{allen1998mechanism}. As a result, the factors and phenomena influencing and driving RIS in BCC Fe-Cr alloys have not been conclusively established. In commercial ferritic/martensitic alloys, Cr segregation has been found to transition from enrichment at low temperatures to depletion at elevated temperatures; notably, for HT9 steel ($\sim$12 at.\% Cr), the transition temperature has been found to occur in the range of 500--515~\textdegree C and for T91 steel ($\sim$9 at.\% Cr) in the range of 600--700~\textdegree C\cite{wharry2013systematic,jiao2018microstructure,lu_RIS_microchem,wharry2014mechanism}.
This transition has been attributed to the competition between SIA-mediated transport dominating at lower temperatures and vacancy-mediated inverse Kirkendall effects controlling behavior at higher temperatures \cite{wharry2014mechanism}.
In addition to the alloy's composition, GB character is also found to exert a strong influence on the segregation behavior, with high-angle random boundaries displaying maximum segregation and coherent $\Sigma3$ boundaries exhibiting variable or heterogeneous segregation \cite{field2013dependence,rahmouni2024radiation}. 

Much of the experimental work on simpler Fe-Cr binary systems has been directed at alloys with relatively high Cr concentration between 5 and 15 at.\%, where clustering or precipitation of Cr-rich phases like $\alpha'$ is increasingly favored.
For instance, Fe-6Cr ($\sim$6 at.\% Cr) alloy irradiated at 290--350~\textdegree C revealed modest Cr enrichment at various GB types \cite{bachhav2014fe6cr}. 
On the other hand, supersaturated Fe-15Cr ($\sim$15 at.\% Cr) neutron-irradiated at 290~\textdegree C to 1.8 dpa exhibited W-shaped profiles (enrichment at the GB and depletion further away) at $\Sigma5$ GB. Here, the segregation profiles became sensitive to carbide/nitride precipitate formation at the GB as Cr depletes from the GB to enrich these precipitates \cite{bachhav2014fe15cr}, complicating the interpretation of the RIS mechanism. 
A similar Fe-15Cr alloy ion irradiated at 350~\textdegree C to 8 dpa exhibited W-shaped Cr profiles with carbon segregation but no precipitate formation, suggesting either the absence or ballistic dissolution of precipitates at this higher dose rate, with carbon possibly contributing to Cr enrichment at the GB \cite{marquis2011systematic}.
 
Computational modeling approaches have sought to bridge these experimental observations and provide insight into underlying atomistic mechanisms behind complex RIS behavior. 
Rate theory models capture RIS over large time and length-scales by describing the rates of point defect reactions in the bulk, their diffusion to and interaction with GBs or dislocation sinks, and their flux couplings with solute atoms \cite{wiedersich1979theory,nastar20121,ardell2016radiation,schuler2022towards}.
These models are capable of predicting the temperature-dependent segregation trends by describing the competing effects of both the vacancy-solute exchange mechanism and SIA-solute drag mechanism. 
However, conventional rate-theory RIS models require physically-accurate temperature and composition-specific transport parameters that must be empirically or theoretically determined, limiting predictive capability for unexplored systems.

Recent advances have addressed these limitation through ab initio physics-based parameterization \cite{barnard2012modeling,thuinet2018multiscale}.
Self-consistent mean field (SCMF) theory \cite{nastar2005mean}, informed by density functional theory (DFT) calculations, provide Onsager coefficients ($L_{ij}$) that quantify the coupling between chemical species (Fe and Cr) and point defect fluxes \cite{messina_OnsTran,shousha2024vacancy}, without the need for experimental tracer diffusivity data or empirical fitting. 
Atomistic kinetic Monte Carlo (AKMC) simulations, parameterized using DFT-derived migration barriers, successfully explain the temperature-dependent crossover from Cr enrichment to depletion. These simulations show that stable mixed Fe-Cr dumbbell migration dominates at low temperatures, while Cr depletion via vacancy-mediated diffusion governs high temperature RIS. The crossover temperature ranges from $\sim$400--600 K and varies across Cr concentration between 5--15 at.\% \cite{senninger_kmc}.
Despite these advances in modeling atomistic transport mechanisms, AKMC simulations remain limited to doses below ~1 dpa and grain sizes less than a few hundred nanometers due to its computational expense, preventing systematic exploration of steady-state RIS across the parameter space of dose and sink density relevant to actual microstructures and irradiation conditions.
Cluster dynamics approaches offer an intermediate-scale alternative for modeling defect cluster evolution in ferritic iron~\cite{kohnert2015clusterI,kohnert2015clusterII}, though their application to alloys and spatially-resolved microstructures remains computationally demanding.

While atomistic transport mechanisms in Fe-Cr have been elucidated using SCMF-derived Onsager coefficients~\cite{messina_OnsTran,shousha2024vacancy}, rate-theory studies that leverage these coefficients to systematically evaluate irradiation dose rate effects, microstructural sink effects, and---critically---point defect flux asymmetries from production and absorption bias have not been performed. 
Since existing RIS models for Fe-Cr neglect these bias effects~\cite{wharry2014mechanism,xia2020radiation,moladje2022radiation,piochaud_PF_RIS}, predictions based solely on Onsager coefficients or partial diffusion coefficient ratios remain incomplete.
Production bias arises when radiation cascades produce unequal numbers of freely migrating vacancies and SIAs that escape recombination, creating asymmetric point defect fluxes to sinks independent of the solute-defect diffusivities and transport mechanisms. 
While well-documented for void swelling~\cite{woo1992production,golubov2000defect}, its impact on RIS has been largely ignored~\cite{ozturk2021_PB,gencturk2023_PB_RIS}.
Molecular dynamics (MD) simulations of ferritic Fe-Cr steels under neutron irradiation reveal that SIAs trapped in clusters surpass vacancies~\cite{terentyev_FeCr_MD,zhang2017molecular}, and rate-theory modeling in Ni-Cr indicates that such production bias can significantly alter RIS under specific conditions~\cite{ozturk2021_PB,gencturk2023_PB_RIS}. 
On the other hand, absorption bias arises from the preferential capture of SIAs by network dislocations and dislocation loops, whose strain fields interact more strongly with SIAs due to their larger relaxation volumes, leading to additional asymmetry in defect fluxes.
Discrete dislocation dynamics (DDD) simulations reveal that sink strengths of network dislocations in BCC Fe depend on point defect relaxation volume, dislocation density, and temperature~\cite{kohnert_DDD}.
Distinguishing these bias effects experimentally remains challenging in the absence of systematic modeling that isolates their individual contributions across irradiation conditions and microstructures.

In this paper, we develop a rate-theory model parameterized with Onsager transport coefficients derived from SCMF theory, production bias informed by MD simulations, and absorption bias from DDD simulations to systematically study RIS in a dilute Fe-0.1Cr (0.1 at.\%) alloy.
Our choice of dilute Fe-Cr composition provides a cleaner system for investigating intrinsic RIS mechanisms: at higher, non-dilute compositions, the occurrence of Cr-rich precipitates or clusters adds complexity to the interpretation of RIS profiles, and concentration-dependent kinetic correlations become significant.
Under unbiased conditions, the transition in RIS from enrichment to depletion is assessed as a function of temperature, and the RIS magnitude is systematically examined as a function of dose rate, grain size, and dislocation density. 
We then examine the effects of cascade production bias and dislocation absorption bias on RIS.
Finally, we discuss the scope and limitations of the classical RIS modeling framework and identify directions for future model development and experimental validation.

\section{Method Description} \label{sec:method}
We employ a rate-theory model incorporating non-equilibrium chemical driving force and flux coupling to describe the spatio-temporal evolution of point defects and chemical species. 
The governing equations are presented in Sec.~\ref{sec:method_eqns}. 
The parameterization of the model for Onsager transport coefficients, production bias and dislocation absorption bias is presented in Sec.~\ref{sec:method_param}.
Numerical implementation of the model is detailed in Sec.~\ref{sec:method_impl}.

\subsection{Governing Equations} \label{sec:method_eqns} 
The evolution of Cr and point defect (vacancies and SIAs) site concentrations in the lattice frame of reference is described by the following coupled partial differential equations for diffusion (Eq.~\ref{eq:time_evol_flux}a) and reaction-diffusion (Eqs.~\ref{eq:time_evol_flux}b and \ref{eq:time_evol_flux}c): 
\begin{subequations}
\begin{flalign}
    &\frac{\partial c_{Cr}}{\partial t} = -\nabla \cdot \boldsymbol{J}_{Cr}, \\ \label{eq:conc-flux_defect}
    &\frac{\partial c_{V}}{\partial t} = -\nabla \cdot \boldsymbol{J}_{V} + P_{V} - R_{VI}\, c_Vc_I - \rho Z_{V} D_{V}(c_{V}-c^e_{V}), \\
    &\frac{\partial c_{I}}{\partial t} = -\nabla \cdot \boldsymbol{J}_{I} + P_{I} - R_{VI}\, c_Vc_I - \rho Z_{I} D_{I} c_{I}, &&
\end{flalign}
\label{eq:time_evol_flux}
\end{subequations}
where $\boldsymbol{J}$ is the flux vector, and $P_V$ and $P_I$ are the vacancy and SIA production rates, respectively. The last terms in the point defect evolution equations describe the absorption of the point defects by mean-field bulk dislocations, of line density $\rho$, with a capture efficiency of $Z$. $c^e_V$ is the equilibrium vacancy concentration at the GB, which is assumed to be the same as that in the bulk.
For the unbiased cases, $P_V=P_I=P$ and $Z_V=Z_I=Z$.
$R_{VI}\, = 4\pi r_\circ(D_V+D_I)/ V_a$ is the recombination rate between vacancies and SIAs, with $r_\circ$ being the recombination radius and $V_a$ the atomic site volume. $D_V$ and $D_I$ are the respective point defect diffusivities. 
Following the model we developed (Kadambi et al.~\cite{kadambi_pf_RIS_FeCrNi}) for RIS in multicomponent alloys, the flux coupling between atoms and point defects is written as:
\begin{subequations}
\begin{flalign}
    &\boldsymbol{J}_{Cr} = - \tilde{L}_{Cr Cr} \nabla \tilde{\mu}_{Cr} - L_{Cr V} \nabla \mu_{V} - L_{Cr I} \nabla \mu_{I}, \\
    &\boldsymbol{J}_{V} = - \tilde{L}_{V Cr} \nabla \tilde{\mu}_{Cr} - L_{VV} \nabla \mu_{V}, \\
    &\boldsymbol{J}_{I} = - \tilde{L}_{I Cr} \nabla \tilde{\mu}_{Cr} - L_{II} \nabla \mu_{I}. &&
\end{flalign}
\label{eq:total_flux}
\end{subequations}
Under symmetric fluxes (satisfied in the absence of cascade production bias or dislocation absorption bias), lattice site conservation ensures $\boldsymbol{J}_{Fe} + \boldsymbol{J}_{Cr} + \boldsymbol{J}_{V} - \boldsymbol{J}_I = 0$. $\tilde{\mu}_{Cr} = \mu_{Cr} - \mu_{Fe} = \partial f_C(c_{Cr})/\partial c_{Cr}$ is the diffusion potential of Cr relative to Fe, with $f(c_{Cr})$ being free energy of Fe-Cr alloy. 
$\mu_{V}=(RT/V_m)\ln(c_V/c^e_V)$ and $\mu_{I}=(RT/V_m)\ln(c_I/c^e_I)$ are the chemical potentials of the point defects, with $c^e$ being the thermal equilibrium point defect concentration and $V_m$ being the constant molar lattice volume. $c^e_V=\exp{\left(-E^f_V/k_BT\right)}\exp{\left(S^f_V/k_B\right)}$ and $c^e_I=\exp{\left(-E^f_I/(k_BT)\right)}$, where $E^f$ is the point defect formation energy and $S^f_V$ is the vacancy formation entropy.
$\tilde{L}$'s are the relative transport coefficients, given in terms of the Onsager coefficients as~\cite{kadambi_pf_RIS_FeCrNi}: 
\begin{subequations}
\begin{flalign}
    &\tilde{L}_{V Cr} = - L^{V}_{Cr Cr} - L^{V}_{Cr Fe} + c_{Cr} L_{VV},\\
    &\tilde{L}_{I Cr} = L^{I}_{Cr Cr} + L^{I}_{Cr Fe} - c_{Cr} L_{II}, \\
    &\tilde{L}_{Cr Cr} = L^{V}_{CrCr} + L^{I}_{CrCr} + \, c_{Cr} L_{Cr {V}} - \, c_{Cr} L_{Cr {I}}. &&
\end{flalign}
\label{eq:relative_Onsager}
\end{subequations}
At dilute Cr concentrations, we make the assumption of neglecting the formation of Cr-rich phases via spinodal decomposition or radiation-induced precipitation. Thus, we simplify the combined form of Eqs.~\ref{eq:time_evol_flux} and \ref{eq:total_flux} by writing the diffusion/chemical potential gradients in terms of concentration gradients as:
\begin{subequations}
\begin{flalign}
    &\frac{\partial c_{Cr}}{\partial t} = \nabla \cdot \tilde{L}_{Cr Cr} \tilde{\theta}_{Cr Cr} \nabla c_{Cr} + \nabla \cdot L_{Cr V} \theta_{VV} \nabla c_{V}  + \nabla \cdot  L_{Cr I} \theta_{II} \nabla c_{I}, \\ 
    &\frac{\partial c_V}{\partial t} = \nabla \cdot \tilde{L}_{V Cr} \tilde{\theta}_{Cr Cr} \nabla c_{Cr} + \nabla \cdot  L_{VV} \theta_{VV} \nabla c_V + P_V - R\, c_Vc_I - \rho Z_V D_V (c_V - c^e_V), \\
    &\frac{\partial c_I}{\partial t} = \nabla \cdot \tilde{L}_{I Cr} \tilde{\theta}_{Cr Cr} \nabla c_{Cr} + \nabla \cdot L_{II} \theta_{II} \nabla c_I + P_I - R\, c_Vc_I - \rho Z_I D_I c_I, &&
\end{flalign}
\label{eq:time_evol_conc}
\end{subequations}
where $\tilde{\theta}_{Cr Cr}=\partial^2 f_C/{\partial c_{Cr}}^2$, ${\theta}_{VV}=(RT/V_m)/c_V$, and ${\theta}_{II}=(RT/V_m)/c_I$.
Also, due to the dilute concentration of Cr in BCC Fe, we can assume $f$ to be described by an ideal solution free energy, resulting in $\tilde{\theta}_{Cr Cr}=(RT/V_m)/(c_{Fe} c_{Cr})$.
We note that non-ideal thermodynamics from CALPHAD-based free energy description can be readily incorporated following the approach outlined in Kadambi et al.~\cite{kadambi_pf_RIS_FeCrNi}, and $\alpha$-$\alpha'$ spinodal decomposition can also be simulated by retaining the chemical/diffusion potential implementation in Eq.~\ref{eq:total_flux}.
Considering that the point defect concentrations are always maintained at very dilute limits (less than $10^{-3}$), $c_{Fe}\approx1-c_{Cr}$ is applied in the description of the thermodynamic free energy and its derivatives.
Finally, we note that the point defect diffusivities are obtained from the Onsager coefficients as $D_V = L_{VV}\theta_{VV}$ and $D_I = L_{II}\theta_{II}$.

\subsection{Parameterization} \label{sec:method_param}

\subsubsection{Onsager transport coefficients}

\begin{figure}[htp!]
    \centering
     \includegraphics[scale=0.4]{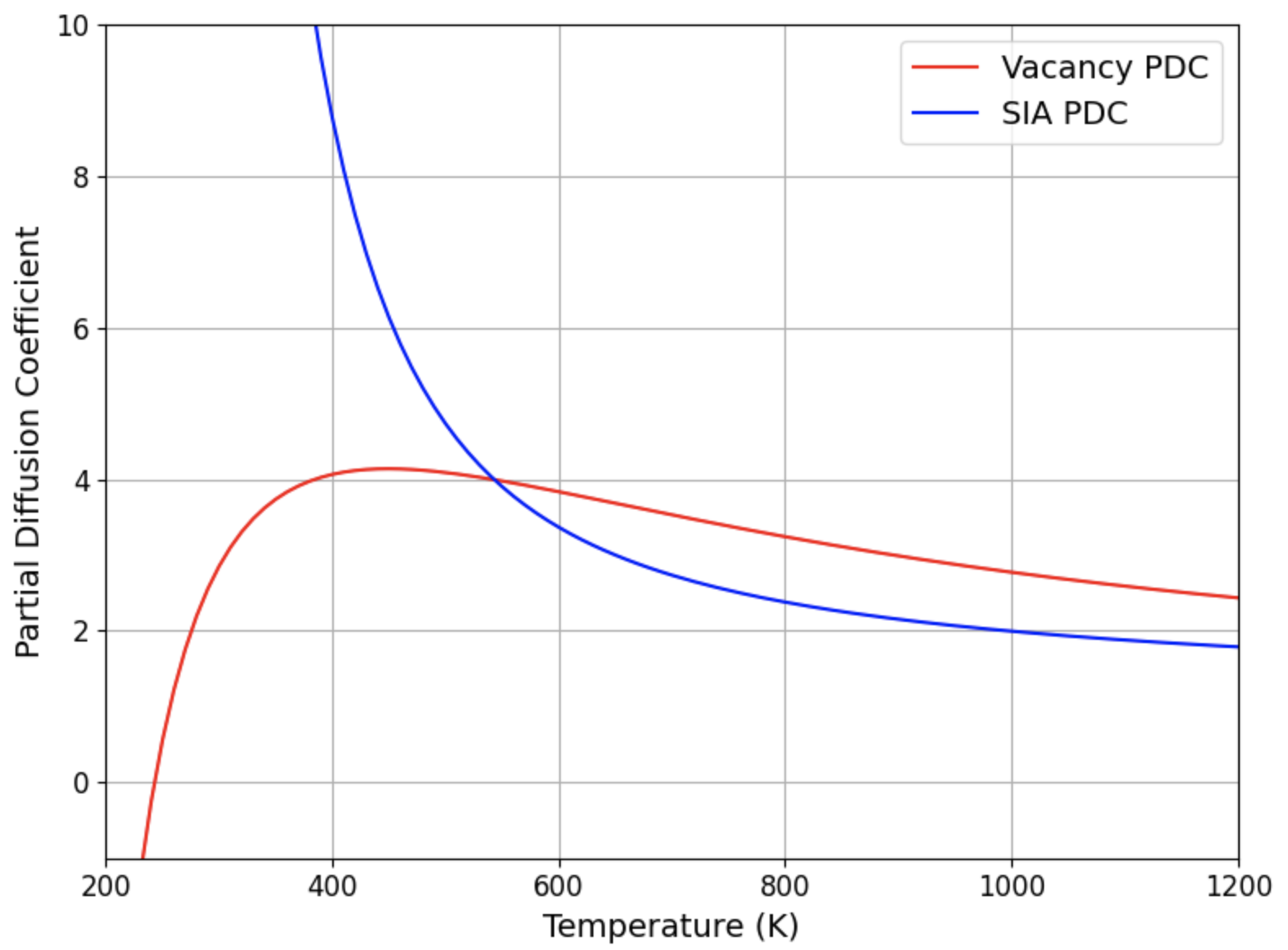}
     \caption{Temperature dependence of partial diffusion coefficient (PDC) ratios for vacancies and SIAs in Fe-0.1Cr. The PDC ratios are plotted from the Onsager coefficients obtained from KineCluE calculations of Messina et al.~\cite{messina_OnsTran}. } 
     \label{fig:PDC_plot}
 \end{figure}

The Onsager phenomenological coefficients quantify the coupling between solute and point defect fluxes. These are critical inputs for our rate-theory model and are key to predicting RIS tendencies under symmetric point defect fluxes. These coefficients (Fig.~\ref{fig:ons_coeff}) were obtained from SCMF calculations performed using the KineCluE code by Messina et al.~\cite{messina2014exact,messina_OnsTran}, who calculated the coefficients for various Fe-based dilute binary alloys including Fe-Cr. Within KineCluE, the Onsager coefficient matrix is computed from atomic jump frequencies, which are parameterized using DFT calculations of solute-defect binding energies, migration barriers, and attempt frequencies. This ab initio-informed approach captures the kinetic correlations arising from solute-vacancy and solute-SIA interactions without relying on empirical fitting to experimental tracer diffusivity data.
The RIS tendency is determined by the partial diffusion coefficient (PDC) ratios:
\begin{subequations}
\begin{flalign}
    D^V_{\text{pd}} &=\frac{(1-c_{Cr})}{c_{Cr}}\frac{L_{VCr}}{L_{VFe}}, \\
    D^I_{\text{pd}} &=\frac{(1-c_{Cr})}{c_{Cr}}\frac{L_{ICr}}{L_{IFe}}, &&
\end{flalign}
\end{subequations}
where $D^I_{\text{pd}}$ and $D^V_{\text{pd}}$ are the PDC ratios for SIAs and vacancies respectively. The sign of $D^V_{\text{pd}} - D^I_{\text{pd}}$ determines the RIS tendency, with a positive value denoting Cr depletion via the vacancy exchange mechanism and a negative value denoting Cr enrichment via the SIA drag mechanism. 
The PDC ratios derived from the Onsager data of Ref.~\cite{messina_OnsTran} are plotted as a function of temperature in Fig.~\ref{fig:PDC_plot}. At low temperatures ($<$550 K) where $D^I_{\text{pd}} > D^V_{\text{pd}}$, the SIA flux preferentially transports Cr toward sinks, causing enrichment. At high temperatures ($>$550 K) where $D^V_{\text{pd}} > D^I_{\text{pd}}$, vacancy flux dominates, leading to Cr depletion. The crossover temperature near 550 K marks the transition between the two regimes and is fundamentally determined by the competition between vacancy and SIA-mediated diffusion mechanisms. 

\subsubsection{Production bias}

Radiation damage begins at the atomic level where the incident irradiating particle impacts the primary knock-on atom (PKA). The initial interaction often results in the formation of Frenkel pairs, which are pairs of vacancy and SIA defects. For impacts of sufficient energy, the PKA will induce a cascade of atomic collisions resulting in a vacancy-rich core and an SIA-rich perimeter surrounding the impact site, commonly seen with protons, neutrons, and heavy ions \cite{wasfundradmatsci2017}. In contrast, electron irradiation produces isolated Frenkel pairs rather than cascades due to limited energy transfer. The cascade process occurs within a span of picoseconds, followed by the collapse of the cascade ($\sim$10 ps), during which the kinetic energy of the displaced atoms dissipates and many of the vacancies and SIAs recombine and annihilate. However, the lattice is rarely restored perfectly with a fraction of point defects surviving to diffuse outside the cascade and through the material. 
In these damage cascades, a fraction of the point defects form clusters, which can be either glissile or sessile. The clustering behavior is often unequal between SIAs and vacancies, resulting in an effective production bias towards one type of point defect.
MD simulations of cascade evolution in Fe and Fe-Cr alloys demonstrate that SIAs exhibit stronger clustering than vacancies~\cite{terentyev_FeCr_MD,zhang2017molecular}, forming mobile one-dimensional glissile clusters that can escape the cascade region. This asymmetric clustering behavior results in an effective production bias: the number of freely migrating vacancies can exceed that of freely migrating SIAs by 20--30\% under neutron irradiation conditions, even though Frenkel pairs are initially created. 
To implement the vacancy-favored production bias in our model, the Frenkel pair production rate $P$ is multiplied by a factor $(1+\epsilon)$, giving $P_V = (1+\epsilon)P$ and $P_I = P$, where $\epsilon$ is the fraction of excess of freely migrating vacancies relative to SIAs (e.g., $\epsilon$ = 0.3 corresponds to a 30\% bias production bias towards vacancies).

\subsubsection{Dislocation sink strength}

\begin{figure}
\centering
\begin{subfigure}{\textwidth}
    \includegraphics[scale=0.5]{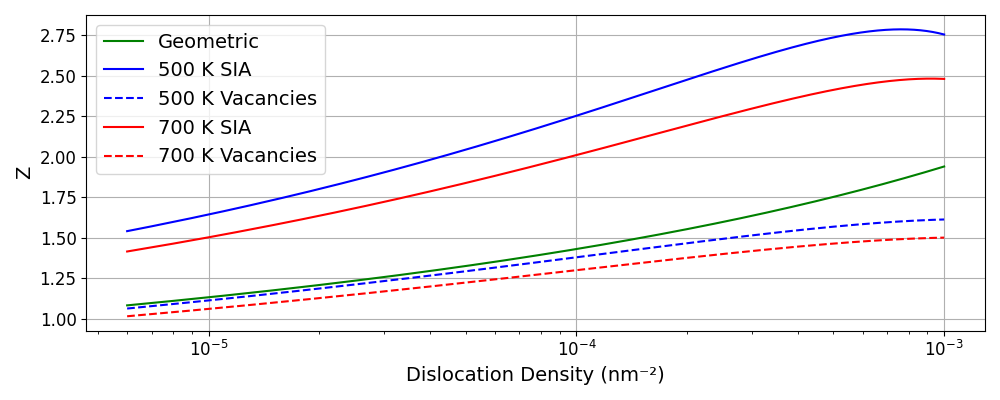}
    \caption{}
    \label{fig:sink_strength}
\end{subfigure}
\begin{subfigure}{\textwidth}
    \includegraphics[scale=0.5]{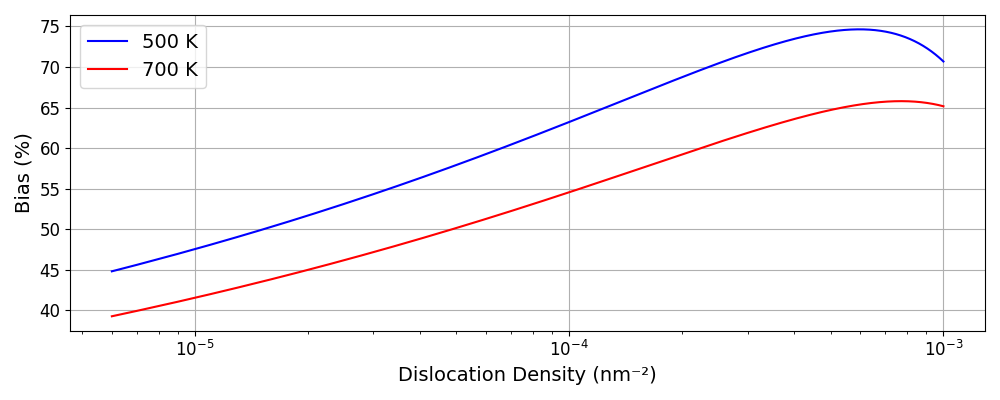}
    \caption{}
    \label{fig:sink_bias}
\end{subfigure}
\caption{(a) Dislocation capture efficiency Z for vacancies and SIAs as a function of dislocation density $\rho$, comparing the classical unbiased model with the DDD-based biased model from Kohnert et al.~\cite{kohnert_DDD} at 500 K and 700 K. (b) SIA absorption bias $B$ predicted by the DDD-based model as a function of dislocation density at 500 K and 700 K. }
\label{fig:sink_info}
\end{figure}

The dislocation sink strength for point defects in Eq.~\ref{eq:time_evol_conc} is given by $k^2_d=\rho Z_d$, where $d$ is the defect type (vacancy or SIA).
To parameterize the capture efficiency $Z_d$ for unbiased and biased cases, the following two models were used. The first is a simple cylindrical cell wall approach that does not account for lattice strain effects and any absorption biases arising from them. In this approach, the capture efficiency ($Z=Z_V=Z_I$)is given by~\cite{kohnert_DDD}:
\begin{flalign}
    Z = \frac{2\pi}{\ln(\frac{R}{r_0})}, &&
\end{flalign}
where $r_0$ is the dislocation core or interaction radius, and $R$ is the outer cell wall radius related to the dislocation density $\rho$ as $R = {1}/{\sqrt{\pi\rho}}$
This model is employed when dislocation absorption bias effects are neglected, assuming equal capture rates for vacancies and SIAs. 
The second model is a surrogate or reduced order model from Kohnert et al.~\cite{kohnert_DDD} who derived it from DDD simulations of dislocation networks in BCC Fe considering lattice strain effects, dislocation {configuration}, and climb kinetics. The capture efficiency is given by:
\begin{flalign}
    Z_d=\frac{2\pi}{\ln(\frac{R}{\delta})}\times\left(A_0+A_1 \frac{r_0}{R}+A_2\frac{\delta-r_0}{R} \right), &&
\label{eq:Z_bias}
\end{flalign}
where $A_0$, $A_1$, and $A_2$ are fitting parameters determined from DDD simulations {of three-dimensional dislocation networks in BCC Fe spanning densities from $5 \times 10^{-6}$ to $10^{-3}$ nm$^{-2}$ and temperatures from 500 to 700 K~\cite{kohnert_DDD}. The parameters used here correspond to the ``full network'', which sampled over 40 statistically independent network instantiations to capture variability in spatial arrangement and character distribution (edge segments, screw segments, and junctions). As a first attempt to systematically demonstrate the effects of production and absorption bias on RIS in Fe-Cr, we adopt this mean-field homogenization of dislocation sink behavior. We recognize that under irradiation, dislocation loops and network lines are generated and evolve dynamically through processes such as loop nucleation, growth, and interaction with the any existing network. Such evolutionary aspects not treated in this work for simplicity.} $R$ is the cell radius (defined identically to the geometric sink strength model), and $\delta_d$ is the effective capture distance accounting for elastic interaction:
\begin{flalign}
\delta_d = \sqrt{r_0^2+\frac{L^2_d}{4}}. &&
\end{flalign}
The effective elastic interaction length $L_d$ is given by:
\begin{flalign}
    L_d=\frac{Kb|\Delta V_d|}{2k_bT}\left(\frac{1-2\nu}{1-\nu}\right), &&
\end{flalign}
where $K$ is the bulk modulus, $b$ is the Burgers vector magnitude, $\Delta V_d$ is the relaxation volume of the point defect, $k_b$ is Boltzmann's constant, and $\nu$ is Poisson's ratio.
The relaxation volume differs significantly between vacancies ($\Delta V_d = -0.5\Omega$, where $\Omega$ is atomic volume) and SIAs ($\Delta V = +1.5\Omega$), leading to preferential SIA capture by dislocations.
Physically, $L_d$ captures the distance over which elastic interactions between point defects and dislocations are significant. The larger relaxation volume of an SIA defect results in a longer interaction distance and thus higher capture efficiency compared to a vacancy.
Since we model a dilute Fe-Cr alloy, these material parameters are taken from $\alpha$-Fe \cite{kohnert_DDD}, with the exception of the relaxation volume of SIA, for which the value for \textlangle110\textrangle Fe-Cr dumbbell determined from ab initio calculation was employed\cite{wrobel_md_relax_vol}. 
Fig.~\ref{fig:sink_strength} compares the capture efficiencies $Z$ predicted by both models as a function of dislocation density $\rho$ at two representative temperatures (500 K and 700 K). There is a gradual increase in $Z$ in both models with $\rho$. The DDD-based model predicts systematically higher $Z$ and exhibits a temperature dependence. Fig.~\ref{fig:sink_bias} shows the resulting absorption bias $B$, defined as $(Z_I-Z_V)/Z_V$. The bias increases with dislocation density, reaching a maximum near  $5\times10\textsuperscript{-4}$ nm$^{-2}$ before decreasing slightly at higher $\rho$. This non-monotonic behavior reflects the competing effects of elastic interaction range and geometric capture probability.

Unless otherwise noted, all simulations in Sec.~\ref{sec:results} employ the default parameters listed in Table~\ref{tab:parameters}.

\begin{table}[htp!]
\centering
\caption{Default model parameters for Fe-0.1Cr RIS simulations.}
\label{tab:parameters}
\begin{tabular}{l l l l}
\toprule
\textbf{Symbol} & \textbf{Description} & \textbf{Value} & \textbf{Ref.} \\
\midrule
$c^\circ_{\text{Cr}}$ & Nominal Cr concentration & 0.1 at.\% & \\
$T$ & Temperature & 500, 700 K & \\
$P$ & Point defect production rate & $10^{-7}$ dpa/s & \\
$L$ & Grain size & 5 $\mu$m & \\
$\rho$ & Dislocation density & $10^{-4}$ nm$^{-2}$ & \\
$E^f_V$ & Vacancy formation energy & 1.8 eV & \cite{messina_OnsTran} \\
$S^f_V$ & Vacancy formation entropy & $2k_B$ & \cite{messina_OnsTran} \\
$E^f_I$ & SIA formation energy & 3.6 eV & \cite{messina_OnsTran} \\
$r_\circ$ & Recombination radius & $2a$ & \cite{messina_OnsTran} \\
$a$ & Lattice parameter & 0.287 nm & \\
$\Omega$ & Atomic volume & $a^3/2$ & \\
$b$ & Burgers vector magnitude & 0.25 nm & \cite{kohnert_DDD} \\
$K$ & Bulk modulus & 170 GPa & \cite{kohnert_DDD} \\
$\nu$ & Poisson's ratio & 0.29 & \cite{kohnert_DDD} \\
$\Delta V_V$ & Vacancy relaxation volume & $-0.5\Omega$ & \cite{kohnert_DDD} \\
$\Delta V_I$ & SIA relaxation volume & $+1.5\Omega$ & \cite{wrobel_md_relax_vol} \\
$r_0$ & Dislocation core radius & $2a$ & \cite{kohnert_DDD} \\
$A_0$ & DDD fitting parameter & 0.87 & \cite{kohnert_DDD} \\
$A_1$ & DDD fitting parameter & $-5.12$ & \cite{kohnert_DDD} \\
$A_2$ & DDD fitting parameter & $-0.77$ & \cite{kohnert_DDD} \\
\bottomrule
\end{tabular}
\end{table}

\FloatBarrier
\subsection{Implementation} \label{sec:method_impl}
The coupled system of partial differential equations was solved numerically using the Multiphysics Object-Oriented Simulation Environment (MOOSE) framework. The weak formulation was discretized spatially using the finite element method with linear Lagrange basis functions. Temporal integration employed an implicit backward differencing formula (BDF2) of second order, providing unconditional stability for the stiff reaction-diffusion system. The resulting nonlinear equations at each time step were solved using the Newton-Raphson iterative scheme with a relative convergence tolerance of $10^{-8}$.

Eqs.~\ref{eq:time_evol_conc}a,~\ref{eq:time_evol_conc}b and~\ref{eq:time_evol_conc}c were solved over a one-dimensional domain of length $L$ representing a grain with GBs at opposite ends ($x = 0$ and $x = L$). The grain size $L$ was varied from 50 nm to 5 $\mu$m to investigate size effects. Dirichlet boundary conditions were applied at the GBs to impose ideal sink behavior:
\begin{flalign}
c_V\big|_{x=0,L} = c_V^e, \quad c_I\big|_{x=0,L} = 0, &&
\label{eq:BC_defects}
\end{flalign}
which represents perfect point defect absorption. Periodic boundary conditions were imposed on the Cr concentration:
\begin{flalign}
c_{Cr}(0) = c_{Cr}(L). &&
\label{eq:BC_Cr}
\end{flalign}
The total Cr site concentration within the grain was found to be conserved.
A uniform spatial mesh of 1~nm size was employed. With the Dirichlet boundary 
condition, mesh refinement leads to progressively steeper point defect concentration gradients near the GB and correspondingly higher predicted RIS magnitudes. A 1~nm mesh size was therefore selected based on two considerations: (i) comparison with a Robin boundary condition accounting for finite sink efficiency~\cite{kadambi_pf_RIS_FeCrNi}, which provides mesh-independent results; and (ii) relevance with the spatial resolution of experimental characterization techniques such as scanning transmission electron microscopy with energy dispersive X-ray spectroscopy (STEM-EDS) and atom probe tomography (APT), which are typically on the order of $\sim$1--2~nm and $<$0.5~nm, respectively~\cite{Parish2015,Lach2021}. 
Time step sizes were automatically adjusted using the IterationAdaptive time-stepping algorithm. Steady-state was typically reached around 10 dpa depending on temperature and irradiation parameters.

\section{Simulation Results} \label{sec:results}

\subsection{Temperature crossover}

\begin{figure}[htp!]
    \centering
    \includegraphics[scale = 0.5]{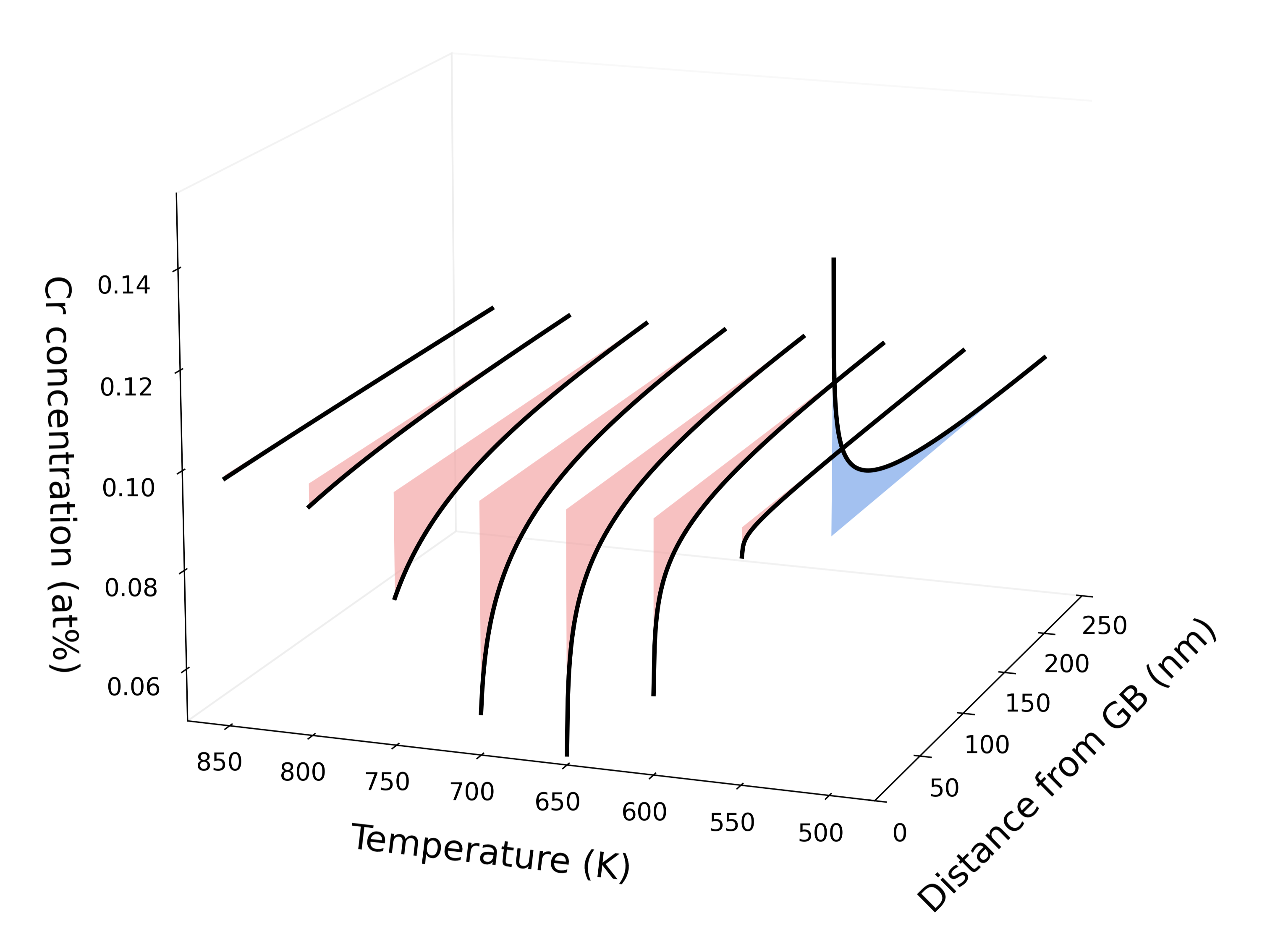}
    \caption{Cr concentration profiles at 10 dpa as a function of normalized distance from the GB and irradiation temperature for a 5 $\mu$m grain with dislocation density of $10^{-4}$ nm$^{-2}$ irradiated at $10^{-7}$ dpa/s under unbiased conditions.}
    \label{fig:temperature}
\end{figure}

To demonstrate the effect of temperature on RIS behavior, simulations were performed on a system of 5 $\mu$m grain size and a dislocation density of $10^{-4}$ nm$^{-2}$ under a dose rate of $10^{-7}$ dpa/s. Production bias and dislocation sink biases were not considered in this baseline case to isolate the intrinsic temperature dependence of RIS driven solely by differences in Onsager transport coefficients. Fig.~\ref{fig:temperature} shows the spatial distribution of Cr at 10 dpa across temperatures ranging from 500 to 850 K. (The corresponding point defect concentration profiles are presented in Fig.~\ref{fig:temp_defect} of the Supplementary Material.) Cr enrichment occurs at lower temperatures ($<$550 K). At temperatures above 550 K, the system transitions to Cr depletion. The characteristic length scale over which RIS develops extends approximately 50$-$100 nm from the GB. 

\begin{figure}[htp!]
    \centering
    \includegraphics[scale = 0.7]{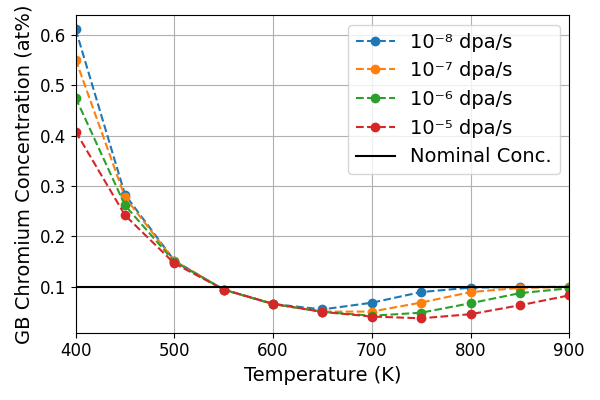}
    \caption{Cr concentration at the GB versus temperature for dose rates ranging from $10^{-8}$ to $10^{-5}$ dpa/s for a system of 5 $\mu$m grain with dislocation density of $10^{-4}$ nm$^{-2}$ irradiated to 10 dpa under unbiased conditions. }
    \label{fig:dose_rate}
\end{figure}

Fig.~\ref{fig:dose_rate} presents the GB Cr concentration at 10 dpa across temperatures ranging from 400 K to 900 K for four different dose rates spanning three orders of magnitude ($10^{-8}$ to $10^{-5}$ dpa/s). All dose rates exhibit the characteristic enrichment-to-depletion transition, demonstrating that dose rate does not alter the fundamental segregation direction, which is governed by the temperature-dependent Onsager transport coefficients. However, dose rate significantly affects both the magnitude of RIS and the temperatures at which maximum RIS occurs. The temperature of maximum Cr depletion is shifts to higher values as the dose rate increases. In all cases, the crossover from enrichment to depletion occurs at around 550 K, consistent with the crossover temperature predicted by the partial diffusion coefficient ratios shown in Fig.~\ref{fig:PDC_plot}. 

\FloatBarrier
\subsection{Effect of sink density}

\begin{figure}[htp!]
    \centering
     \includegraphics[scale=0.55]{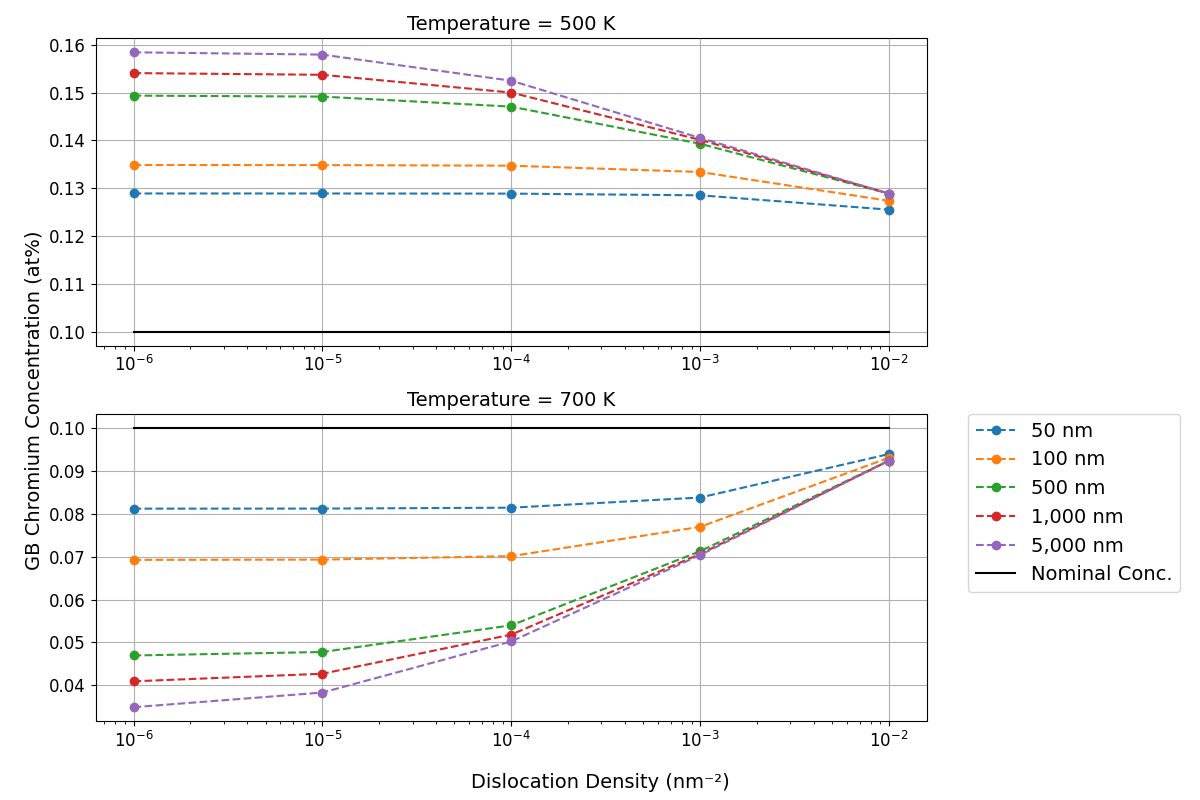}
     \caption{GB Cr concentration as a function of dislocation density $\rho$ for grain sizes ranging from 50 to 5 $\mu$m at (top) 500 K and (bottom) 700 K. Results correspond to irradiation up to 10 dpa at a dose rate of $10^{-7}$ dpa/s under unbiased conditions.} 
     \label{fig:grain_sink}
 \end{figure}

\begin{figure}
\centering
\begin{subfigure}{0.6\textwidth}
    \includegraphics[width=\textwidth]{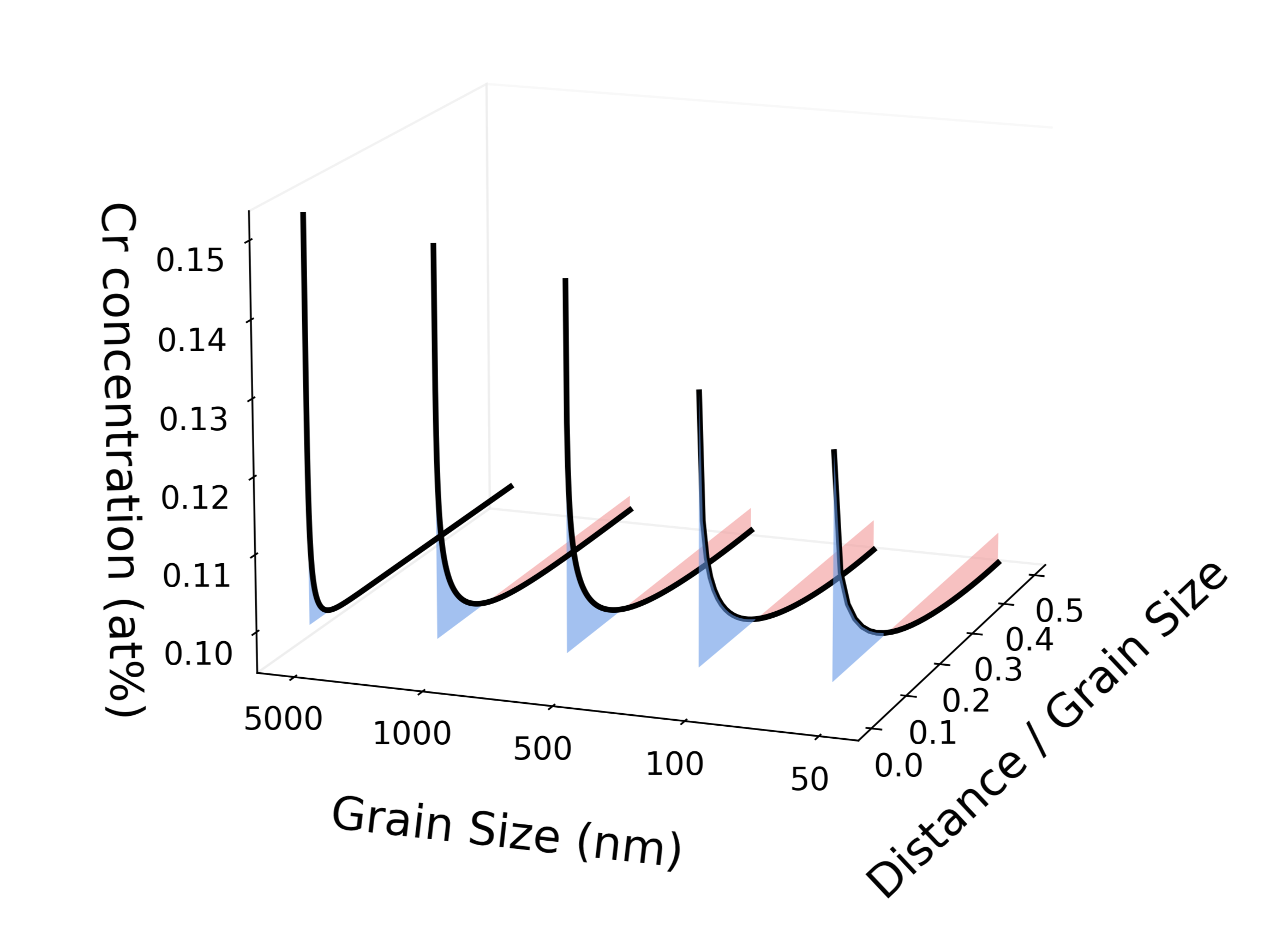}
    \caption{500 K}
    \label{fig:grain_500}
\end{subfigure}
\begin{subfigure}{0.6\textwidth}
    \includegraphics[width=\textwidth]{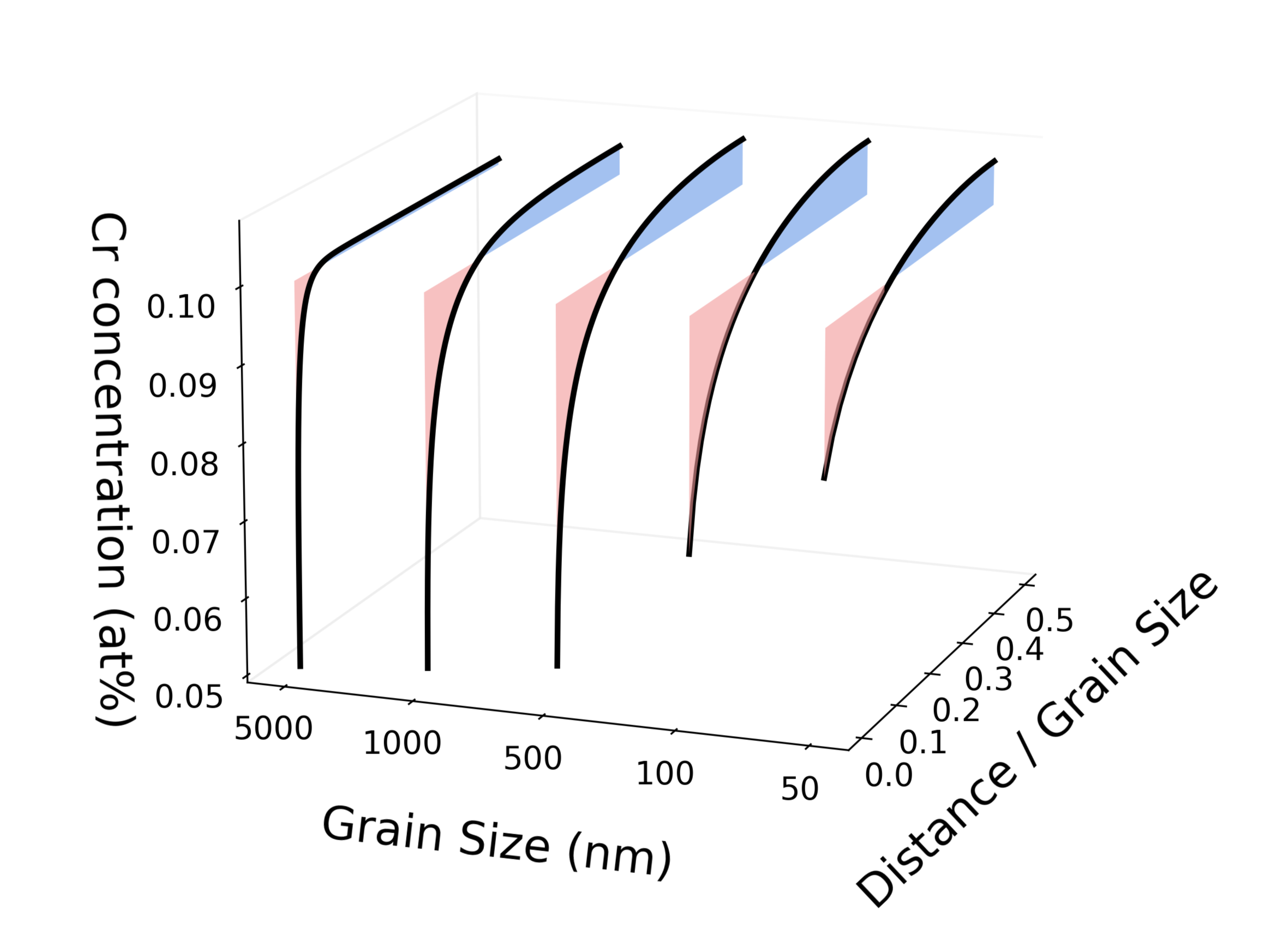}
    \caption{700 K}
    \label{fig:grain_700}
\end{subfigure}
\caption{Cr concentration profiles showing RIS penetration depth as a function of grain size at (a) 500 K (enrichment) and (b) 700 K (depletion). Results correspond to irradiation up to 10 dpa at a dose rate of $10^{-7}$ dpa/s under unbiased conditions.}
\label{fig:grain_variations}
\end{figure}

The effects of grain size and dislocation density on RIS behavior are studied at a dose rate of $10^{-7}$ dpa/s with grain sizes ranging from 50 nm to 5 $\mu$m. As before, production and absorption biases were excluded to isolate microstructural effects under unbiased conditions. Fig.~\ref{fig:grain_sink} illustrates the combined effects of grain size and dislocation density on RIS behavior at two representative temperatures: 500 K (enrichment regime as per the unbiased baseline case) and 700 K (depletion regime) and dislocation densities spanning $10^{-6}$ to $10^{-2}$ nm$^{-2}$. The results at 10 dpa show that smaller grains exhibit lower RIS magnitudes at both temperatures. However, the grain size effect becomes less pronounced at higher dislocation densities. At $10^{-2}$ nm$^{-2}$, the difference in RIS between 50 nm and 5 $\mu$m grains diminishes substantially as the total dislocation sink strength overwhelms the GB contribution.
Fig.~\ref{fig:grain_variations} presents visualizations of Cr concentration profiles extending from the grain boundary into the grain interior for both temperature regimes. While smaller grains exhibit lower absolute RIS magnitudes, segregation affects a larger fraction of the grain volume.
The underlying point defect concentration profiles are provided in Figs.~\ref{fig:grain_defect_500} and \ref{fig:grain_defect_700} of the Supplementary Material.)

\begin{figure}
\centering
\begin{subfigure}{0.6\textwidth}
    \includegraphics[width=\textwidth]{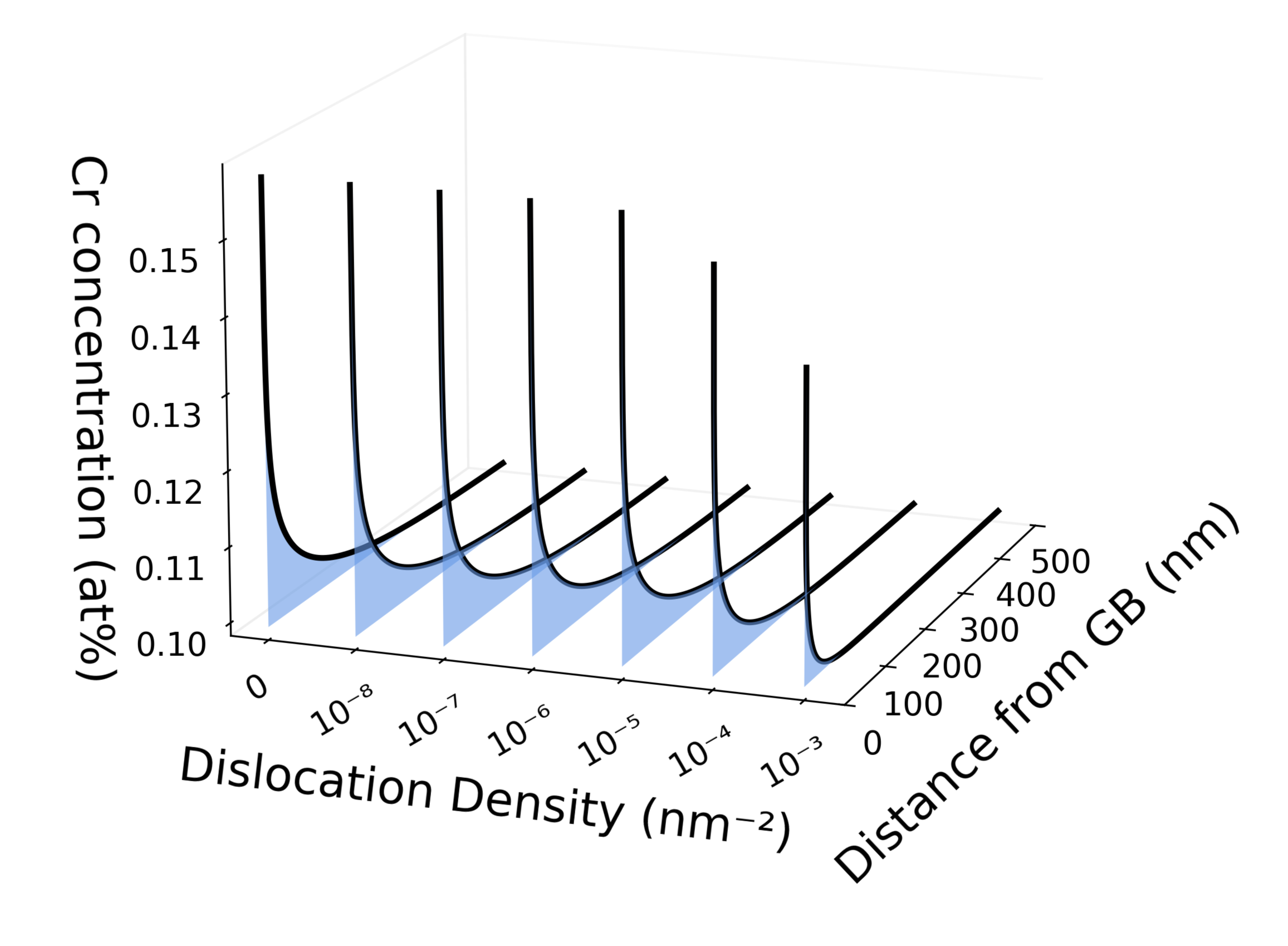}
    \caption{500 K}
    \label{fig:sink_500}
\end{subfigure}
\begin{subfigure}{0.6\textwidth}
    \includegraphics[width=\textwidth]{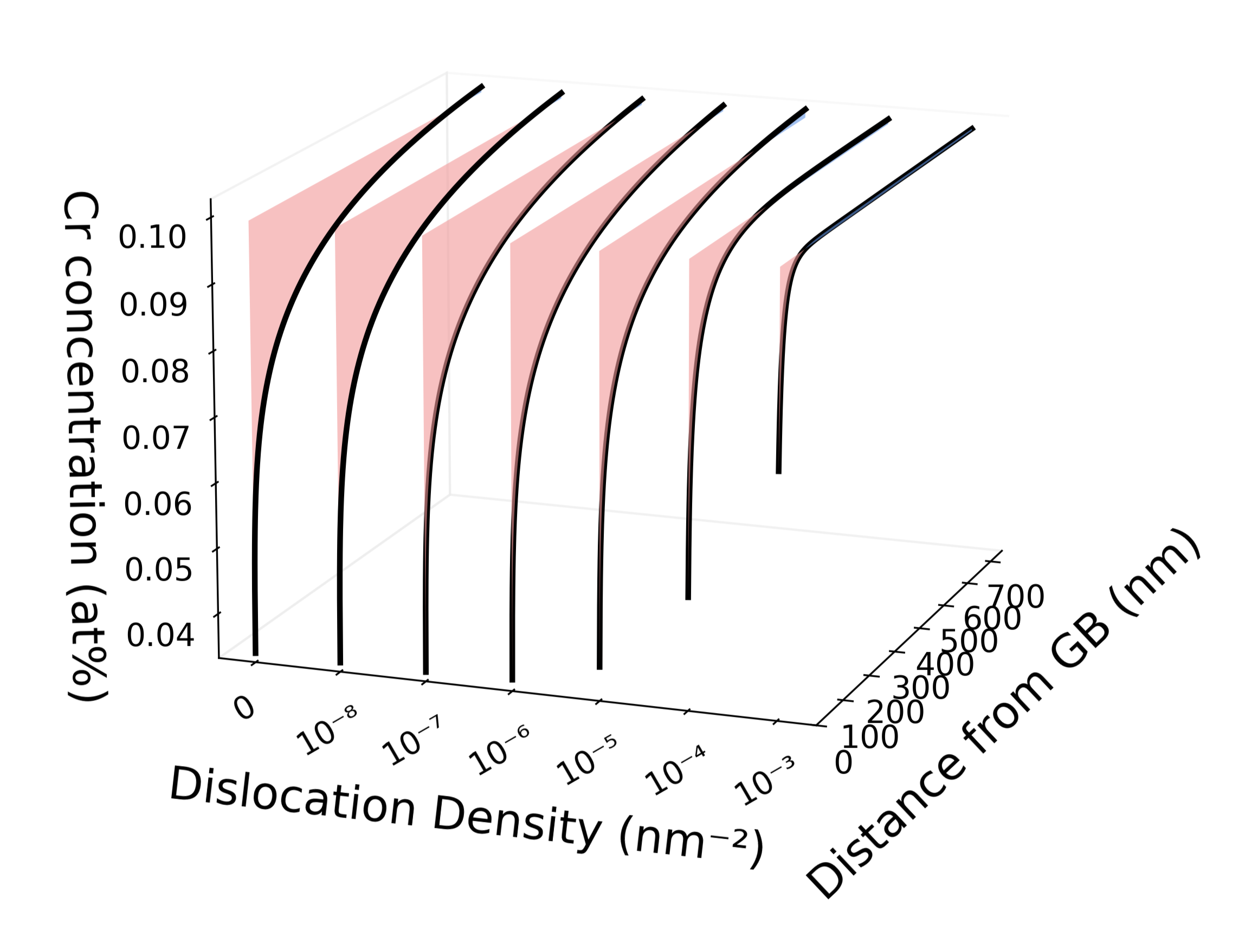}
    \caption{700 K}
    \label{fig:sink_700}
\end{subfigure}
\caption{Effect of dislocation density $\rho$ on Cr concentration profiles for a 5 $\mu$m grain at (a) 500 K and (b) 700 K. Results correspond to irradiation up to 10 dpa at a dose rate of $10^{-7}$ dpa/s under unbiased conditions.}
\label{fig:sink_variations}
\end{figure}

Fig.~\ref{fig:grain_sink} reveals that increasing dislocation density generally suppresses RIS magnitude at both temperatures, though the effect depends on grain size. This suppression occurs because dislocations act as additional point defect sinks distributed throughout the grain interior, capturing defects before they can reach GBs and reducing the net flux available to drive segregation. The magnitude of dislocation-induced suppression scales inversely with grain size. For 50 nm grains at 500 K, increasing dislocation density across the same range produces only minimal change in enrichment, indicating that GB sink strength already dominates point defect kinetics in these small grains. For 5 $\mu$m grains, where GB sink strength per unit volume is much lower, dislocations provide the dominant sink contribution at densities above $10^{-4}$ nm$^{-2}$.
Fig.~\ref{fig:sink_variations} provides additional detail on dislocation density effects across varying grain sizes at both 500 K and 700 K. The profiles confirm that dislocation density not only reduces RIS magnitude but also decreases the RIS width. At high dislocation densities ($>10^{-3}$ nm$^{-2}$), Cr profile widths are reduced to 5--10 nm due to steep point defect concentration gradients maintained by dislocation sinks distributed throughout the grain interior. (The point defect concentration profiles demonstrating the dislocation density effects on point defect distributions are presented in Figs.~\ref{fig:sink_defect_500} and \ref{fig:sink_defect_700} of the Supplementary Material.)

\subsection{Effect of production bias}

\begin{figure}
\centering
\begin{subfigure}{0.6\textwidth}
    \includegraphics[width=\textwidth]{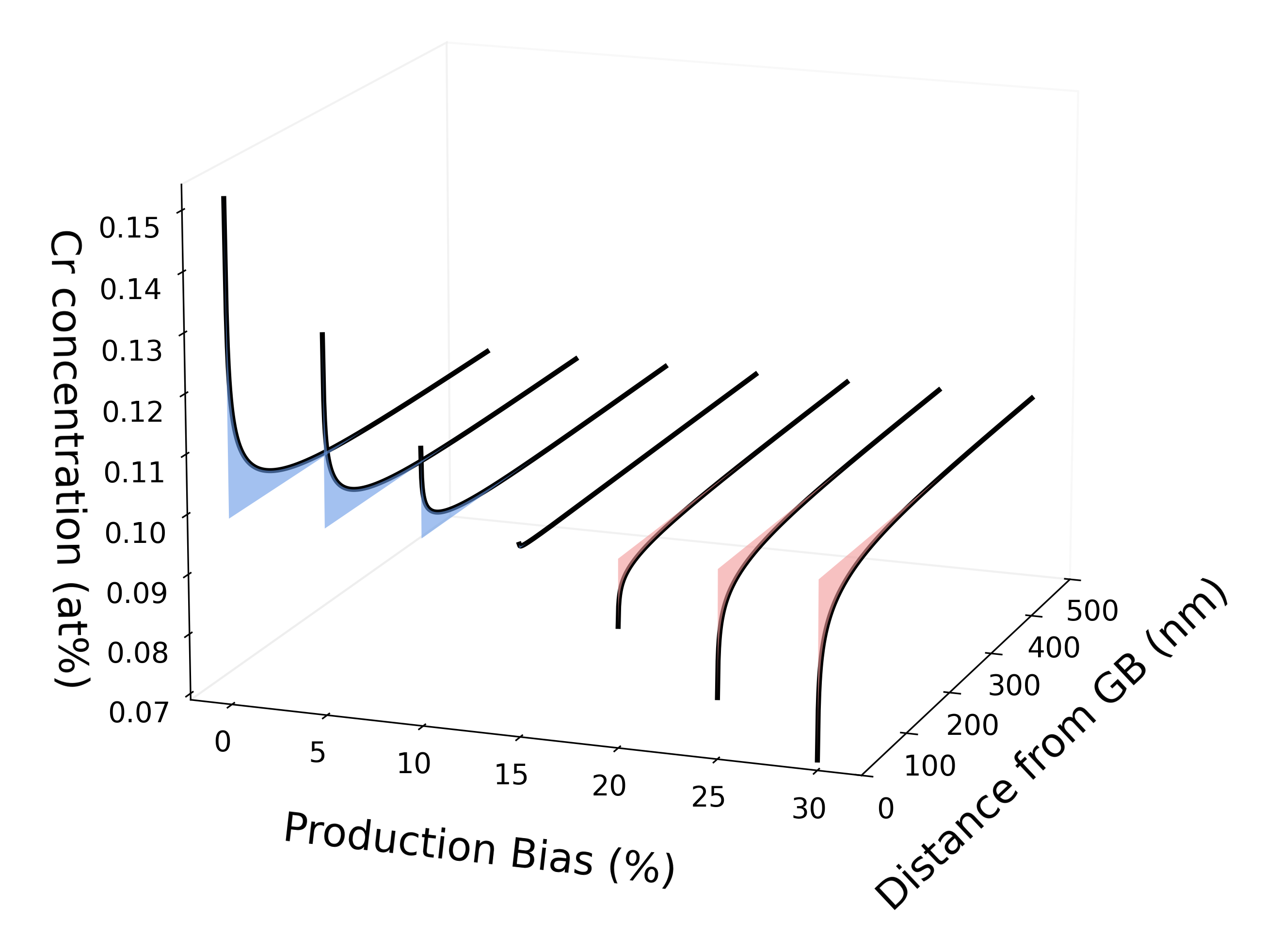}
    \caption{500 K}
    \label{fig:pb_500_hs}
\end{subfigure}
\begin{subfigure}{0.6\textwidth}
    \includegraphics[width=\textwidth]{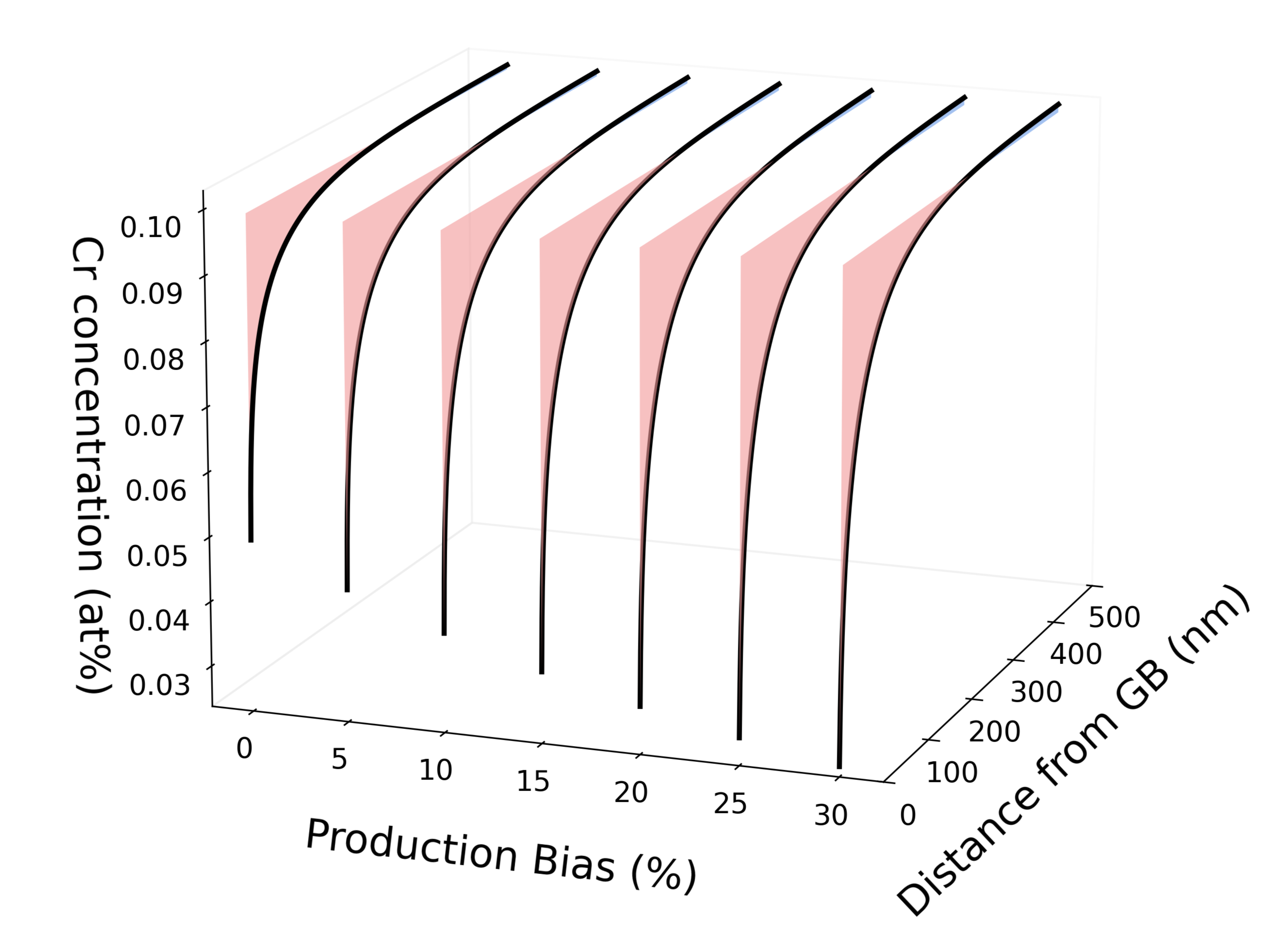}
    \caption{700 K}
    \label{fig:pb_700_hs}
\end{subfigure}
\caption{Impact of production bias (0--30\% excess vacancy production) on Cr concentration profiles at (a) 500 K and (b) 700 K for a 5 $\mu$m grain with dislocation density $10^{-4}$ nm$^{-2}$. }
\label{fig:production_bias_high_sink}
\end{figure}

To investigate the impact of production bias on RIS, simulations were performed at 500~K (enrichment regime) and 700~K (depletion regime) using a grain size of 5 $\mu$m, a dose rate of $10^{-7}$ dpa/s, and a dislocation density of $10^{-4}$ nm$^{-2}$. The production bias favoring vacancies was varied from 0\% (neutral production) to 30\% (vacancies produced at 1.3 times the SIA production rate).
Fig.~\ref{fig:pb_500_hs} presents the Cr concentration profiles at 500 K across varying production bias magnitudes. Introducing even modest production bias (5\%) toward vacancies reduces the magnitude of GB enrichment compared to the neutral production case (0\%). At 10--15\% bias, enrichment decreases further, approaching the nominal bulk concentration. Beyond $\sim$15\% bias, the system reverses completely to Cr depletion. This dramatic reversal---from $\sim$40\% enrichment under unbiased conditions to $\sim$20\% depletion under 30\% bias---demonstrates that production bias effects can override the RIS tendency predicted by Onsager transport coefficients alone. 
Fig.~\ref{fig:pb_700_hs} presents corresponding results at 700 K where unbiased conditions already favor Cr depletion. Introducing production bias toward vacancies amplifies this depletion. Unlike the enrichment regime where production bias induces reversal, here production bias simply enhances the existing vacancy-driven depletion mechanism.

\begin{figure}[htp!]
    \centering
     \includegraphics[scale=0.5]{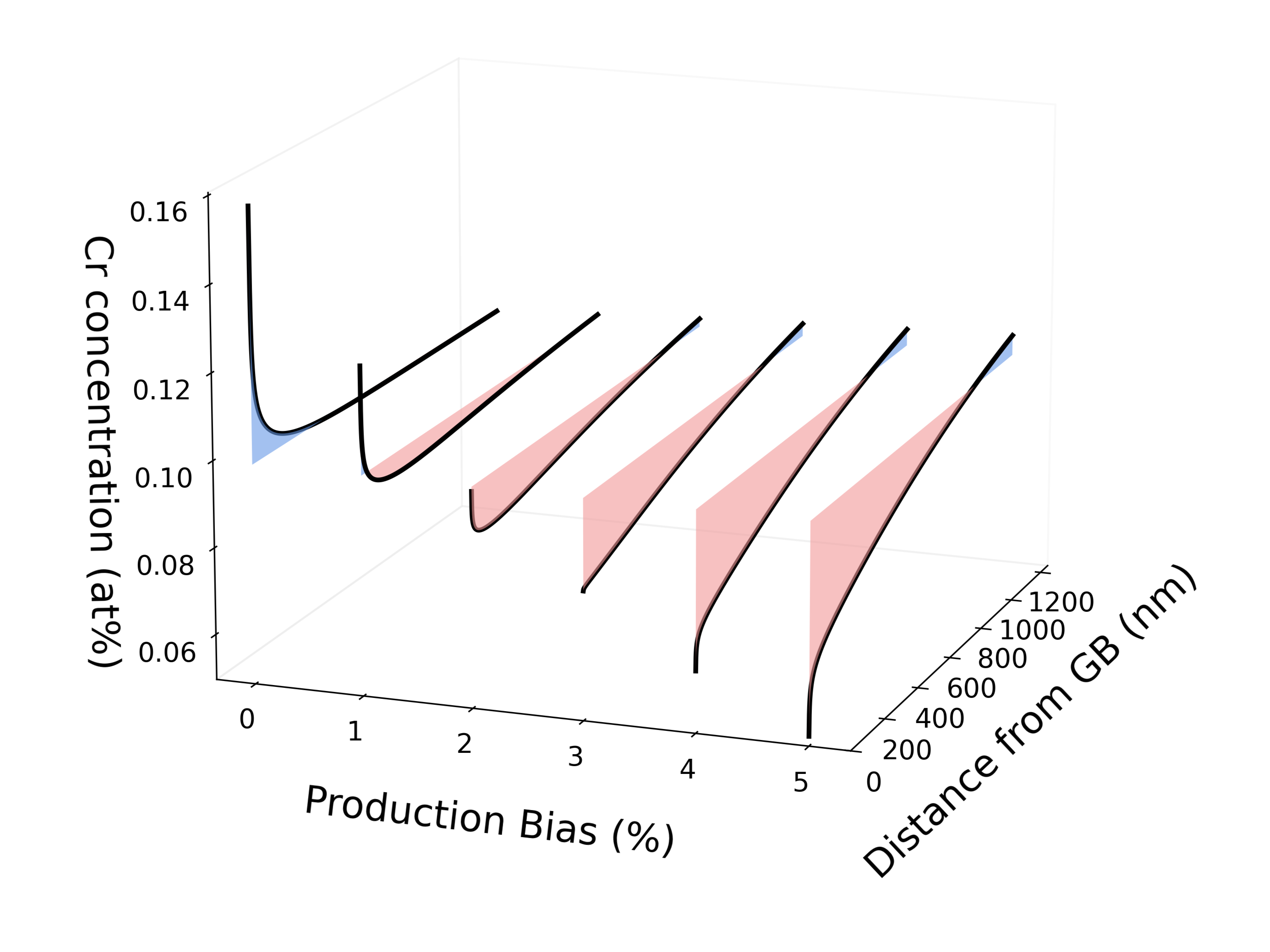}
     \caption{Cr concentration profiles at 500 K showing the effect of production bias (0--5\%) for a 5 $\mu$m grain with reduced dislocation density ($\rho = 10^{-6}$ nm$^{-2}$). Non-monotonic W-shaped profiles appear at intermediate bias levels (1--2\%).}
     \label{fig:pb_500_ls}
 \end{figure}

Fig.~\ref{fig:pb_500_ls} examines production bias effects at a reduced dislocation density of $10^{-6}$ nm$^{-2}$, two orders of magnitude lower than the previous cases. At this lower sink density, production bias effects become more pronounced: only 1--2\% bias proves sufficient to induce a detectable reduction in enrichment, and the segregation transition occurs at approximately 2\% bias rather than $\sim$15\% at $10^{-4}$ nm$^{-2}$.
Most significantly, the results reveal the development of non-monotonic Cr concentration profiles during the enrichment-to-depletion transition at this reduced sink density. 
{At production bias values between 1\% and 2\%, the spatial profiles exhibit non-monotonic ``W-shaped'' behavior with localized Cr enrichment near the GB, while Cr depletion occurs over a broader region extending 1000 nm from the GB into the grain interior. At approximately 3\% bias, the near-GB enrichment disappears, marking the transition to uniform Cr depletion profiles. Beyond 3\% bias, spatially uniform Cr depletion develops throughout the near-GB region.}

The point defect concentration profiles corresponding to the above results are presented in Figs.~\ref{fig:pb_defect_500_hs}, \ref{fig:pb_defect_700_hs} and \ref{fig:pb_defect_500_ls} of the Supplementary Material.
The contrasting sensitivity to production bias at different dislocation densities can be understood through the governing kinetic regime. At high dislocation density ($\rho = 10^{-4}$~nm$^{-2}$), the system operates in the sink-dominated regime where $D_V k_V^2 c_V \gg R_{VI}c_Vc_I$. Under these conditions, the steady-state vacancy-to-SIA concentration ratio for the biased case relative to that of the unbiased case scales linearly with $(1+\epsilon)$ (Figs.~\ref{fig:cV_cI_500_pb_hs} and~\ref{fig:cV_cI_700_pb_hs}). In contrast, at reduced dislocation density ($\rho = 10^{-6}$~nm$^{-2}$), the system shifts toward recombination-dominated kinetics. Here, a mere 5\% production bias amplifies the $c_V/c_I$ ratio by a factor of $\sim$2.75, far exceeding the factor of 1.05 predicted by the sink-dominated expression (Fig.~\ref{fig:cV_cI_500_pb_ls}). This nonlinear scaling explains the enhanced sensitivity of RIS to production bias at lower sink densities.

\subsection{Effect of absorption bias}

\begin{figure}
\centering
\begin{subfigure}{0.6\textwidth}
    \includegraphics[width=\textwidth]{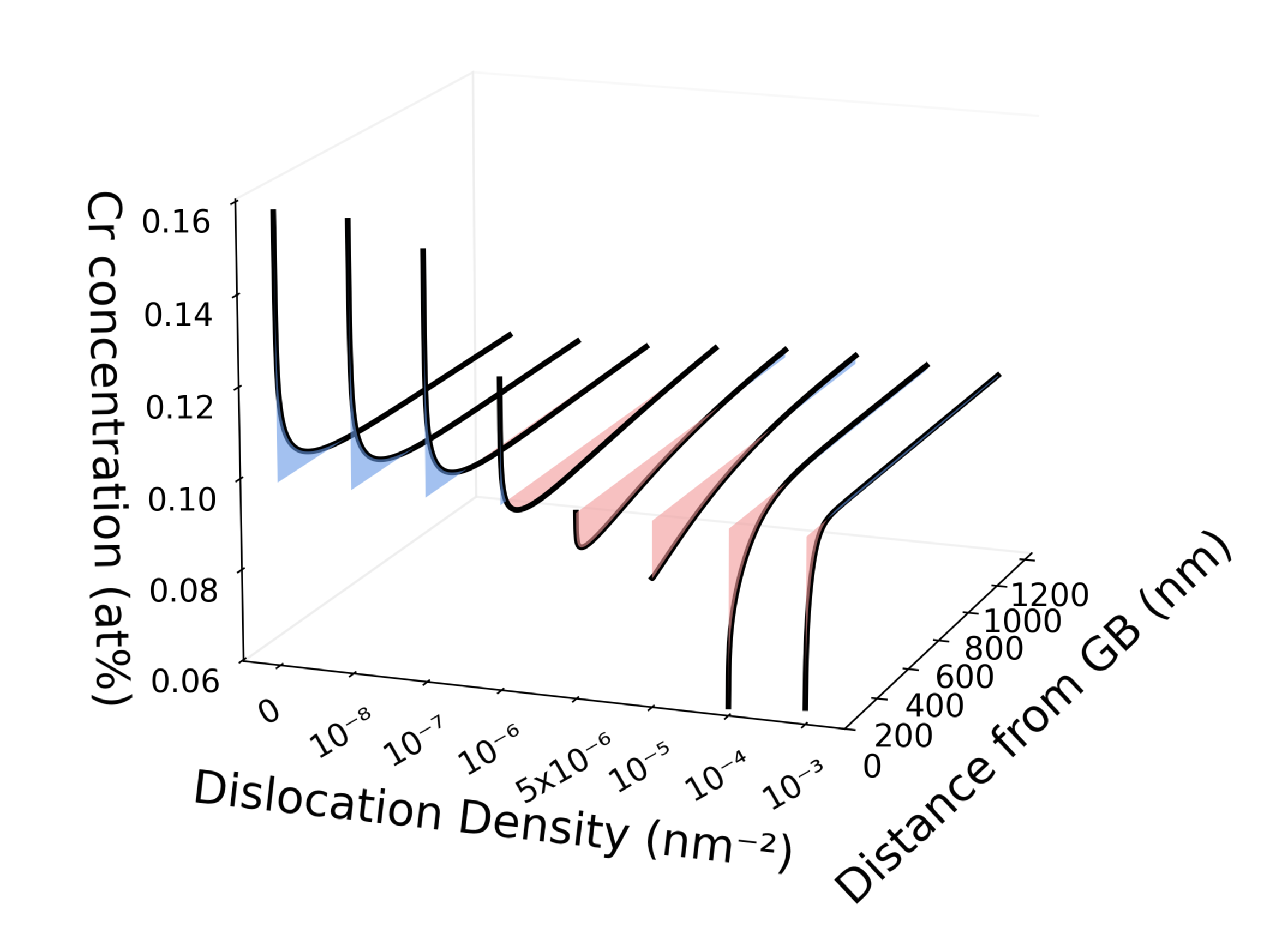}
    \caption{500 K}
    \label{fig:bias_sink_T500}
\end{subfigure}
\begin{subfigure}{0.6\textwidth}
    \includegraphics[width=\textwidth]{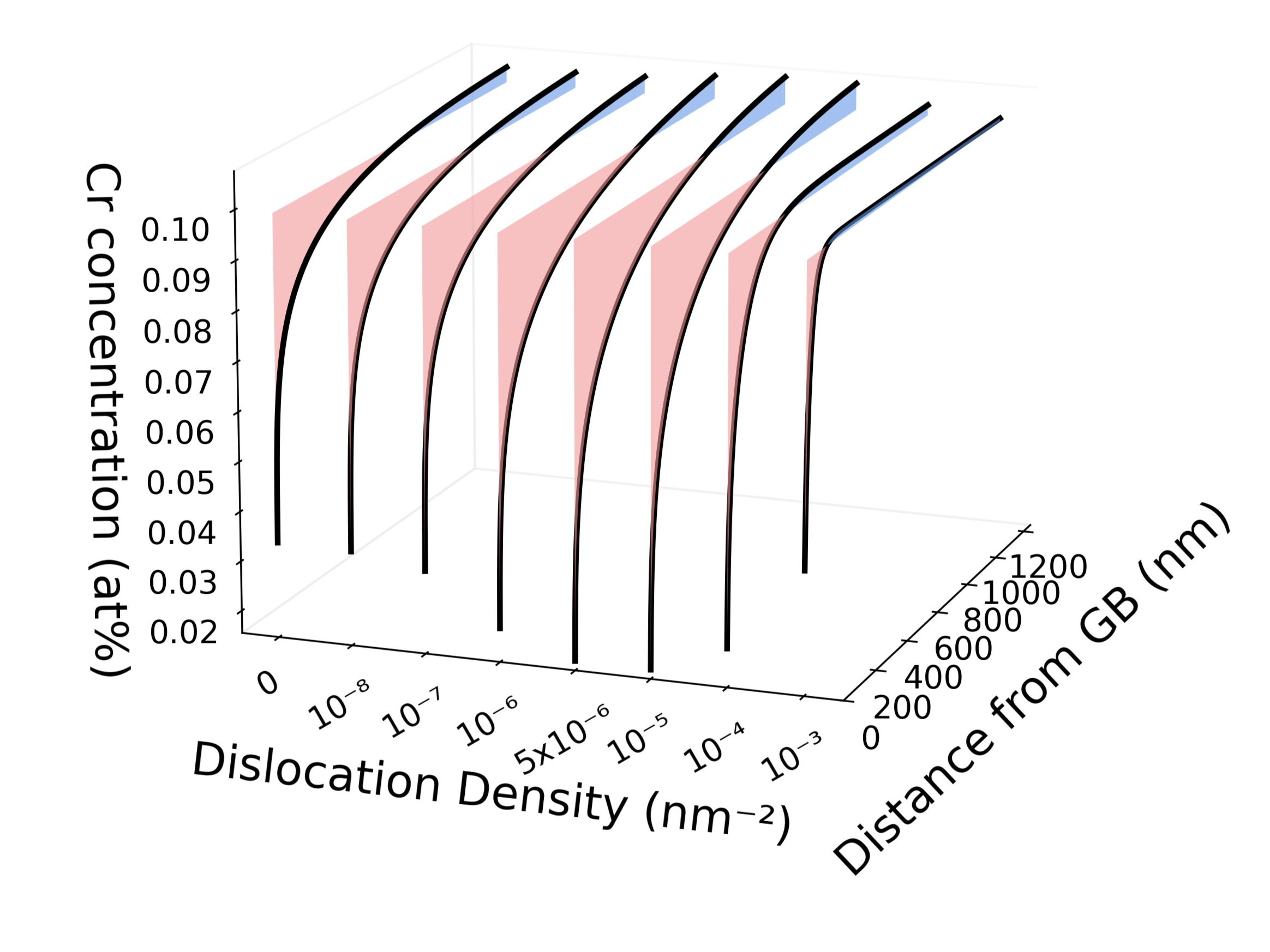}
    \caption{700 K}
    \label{fig:bias_sink_T700}
\end{subfigure}
\caption{Cr concentration profiles showing the effect of SIA absorption bias by dislocations as a function of dislocation density at (a) 500 K and (b) 700 K for a 5 $\mu$m grain.}
\end{figure}

To investigate the effects of absorption bias, the reduced order model from Kohnert et al.~\cite{kohnert_DDD} described in Eq.~\ref{eq:Z_bias} is utilized. As shown in Fig.~\ref{fig:sink_bias}, the absorption bias depends on both temperature and dislocation density. Simulations were performed at 500 K (enrichment regime as per baseline unbiased case) and 700 K (depletion regime) for a grain size of 5 $\mu$m and a dose rate of $10^{-7}$ dpa/s. Dislocation density was varied from $10^{-6}$ nm$^{-2}$ to $10^{-2}$  nm$^{-2}$, and production bias was excluded to isolate absorption effects.
Fig.~\ref{fig:bias_sink_T500} presents steady-state Cr concentration profiles at 500 K across varying dislocation densities. At the lowest dislocation density ($10^{-6}$ nm$^{-2}$), where absorption bias remains negligible, Cr enrichment magnitude is consistent with results from the unbiased geometric sink model shown in Fig.~\ref{fig:sink_info}. As dislocation density increases, Cr enrichment progressively diminishes due to the increasing absorption bias. 
With further increase in dislocation density, the system begins transitioning from enrichment toward depletion despite operating at 500 K where unbiased conditions strongly favor enrichment. Between $10^{-6}$ and $10^{-5}$ nm$^{-2}$, non-monotonic W-shaped profiles are observed, with localized Cr enrichment near the GB and a wide region of depletion further away from the GB and into the bulk grain. Beyond a dislocation density of $10^{-5}$ nm$^{-2}$, the system fully transitions to Cr depletion near the GB. Similar to production bias, absorption bias can induce RIS reversal at 500 K; however, the mechanism differs as preferential SIA capture by dislocations enhances the relative vacancy flux to the GB.

Fig.~\ref{fig:bias_sink_T700} presents steady-state Cr concentration profiles at the higher temperature of 700 K. At the lowest dislocation density ($10^{-6}$ nm$^{-2}$), Cr depletion is observed, with magnitude identical to that obtained for the unbiased geometric sink model in Fig.~\ref{fig:sink_700}. As dislocation density increases, Cr depletion is enhanced. 
This observation correlates with an increase in preferential SIA absorption by dislocations or bias (Fig.~\ref{fig:sink_bias}). 
This enhancement in depletion continues as dislocation density increases up to $5\times10^{-6}$ nm$^{-2}$. Beyond this the depletion decreases, correlating with the slight decrease in the sink capture efficiency beyond $5\times10^{-6}$ nm$^{-2}$ (Fig.~\ref{fig:sink_bias}). 
(The point defect concentration profiles corresponding to the above results are presented in Figs.~\ref{fig:sb_defect_500} and \ref{fig:sb_defect_700} of the Supplementary Material.)
The quantitative impact of absorption bias on point defect concentration can be understood by the steady-state vacancy-to-SIA concentration ratios. At dislocation densities $\rho \geq 10^{-3}$~nm$^{-2}$, the bias factors of 65--75\% (Fig.~\ref{fig:sink_bias}) yield $k_I^2/k_V^2 \approx 1.7$, increasing the bulk $c_V/c_I$ ratio by a corresponding factor relative to the unbiased case (Figs.~\ref{fig:cV_cI_500_sink} and \ref{fig:cV_cI_700_sink}).

\section{Discussion}

\subsection{Temperature-dependent transport mechanisms}

The simulation results elucidate how irradiation parameters, grain size, and dislocation density independently and collectively influence RIS in dilute Fe-Cr steel through distinct but coupled mechanisms. 
Under unbiased conditions where point defect production and absorption remain symmetric across defect types, RIS is determined by the transport mechanisms dominating at a given irradiation temperature. 
Vacancy-solute exchange dominates at elevated temperatures resulting in depletion of Cr, whereas preferential SIA-mediated transport dominates at lower temperatures resulting in Cr enrichment.
The origin of the temperature-dependent reversal lies in the competing PDC ratios shown in Fig.~\ref{fig:PDC_plot}. 
At temperatures below the crossover ($\sim$550 K), the SIA partial diffusion coefficient ratio ($D^I_{\text{pd}}$) exceeds the vacancy ratio ($D^V_{\text{pd}}$), indicating that SIA fluxes couple more strongly with Cr transport toward GBs than vacancy fluxes, resulting in Cr enrichment.
Above the crossover temperature, this relationship inverts: $D^V_{\text{pd}}$ exceeds $D^I_{\text{pd}}$ driving Cr depletion. 
This crossover represents a fundamental characteristic of the Fe-Cr system that remains independent of microstructural details, provided point defect fluxes remain balanced. 

The predicted crossover temperature of approximately 550 K aligns quantitatively with SCMF theory calculations \cite{messina_OnsTran,senninger_kmc}, supporting the consistency of our modeling approach.
{However, this temperature prediction for dilute Fe-0.1Cr is lower than experimental observations in commercial ferritic steels: 773--790 K (500--515\textdegree C) for HT9 ($\sim$12 at.\% Cr) and 873--973 K (600--700\textdegree C) for T91 ($\sim$9 at.\% Cr)~\cite{wharry2013systematic,jiao2018microstructure}. This discrepancy can be attributed primarily to the strong composition dependence of the transport mechanisms and their crossover. AKMC simulations by Senninger et al.~\cite{senninger_kmc} computed Onsager transport coefficient ratios across a range of Cr concentrations and demonstrated that the crossover temperature increases with Cr content in a non-monotonic manner: from $\sim$570 K at 0.25 at.\% Cr to $\sim$680 K at 1 at.\% Cr, reaching a maximum of $\sim$880 K at 5 at.\% Cr before decreasing slightly to $\sim$850 K at 10 at.\% Cr, while at 15 at.\% Cr no crossover was observed below 1500 K. Their predictions were reported to be in qualitative agreement with experimental observations in Fe-Cr alloys with 8--12 at.\% Cr~\cite{wharry2014mechanism}. With concentration-dependent Onsager coefficients from such atomistic calculations, our rate-theory framework can be readily extended to higher Cr concentrations relevant to commercial alloys. 
Accurate comparison with experimental RIS measurements in commercial steels would additionally require Onsager coefficients that account for the effects of impurities such as carbon~\cite{marquis2011systematic,hu2013effect} and other alloying elements present in these multicomponent systems.}

\subsection{Effect of dose rate, grain size and dislocation density}

In addition to temperature, several parameters control RIS magnitude and spatial characteristics without altering the fundamental segregation direction under symmetric flux conditions. 
Dose rate affects the point defect concentration supersaturation that drives segregation, with higher dose rates producing greater supersaturation levels at steady state.
Thus, the crossover temperature itself remains invariant across dose rates as seen in Fig.~\ref{fig:dose_rate}, confirming that temperature alone controls the segregation direction when fluxes are symmetric.
Grain size governs the effectiveness of GBs as point defect sinks and strongly influences RIS behavior. 
As shown in Figs.~\ref{fig:grain_500} and \ref{fig:grain_700}, larger grains exhibit greater RIS magnitudes (up to a certain grain size) because reduced GB sink density per unit volume allows higher point defect supersaturation, increasing the driving force for segregation.
Conversely, smaller grains show reduced RIS magnitudes because higher GB sink density increases point defect absorption rates, reducing point defect supersaturation and thereby defect-mediated fluxes to the GB.

Dislocation density influences RIS through competing mechanisms by providing additional sinks that compete with GBs for point defect absorption. As demonstrated in Figs.~\ref{fig:grain_sink} and \ref{fig:sink_variations}, increased dislocation density generally suppresses overall RIS magnitude, particularly in larger grains where GB sink density alone is insufficient to control point defect kinetics.
This suppression of RIS occurs because dislocations capture point defects before they can reach the GBs, reducing the supersaturation and the net flux available to drive segregation. 
However, the magnitude of this effect scales inversely with grain size: in smaller grains, GBs already provide high sink density, making additional dislocation sinks less influential. 
Critically, under symmetric flux conditions where dislocations absorb vacancies and SIAs equally, these parameters (dose rate, grain size and dislocation density) affect only the RIS magnitude and but not the fundamental enrichment versus depletion tendency, which remains temperature-controlled through the PDC ratio.

\subsection{Flux asymmetry due to production and absorption bias}

When point defect production exhibits bias toward vacancies---a scenario that atomistic simulations have shown to occur in BCC Fe under neutron irradiation where SIAs preferentially cluster---the initially symmetric flux balance becomes disrupted, fundamentally altering RIS. MD simulations~\cite{terentyev_FeCr_MD,zhang2017molecular} of cascade evolution in Fe and Fe-Cr reveal that free, unclustered vacancies are effectively produced in numbers up to 30\% higher than free SIAs following cascade collapse, as SIAs preferentially form clusters. 
This production asymmetry breaks the flux balance assumed in conventional RIS models, creating conditions where one defect type contributes disproportionately to solute transport. 
At 500 K where unbiased conditions predict enrichment, introducing vacancy production bias progressively reduces enrichment magnitude (Fig.~\ref{fig:production_bias_high_sink}).
This reversal occurs because enhanced vacancy production increases the vacancy flux to GBs. Since vacancies preferentially deplete Cr, the amplified vacancy contribution overcomes the SIA-induced enrichment that would otherwise dominate at this temperature.

RIS under biased conditions depends critically on the kinetic regime governing point defect reactions. In the dislocation sink-dominated regime (see Figs.~\ref{fig:cV_cI_500_pb_hs},~\ref{fig:cV_cI_700_pb_hs},~\ref{fig:cV_cI_500_sink}, and~\ref{fig:cV_cI_700_sink} in the Supplementary Material), the steady-state balance simplifies to $c_V/c_I = D_I k_I^2(1+\epsilon)(1+B)/(D_V k_V^2)$. The $c_V/c_I$ ratio, normalized with respect to the ratio ($D_I/D_V)$ for the unbiased case, increases linearly with production bias as $(1+\epsilon)$ or with dislocation absorption bias as $(1+B)$. However, at lower sink densities where recombination becomes significant, this linear relationship breaks down and the steady-state ratio exhibits significantly higher values. This transition to recombination-dominant regime causes even small production biases ($\sim$2--3\%) to reverse RIS at low dislocation density (Fig.~\ref{fig:pb_500_ls}), while substantially larger production biases ($\sim$15\%) are required at high dislocation density.
In this transition regime, non-monotonic ``W"-shaped profiles characterized by localized enrichment at the GB are followed by depletion farther away from the GB.

Complementary to production bias, absorption bias arises from the preferential capture of one point defect type by dislocations due to their distinct strain fields and resulting elastic interactions. 
DDD simulations \cite{kohnert_DDD} reveal that network dislocations in BCC Fe preferentially absorb SIAs over vacancies, with the absorption bias magnitude depending on point defect relaxation volume, dislocation density, and temperature.
This preferential absorption creates an effective imbalance in point defect fluxes similar to production bias.
Preferential SIA absorption enhances the relative vacancy flux to GBs, shifting segregation behavior toward vacancy-dominated depletion.
At 500 K, increasing dislocation density with accompanying absorption bias progressively reduces Cr enrichment magnitude (Fig.~\ref{fig:bias_sink_T500}). 
Similar to production bias, the absorption-enhanced vacancy flux eventually overwhelms the SIA-induced enrichment mechanism, leading to Cr depletion at high dislocation densities.
At 700 K (Fig.~\ref{fig:bias_sink_T700}), {absorption bias produces non-monotonic behavior: as dislocation density increases, the magnitude of Cr depletion initially increases (i.e. Cr concentration at the GB decreases progressively from its nominal value) before eventually decreasing in depletion magnitude. This non-monotonicity arises from competing mechanisms: reduced defect supersaturation (from increased sink density) versus flux asymmetry (preferential SIA absorption by dislocations increases the vacancy-to-SIA flux to GBs). At densities up to $\sim 5 \times 10^{-6}$~nm$^{-2}$, flux asymmetry dominates and amplifies the depletion magnitude; beyond this, reduced defect supersaturation prevails and depletion magnitude decreases. This contrasts sharply with the unbiased case (Figs.~\ref{fig:sink_500},~\ref{fig:sink_700}), where increasing dislocation density monotonically suppresses RIS magnitude at both temperatures (trend commonly expected from conventional RIS predictions) and demonstrates that absorption bias fundamentally alters the relationship between sink density and segregation behavior.}

\subsection{Modeling framework and scope for experimental validation}

Several assumptions are inherent in our model and its parameters. The dilute limit approximation (0.1 at.\% Cr) of the Onsager transport coefficients restricts applicability to dilute Fe-Cr compositions. Concentration-dependent kinetic correlations are expected to become significant at higher Cr contents ($>$5 at.\%), which are typical of commercial ferritic steels such as HT9 and T91.
Concentration-dependent Onsager coefficients, such as those obtained from AKMC or MD calculations of Fe-Cr, are necessary. Onsager coefficient-based models, like the one proposed in our study, would benefit from such Onsager coefficients being made available in the form of published data sets or fitted expressions.
Our choice of dilute Fe-0.1Cr composition provides a cleaner system for investigating intrinsic RIS mechanisms. 
At higher, non-dilute compositions, the occurrence of Cr-rich precipitates or clusters add complexity to the interpretation of RIS profiles~\cite{aydogan2019alpha,kuksenko2013cr}. 
Additionally, our model simulated an ideal GB by imposing equilibrium point defect concentrations via Dirichlet boundary conditions. Different GB types, such as high-angle random GBs and coherent $\Sigma$3 GBs, experimentally exhibit vastly different segregation magnitudes due to distinct sink behaviors. Future work could employ Robin boundary condition, which relates the flux to the local concentration deviation from equilibrium, to capture GB-structure-dependent sink efficiencies~\cite{gu2017point}.
The phase-field framework developed by Kadambi et al.~\cite{kadambi_pf_RIS_FeCrNi} implements a diffuse-interface sink formulation that has been verified against sharp-interface Robin boundary conditions, enabling non-ideal sink behavior to be captured through adjustable sink strength parameters. 
Our current model can be directly extended, using the phase-field approach or Robin boundary conditions formulation, to incorporate GB structure dependence in Fe-Cr systems. 
Such an extension would enable systematic investigation of how GB misorientation and character influence Cr RIS magnitude, addressing the experimentally observed differences between high-angle random boundaries and coherent $\Sigma3$ boundaries~\cite{field2013dependence,rahmouni2024radiation}.

Direct quantitative comparison with experimental RIS measurements in Fe-Cr alloys is complicated by several other factors. 
Carbon and nitrogen impurities, present even in high-purity model alloys, interact with Cr to form carbides and nitrides that obscure intrinsic RIS profiles~\cite{marquis2011systematic}. 
Recent APT studies have revealed complex W-shaped Cr profiles arising from Cr-C co-segregation and precipitate formation at GBs under irradiation~\cite{wang2017carbon}. 
Furthermore, ion irradiation experiments are susceptible to carbon contamination from cracking of hydrocarbons in the vacuum system and on sample surfaces, and  uptake into the sample extending to 1~$\mu$m or more in depth~\cite{was2017carbon}.
Future experiments employing plasma cleaning and liquid nitrogen cold traps can mitigate this contamination~\cite{was2017carbon}, enabling direct comparison with our modeling predictions.

Our model isolates the fundamental competition between vacancy-mediated and SIA-mediated transport mechanisms towards RIS.
The asymmetry in point defect fluxes arising from production bias and absorption bias is a physical reality of radiation damage, and not a modeling artifact. Production bias emerges from the fundamental physics of displacement cascade collapse, where SIAs preferentially cluster while vacancies remain as freely migrating point defects~\cite{woo1992production,osetsky2000structure}. Similarly, absorption bias results from the elastic interaction between point defects and dislocation strain fields, governed by the distinct relaxation volumes of vacancies and SIAs~\cite{singh1997aspects,golubov2000defect}. The resulting asymmetric flux thus constitutes a primary driving force for microstructural evolution under irradiation, including void swelling, loop formation, and RIS.

When significant asymmetries in point defect fluxes ($J_V \ne J_I$) exist, the preferential annihilation of one defect type (e.g., SIAs at dislocations) would physically result in the net creation or destruction of lattice sites or planes, causing the crystal lattice itself to move relative to a fixed observer.
Our present work utilizes the classical RIS model formulated on a stationary lattice frame, which has remained the standard theoretical framework for RIS prediction since its establishment nearly five decades ago~\cite{wiedersich1979theory,marwick1978segregation}. 
This standard rate-theory-based formulation implicitly assumes that RIS occurs on timescales shorter than those governing broader microstructural evolution processes such as GB motion, void nucleation and growth, or dislocation loop evolution~\cite{sizmann1978effect,marwick1978segregation}. Under this assumption, the lattice can be treated as stationary because these slower processes do not significantly progress during the timescales relevant to RIS profile development. This separation of timescales is consistent with the original rate-theory RIS formulations~\cite{wiedersich1979theory,marwick1978segregation} and remains valid when interface motion follows interface-controlled kinetics rather than diffusion-controlled kinetics~\cite{magri2020coupled}.
However, the stationary lattice approximation could have potential limitations when applied to strongly biased conditions~\cite{fischer2010substitutional}. Specifically, it neglects the secondary, advective contribution to solute transport due to lattice movement~\cite{marwick1978segregation,gheno2022simulation}.
An important consequence (see Appendix) of the classical RIS formulation is the prediction of Cr depletion even in theoretical test cases lacking any preferential transport or differences in elemental diffusivities. The result arises from the enforcement of lattice site conservation in the presence of a net defect flux without allowing for the corresponding physical lattice movement. This limitation inherent in the classical formulation needs to be further investigated.
The predictions presented here for bias-driven or flux asymmetry effects are expected to be most accurate at shorter irradiation times before significant microstructural evolution occurs, or in systems where interface motion remains interface-controlled.

{Several simplifications are inherent in our treatment of dislocations as sinks for point defects. Firstly, while the DDD simulations~\cite{kohnert_DDD} generated network geometries from stress-induced evolution of dislocation loops, the extent to which these represent dislocation microstructures that evolve under irradiation (where defect clusters nucleate, transform into loops, and integrate into networks) remains uncertain. Secondly, whether the predicted SIA absorption bias for network dislocations (65--75\% in Fig.~\ref{fig:sink_bias}) applies to small prismatic loops formed during irradiation has not been unestablished. Thirdly, the mean-field homogenization does not capture the local variations in capture efficiency ($\pm$10--15\%)~\cite{kohnert_DDD}, effects of edge versus screw character on bias, or modifications to sink strength from RIS at dislocation cores~\cite{huang2025}. Here, Cr segregation at dislocations due to RIS and/or equilibrium segregation would both alter Cr available in the matrix and compete with RIS to GBs. Finally, dynamic dislocation evolution (loop nucleation, growth, unfaulting, network incorporation) is not treated. Capturing coupled effects of mechanical strain and compositional alterations on absorption bias, along with evolving sink microstructure, should be the focus of future mesoscale RIS modeling efforts.}

Developing a generalized RIS model that fully addresses these limitations presents substantial theoretical and computational challenges~\cite{golubov20121}. Such a model would need to simultaneously account for lattice motion in a laboratory reference frame, void and dislocation loop nucleation and growth, and the coupled evolution of multiple extended defect populations. The advective flux terms necessary to capture lattice motion effects must be consistently coupled with sink evolution equations, and the resulting system becomes significantly more complex than the classical formulation~\cite{marwick1978segregation,fischer2010substitutional,gheno2022simulation}. A complete treatment represents a significant undertaking beyond the scope of the present work. Future efforts may address these aspects systematically to improve model fidelity, building upon the physics-based framework established in this work.

\section{Conclusions}

This study presents a comprehensive physics-based rate-theory framework for predicting RIS in dilute (0.1 at.\% Cr) ferritic Fe-Cr alloy by systematically incorporating production and absorption bias effects alongside transport mechanisms. 

\begin{enumerate}
    \item Under conditions where point defect fluxes remain symmetric, the crossover in Cr RIS from enrichment to depletion is governed by the competition between SIA and vacancy transport mechanisms. The crossover temperature is found to be approximately 550 K for dilute Fe-0.1Cr.
    \item Irradiation dose rate and microstructural parameters---grain size and dislocation density---significantly influence the magnitude and spatial extent of RIS but do not alter the temperature-dependent segregation direction under symmetric flux conditions. Higher dose rates shift the temperature of maximum depletion to higher temperatures, while larger grain sizes exhibit greater RIS magnitudes and increased dislocation densities suppress RIS.
    \item Breaking the symmetry of point defect fluxes through either production bias or absorption bias fundamentally alters RIS behavior beyond what partial diffusivity differences predict. Production bias favoring vacancies and preferential absorption of SIAs by dislocations, both enhance vacancy-mediated chromium depletion, effectively shifting the enrichment-to-depletion transition to lower temperatures. 
    \item Within the classical RIS modeling framework in the lattice frame of reference, both bias mechanisms induce RIS even when Onsager transport coefficients are identical (Appendix), demonstrating that flux asymmetry alone can drive RIS. The predictions presented here for bias-driven or flux asymmetry effects are expected to be most accurate at shorter irradiation times before significant microstructural evolution occurs. They also apply to systems where interface motion remains interface-controlled rather than diffusion-controlled.
\end{enumerate}

Our computational findings indicate that reliable prediction of RIS requires accounting for the full coupling between transport coefficients, point defect flux asymmetries from both production and absorption biases, and microstructural factors. The physics-based rate-theory RIS framework, when validated against experimental predictions, can provide a foundation for understanding RIS behavior and guidance for designing radiation-tolerant ferritic alloys for advanced nuclear energy systems. Future work should focus on extending this framework to higher Cr concentrations relevant to commercial ferritic steels and validating predictions against systematic experimental measurements.

\appendix
\section{RIS driven solely by flux asymmetry}
\begin{figure}[htp!]
    \centering
     \includegraphics[scale=0.45]{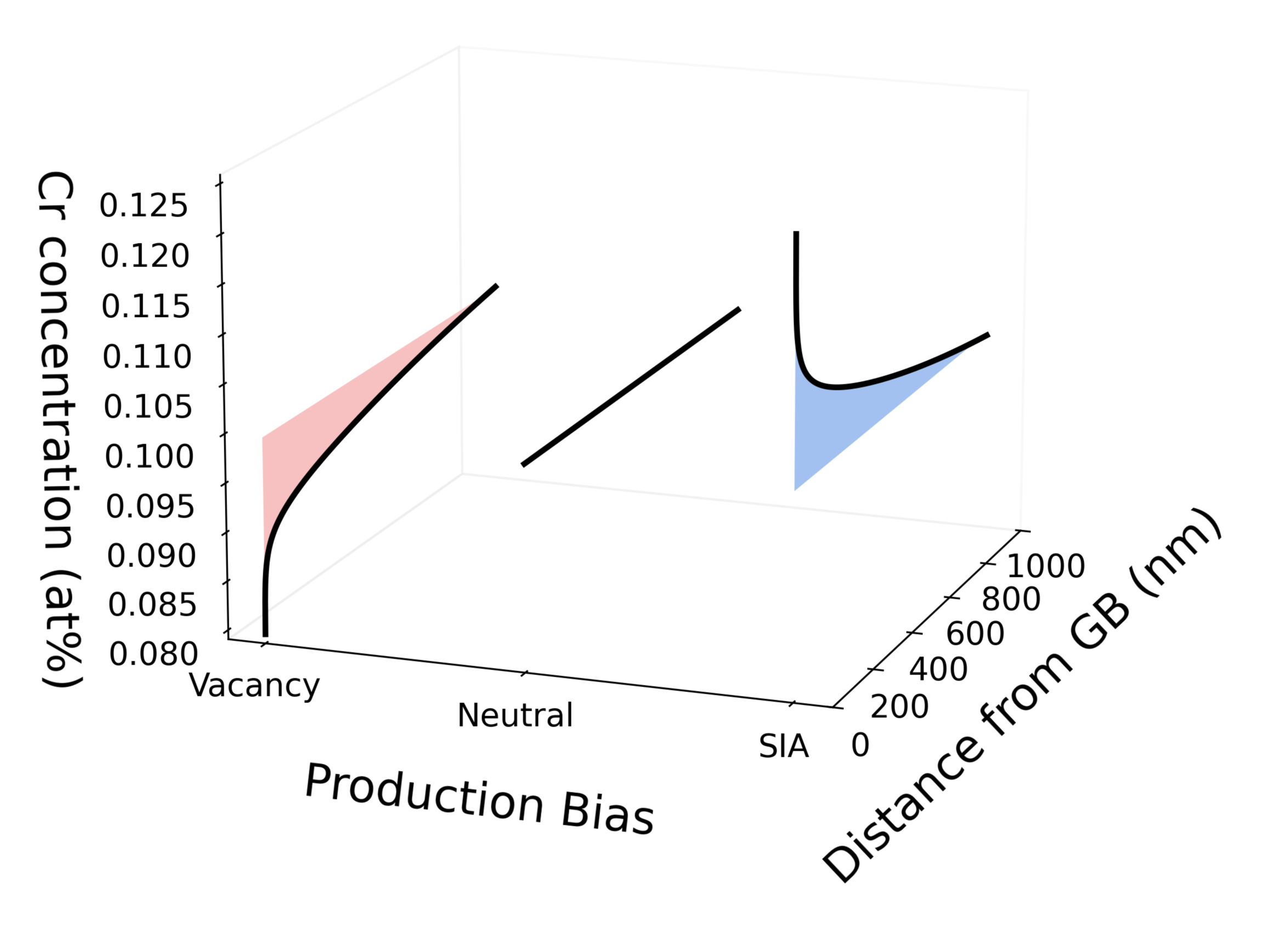}
     \caption{Steady-state Cr concentration profiles demonstrating RIS when PDC ratios are one (i.e., $D^V_{\text{pd}}=D^I_{\text{pd}}=1$) for various cases of production bias: neutral or 0\% bias, {3\% bias in favor of SIAs, and 3\% bias in favor of vacancies.} }
     \label{fig:prod_bias_append}
 \end{figure}

To isolate the effect of flux asymmetry created by production bias from transport coefficient  (diffusivity) differences, simulations were performed for a theoretical test case where the Onsager transport coefficients were set such that $D^V_\mathrm{pd}=D^I_\mathrm{pd}$. This was achieved by replacing the Onsager transport coefficients of the SIA defects with those of the vacancies ensuring both had the same diffusivity. This case ensures no preferential coupling between either point defect type and the solute. Simulations were performed at a dose rate of $10^{-7}$ dpa/s, temperature of 500 K on a system with grain size of 5 $\mu$m and dislocation density of $10^{-6}$ nm$^{-2}$. The production bias was varied from 3\% in favor of vacancies ($\epsilon=0.03$) to neutral (0\%) to 3\% in favor of SIAs ($\epsilon=-0.03$, corresponding to excess SIA production).
{The 3\% production bias was selected as it represents the threshold where the enrichment-to-depletion transition completes without non-monotonic profiles (Fig.~\ref{fig:pb_500_ls}). 
Equal-magnitude of bias in the opposite directions ($\pm$3\%) demonstrate that the bias direction alone determines segregation tendency when transport coefficients are identical.}
Fig.~\ref{fig:prod_bias_append} shows that no RIS is observed for the neutral case, as expected when both transport coefficients and point defect fluxes and symmetric. However, when a production bias is applied, Cr depletion occurs when vacancies are favored in point defect production rate and enrichment occurs when SIAs are favored.
This segregation arises because the biased production creates an imbalance in point defect fluxes to the GB. That is, when vacancies are produced in excess, a net vacancy flux develops toward the GB, and the resulting lattice site conservation requirement drives Cr away from the GB leading to depletion. Conversely, when SIAs are produced in excess, the net SIA flux toward the GB drags Cr along leading to enrichment. 
{The segregation direction depends on which defect type is favored; that is, excess vacancies drive Cr depletion, while excess SIAs drive Cr enrichment, with bias magnitude controlling only the extent of segregation.}
 We note that similar trends are obtained when the flux asymmetry is driven by absorption biases, confirming that flux asymmetry alone can drive RIS within the classical modeling framework.

\section*{Supplementary Material}
This supplementary material presents the Onsager transport coefficient data used in the model, as well as the point defect concentration profiles corresponding to each of the RIS results shown in the main text.

\section*{ACKNOWLEDGMENTS}

This work was supported by the Laboratory-Directed Research and Development (LDRD) program of Idaho National Laboratory (INL) under the project 23A1070-092FP.
K.A. acknowledges the support from the U.S. Nuclear Regulatory Commission (NRC) under award agency number 31310024M0002 at Texas A\&M University. 
This research made use of Idaho National Laboratory's High Performance Computing systems located at the Collaborative Computing Center and supported by the Office of Nuclear Energy of the U.S. Department of Energy and the Nuclear Science User Facilities under Contract No. DE-AC07-05ID14517. 

\bibliography{aip_references.bib}

\clearpage
\newpage
\setcounter{section}{0}
\setcounter{page}{1}
\setcounter{figure}{0}
\setcounter{equation}{0}
\renewcommand{\appendixname}{}
\renewcommand{\thesection}{S\arabic{section}}
\renewcommand{\theequation}{S\arabic{equation}}
\renewcommand{\thepage}{S\arabic{page}}
\renewcommand{\thetable}{S\arabic{table}}
\renewcommand{\thefigure}{S\arabic{figure}}
\numberwithin{equation}{section}

\begin{center}
  \large \textbf{Supplementary Material} \\
  \vspace{0.5cm}

\large \textbf{Radiation-induced segregation in dilute Fe-Cr: A rate-theory framework for the Cr enrichment-depletion transition at the grain boundary} \\
  \vspace{0.5cm}
  \normalsize Russell Oplinger \textsuperscript{1,2}, Mukesh Bachhav \textsuperscript{1}, Karim Ahmed \textsuperscript{2}, Sourabh Bhagwan Kadambi \textsuperscript{1} \\
  \vspace{0.2cm}
  \textsuperscript{1}\textit{Idaho National Laboratory, Idaho Falls, Idaho 83415, USA} \\
  \textsuperscript{2}\textit{Texas A\&M University, College Station, Texas 77843, USA} \\
  \vspace{0.2cm}
  \today
\end{center}

This supplementary section presents the Onsager transport data sourced from Messina et al\cite{messina_OnsTran} that is used by the model as well as the underlying vacancy and SIA concentration profiles that develop under irradiation corresponding to each of the Cr RIS plot for the Fe-0.1Cr system presented in the main paper.
The shaded regions represent the supersaturation in the defect concentration over the thermal equilibrium value.

\FloatBarrier
\section{Onsager Transport Coefficients}

The temperature dependent Onsager transport coefficients for both the interstitial and vacancy point defects is shown below in Fig. \ref{fig:ons_coeff}. Note that the FeFe transport coefficients shown in Fig. \ref{fig:ons_vac} were multiplied by -1 to facilitate plotting and presentation. All Onsager transport data was obtained from the work of Messina et al\cite{messina_OnsTran}.

\begin{figure}[ht!]
\centering
\begin{subfigure}{0.49\textwidth}
    \includegraphics[width=\textwidth]{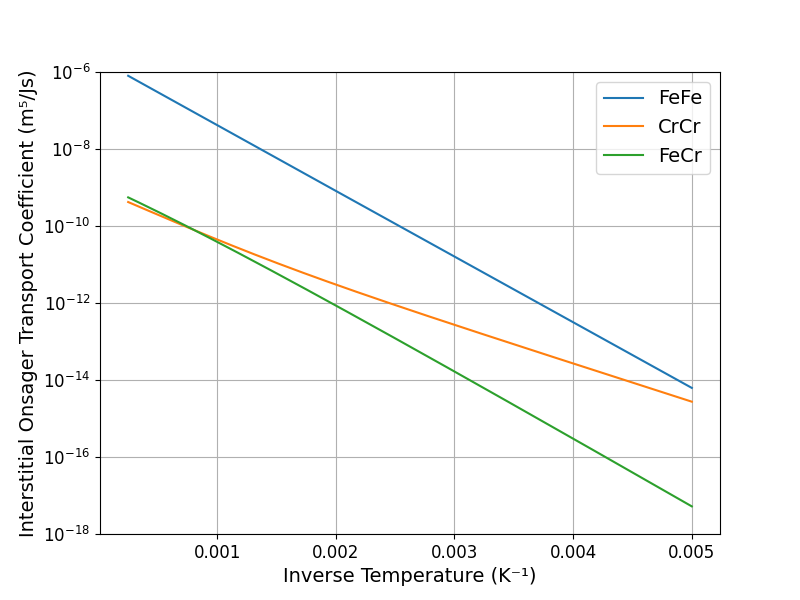}
    \caption{SIA transport coefficients}
    \label{fig:ons_sia}
\end{subfigure}
\begin{subfigure}{0.49\textwidth}
    \includegraphics[width=\textwidth]{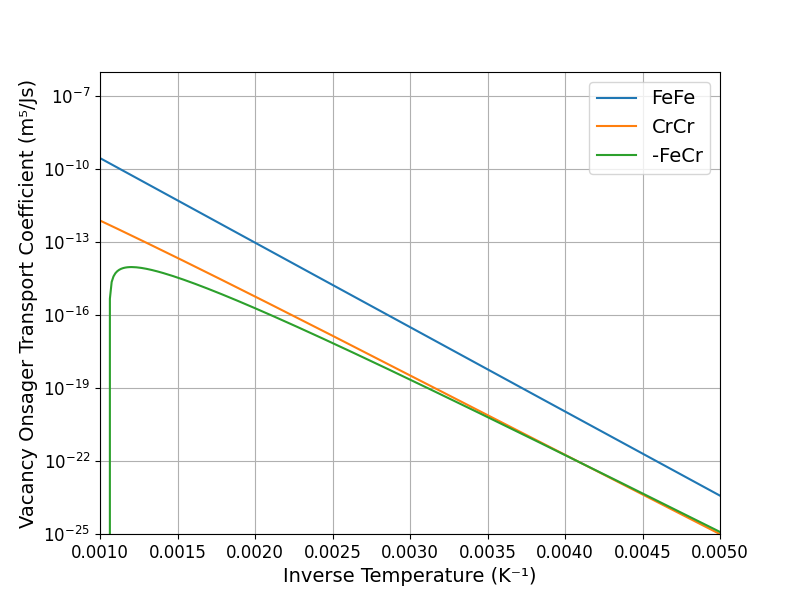}
    \caption{Vacancy transport coefficients}
    \label{fig:ons_vac}
\end{subfigure}
\caption{Temperature dependent Onsager transport coefficients for both point defect types obtained from Messina et al\cite{messina_OnsTran}.}
\label{fig:ons_coeff}
\end{figure}

\section{Temperature dependence}

\begin{figure}[ht!]
\centering
\begin{subfigure}{0.49\textwidth}
    \includegraphics[width=\textwidth]{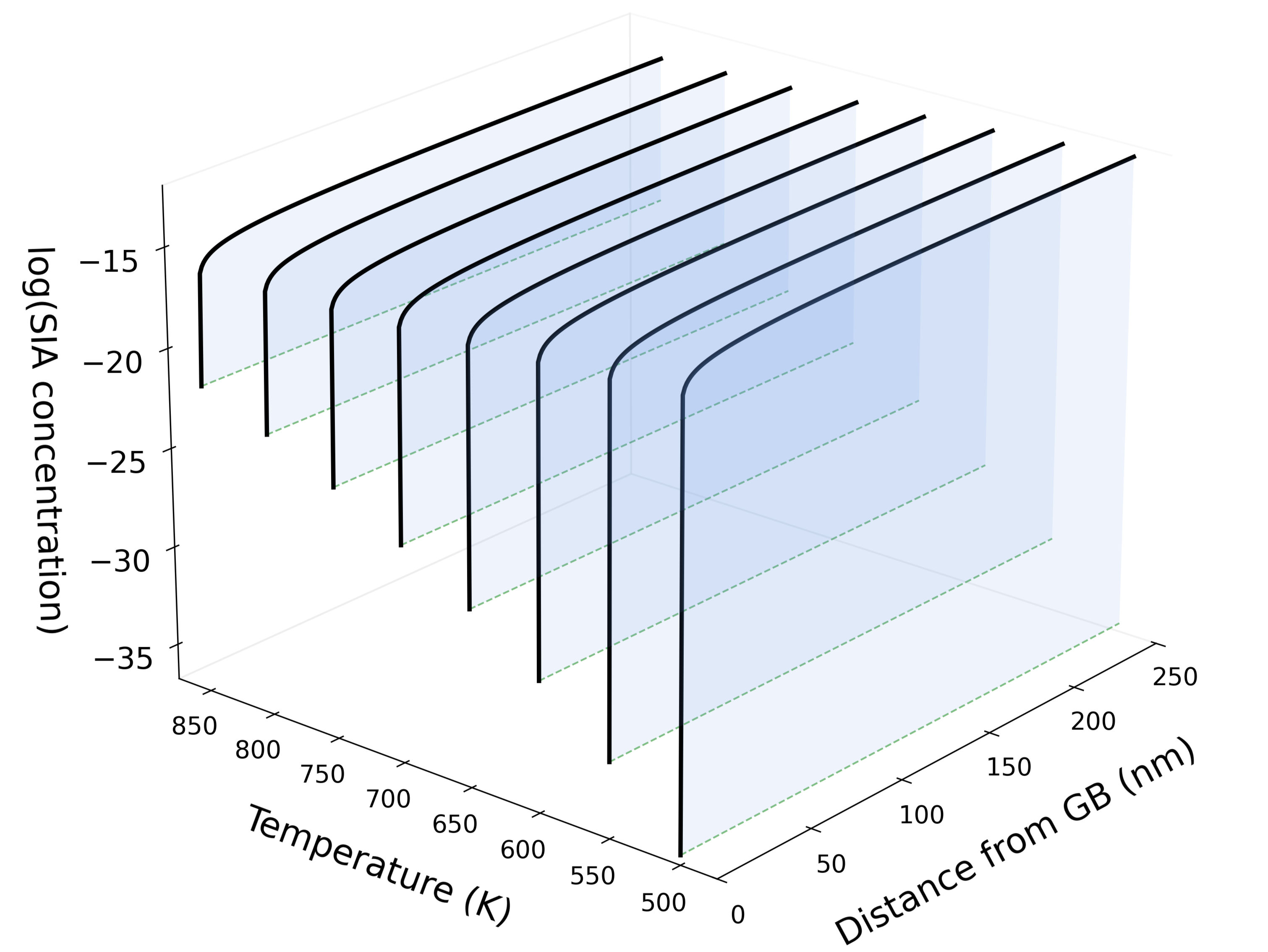}
    \caption{SIA concentration profiles}
    \label{fig:temp_sia}
\end{subfigure}
\begin{subfigure}{0.49\textwidth}
    \includegraphics[width=\textwidth]{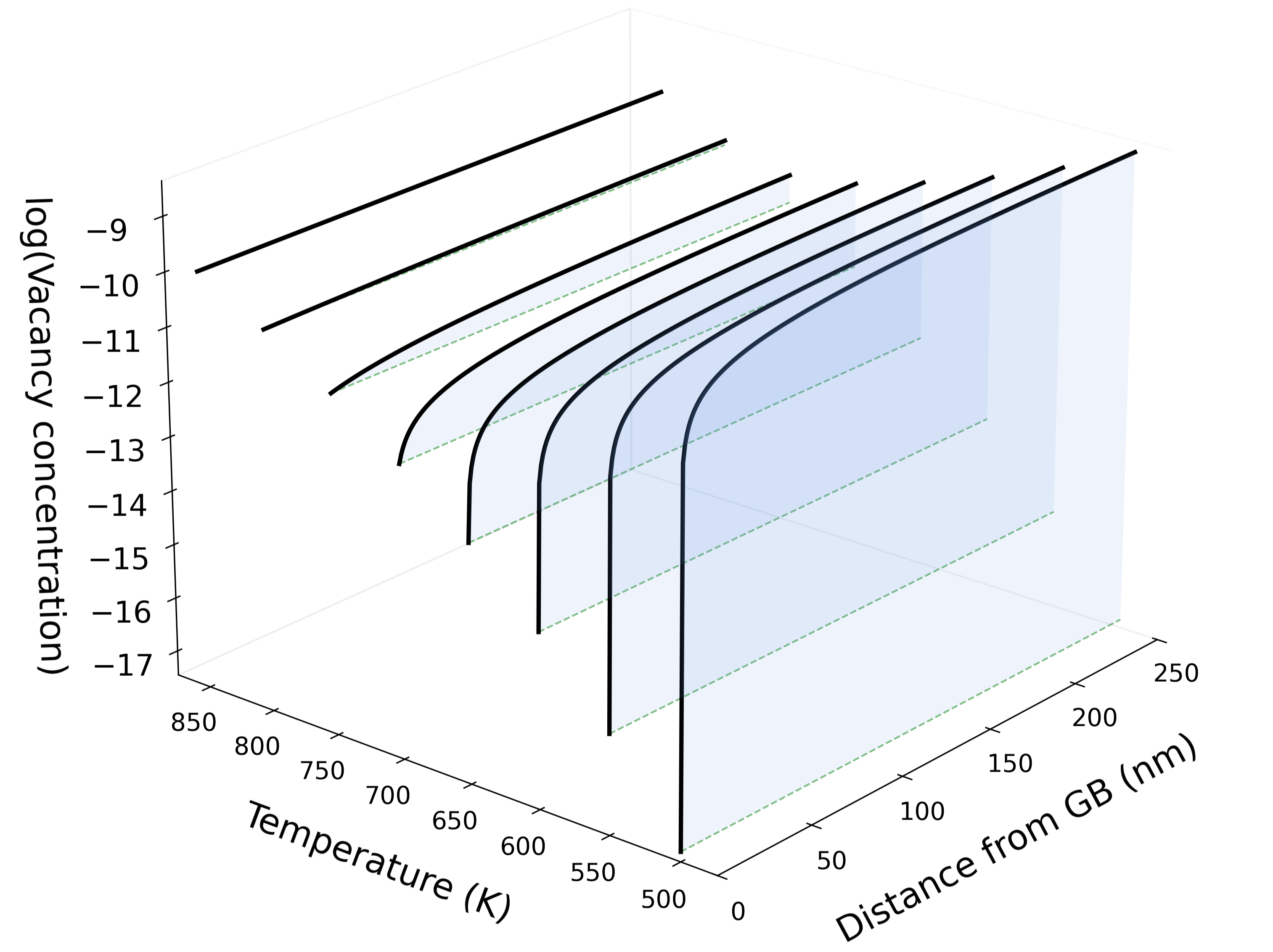}
    \caption{Vacancy concentration profiles}
    \label{fig:temp_vac}
\end{subfigure}
\caption{Point defect concentration profiles corresponding to the RIS profiles shown in Fig.~\ref{fig:temperature}.}
\label{fig:temp_defect}
\end{figure}

Figs.~\ref{fig:temp_sia} and \ref{fig:temp_vac} present the concentration profiles of SIAs and vacancies, respectively, corresponding to Fig.~\ref{fig:temperature}. Vacancy concentration profiles exhibit increasing bulk concentration accumulation at lower temperatures due to reduced thermal vacancy transport; under these conditions, recombination-controlled kinetics dominate as limited vacancy mobility leads to bulk supersaturation. Conversely, at high temperatures, sink-controlled kinetics prevail and vacancies diffuse rapidly to the GB. This behavior, in combination with the exponentially increasing thermal equilibrium concentration, reduces supersaturation to negligible levels. In contrast, the low migration energy of SIAs ensures fast diffusion and sink-controlled kinetics across all temperatures, resulting in relatively temperature-independent bulk concentrations.

\section{Grain Size Variation}

\begin{figure}[ht!]
\centering
\begin{subfigure}{0.49\textwidth}
    \includegraphics[width=\textwidth]{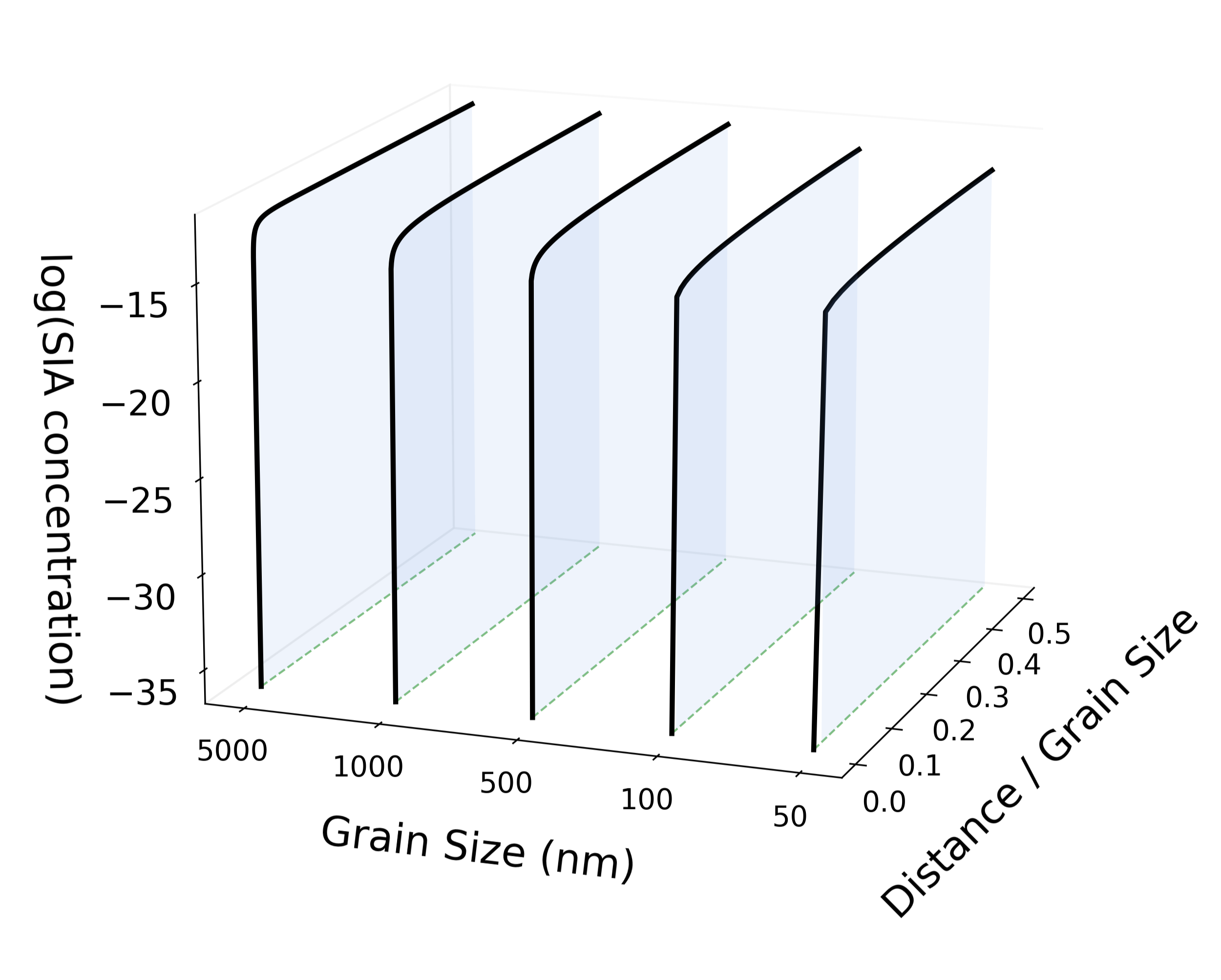}
    \caption{SIA concentration profiles}
    \label{fig:grain_sia_500}
\end{subfigure}
\begin{subfigure}{0.49\textwidth}
    \includegraphics[width=\textwidth]{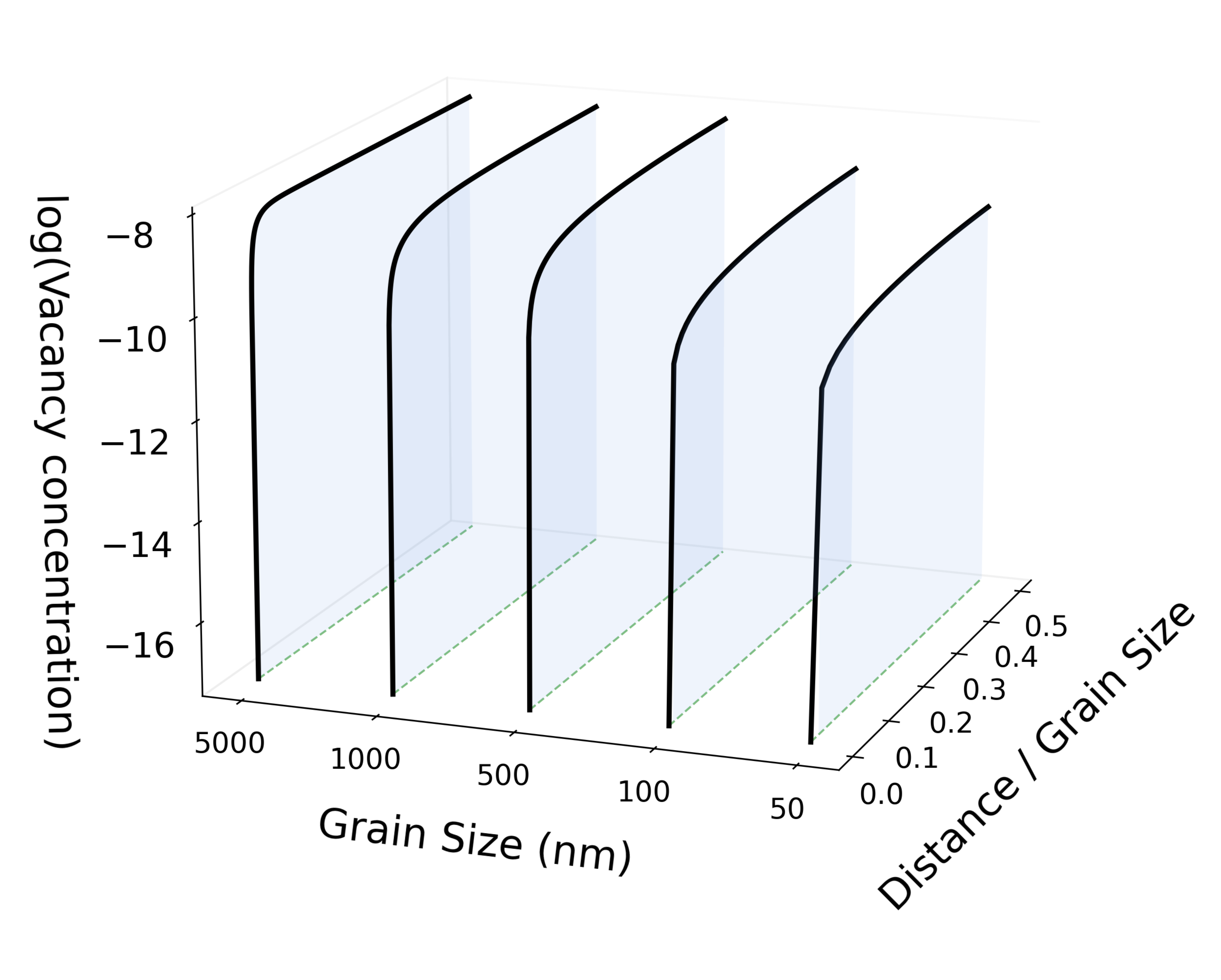}
    \caption{Vacancy concentration profiles}
    \label{fig:grain_vac_500}
\end{subfigure}
\caption{Point defect concentration profiles at 500~K corresponding to the Cr RIS profiles shown in Fig.~\ref{fig:grain_500} (enrichment regime).}
\label{fig:grain_defect_500}
\end{figure}

\begin{figure}[ht!]
\centering
\begin{subfigure}{0.49\textwidth}
    \includegraphics[width=\textwidth]{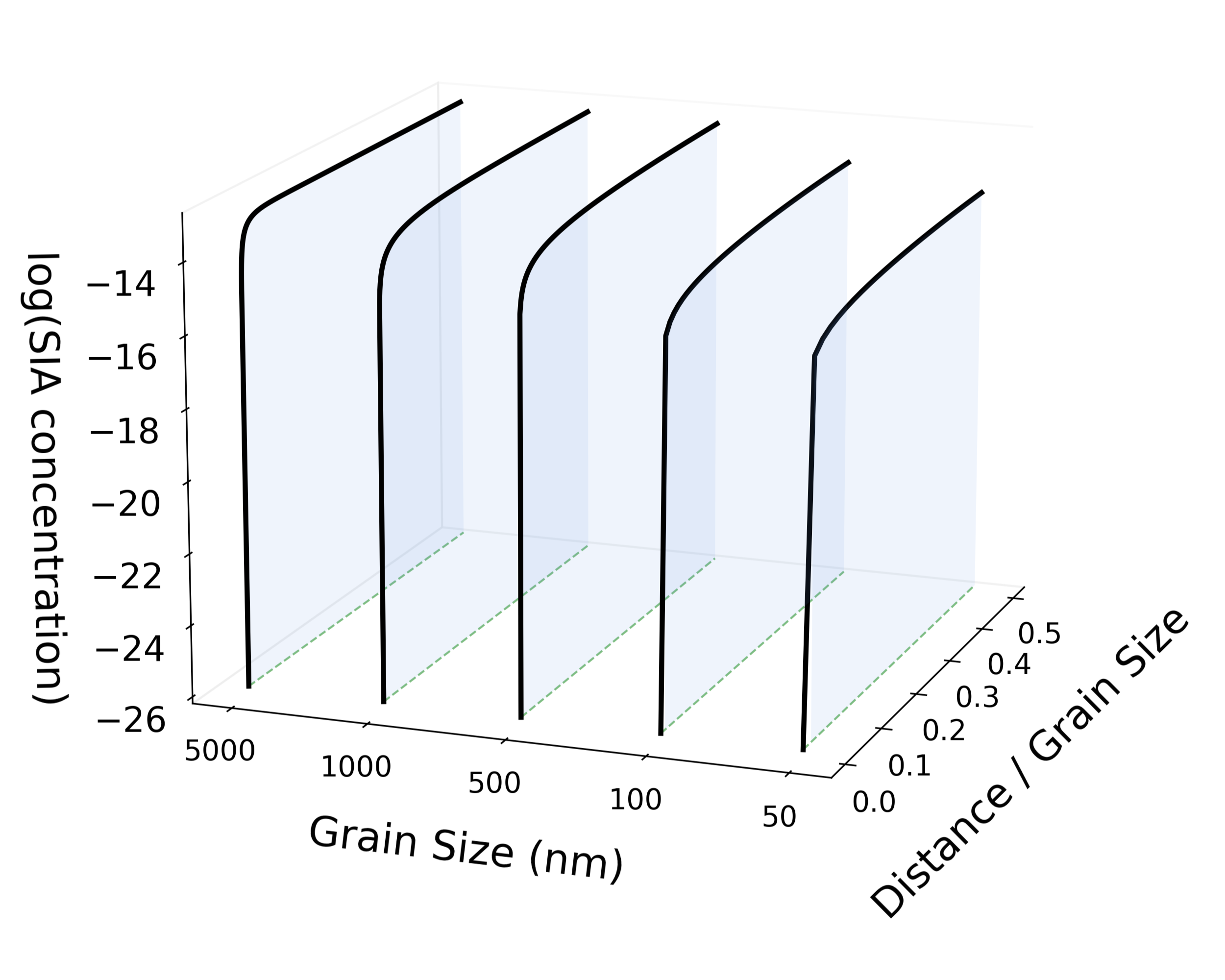}
    \caption{SIA concentration profiles}
    \label{fig:grain_sia_700}
\end{subfigure}
\begin{subfigure}{0.49\textwidth}
    \includegraphics[width=\textwidth]{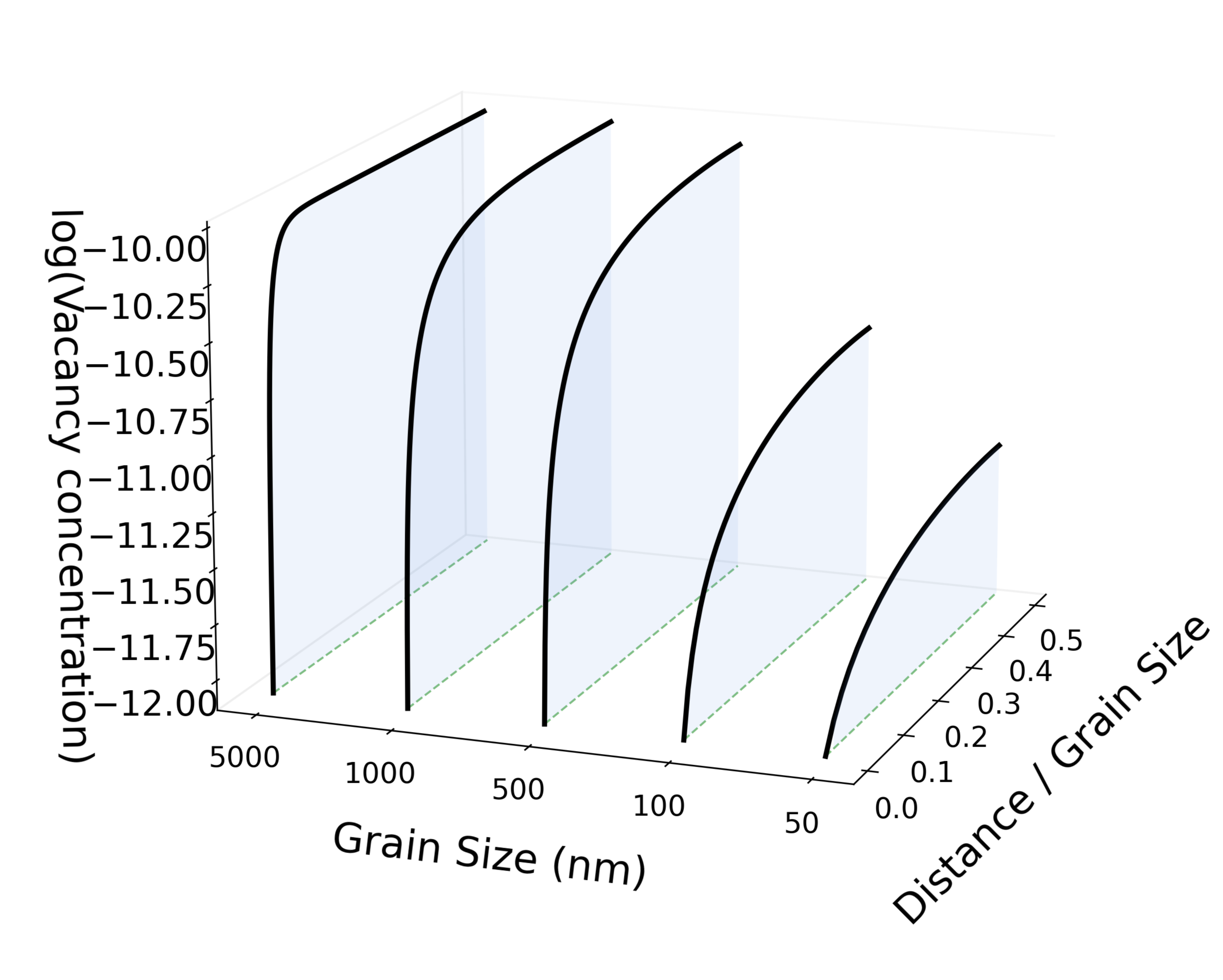}
    \caption{Vacancy concentration profiles}
    \label{fig:grain_vac_700}
\end{subfigure}
\caption{Point defect concentration profiles at 700~K corresponding to the Cr RIS profiles shown in Fig.~\ref{fig:grain_700} (depletion regime).}
\label{fig:grain_defect_700}
\end{figure}

Figs.~\ref{fig:grain_sia_500} and \ref{fig:grain_vac_500} present the SIA and vacancy concentration profiles at 500~K corresponding to Cr enrichment regime in Fig.~\ref{fig:grain_500}, while Figs.~\ref{fig:grain_sia_700} and \ref{fig:grain_vac_700} present the profiles at 700~K corresponding to depletion regime in Figs.~\ref{fig:grain_700}. 
In smaller grains, the higher GB sink density limits point defect accumulation, resulting in lower bulk supersaturation. Conversely, larger grains permit greater defect accumulation in the grain interior before annihilation at GBs. The enhanced thermal mobility at 700~K produces broader defect profiles compared to the 500~K case.
Decreasing grain size increases the effective  GB sink strength ($\propto 1/d^2$), shifting point defect kinetics toward the sink-controlled regime. At 500~K, this transition is more pronounced for vacancies, which exhibit reduced bulk concentrations as grain boundaries become more accessible.

\section{Dislocation sink density}
\begin{figure}[ht!]
\centering
\begin{subfigure}{0.49\textwidth}
    \includegraphics[width=\textwidth]{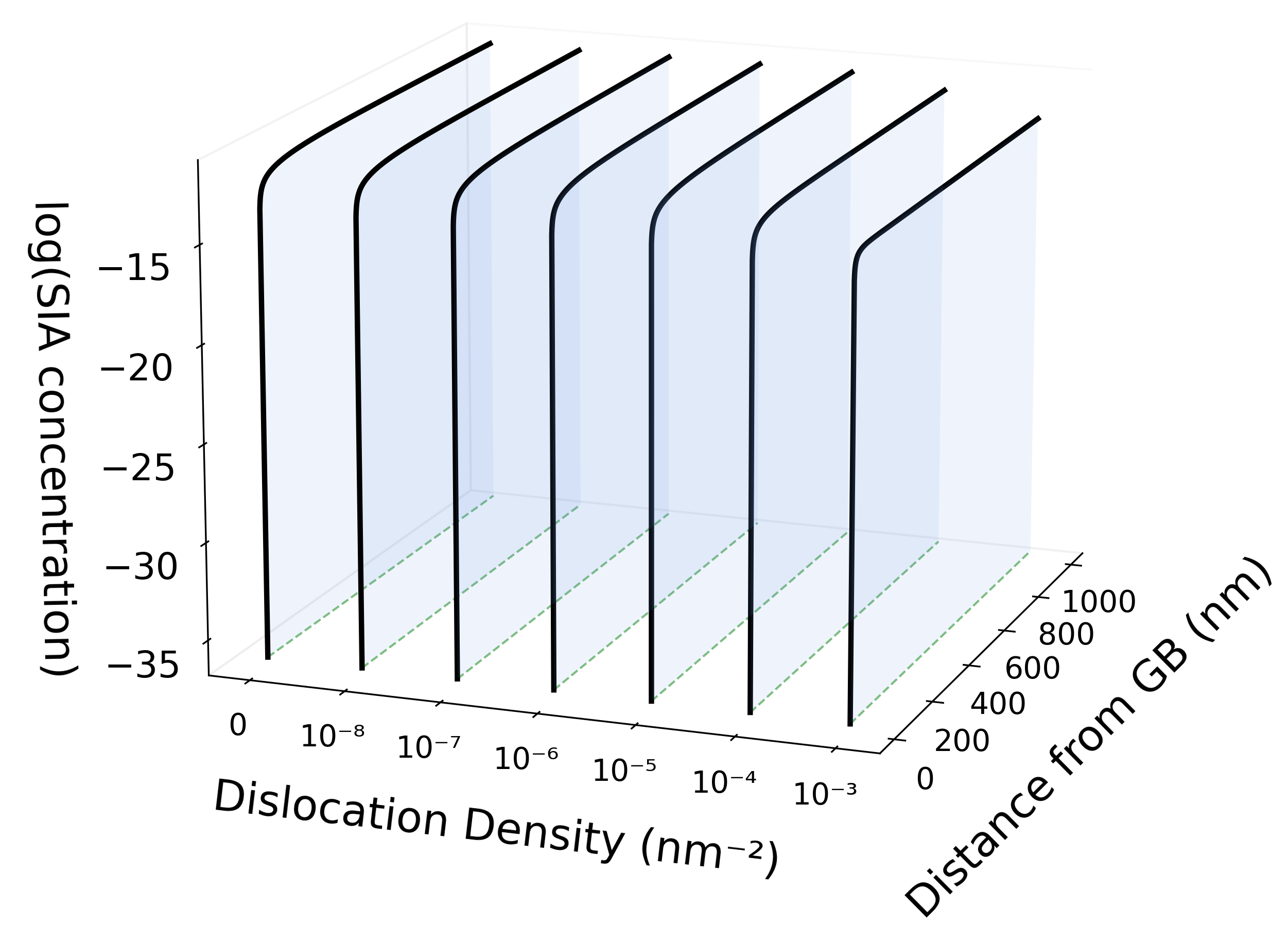}
    \caption{SIA concentration profiles}
    \label{fig:sink_sia_500}
\end{subfigure}
\begin{subfigure}{0.49\textwidth}
    \includegraphics[width=\textwidth]{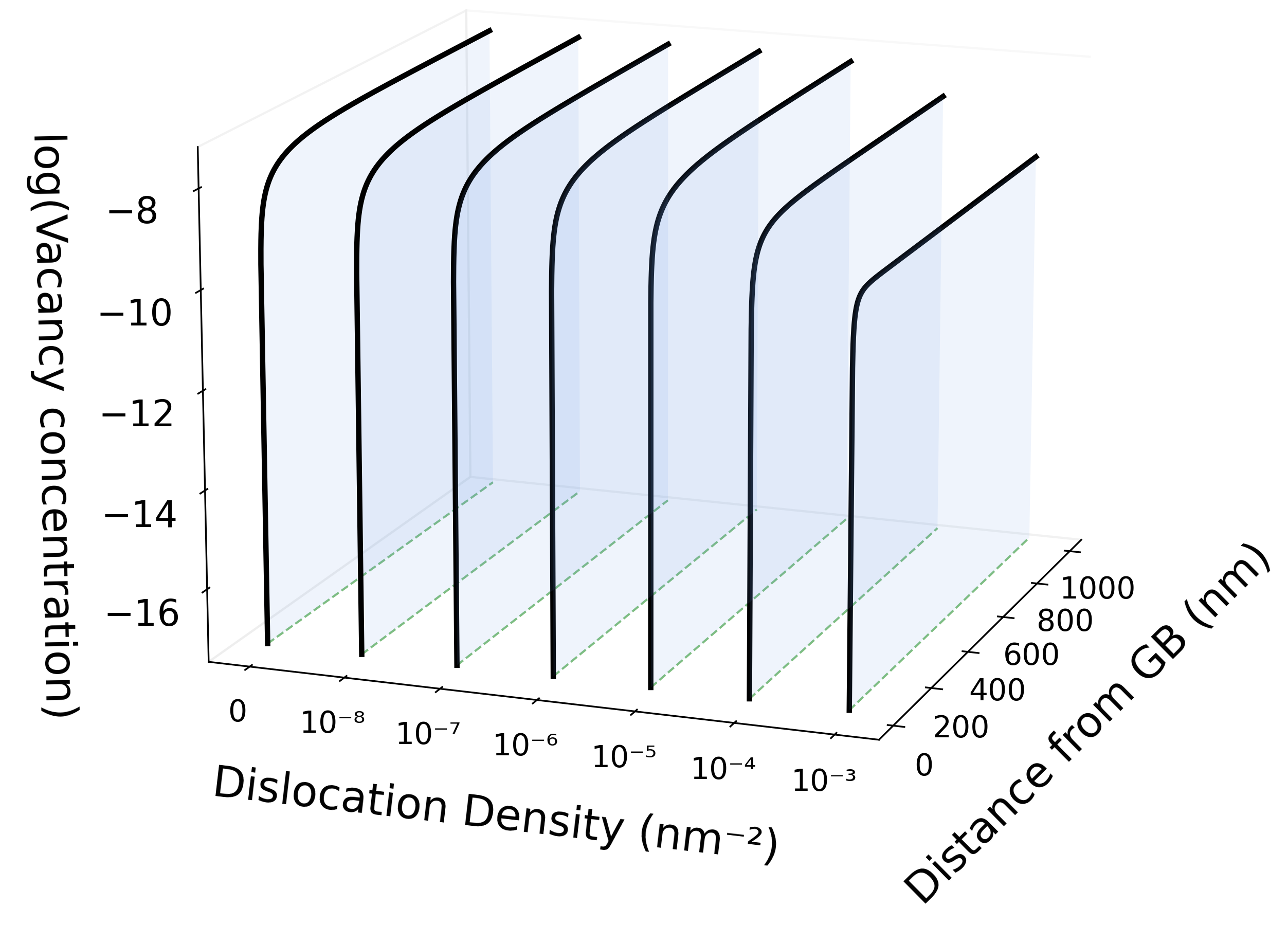}
    \caption{Vacancy concentration profiles}
    \label{fig:sink_vac_500}
\end{subfigure}
\caption{Point defect concentration profiles at 500~K corresponding to the Cr RIS profiles shown in Fig.~\ref{fig:sink_500} (enrichment regime).}
\label{fig:sink_defect_500}
\end{figure}

\begin{figure}[ht!]
\centering
\begin{subfigure}{0.49\textwidth}
    \includegraphics[width=\textwidth]{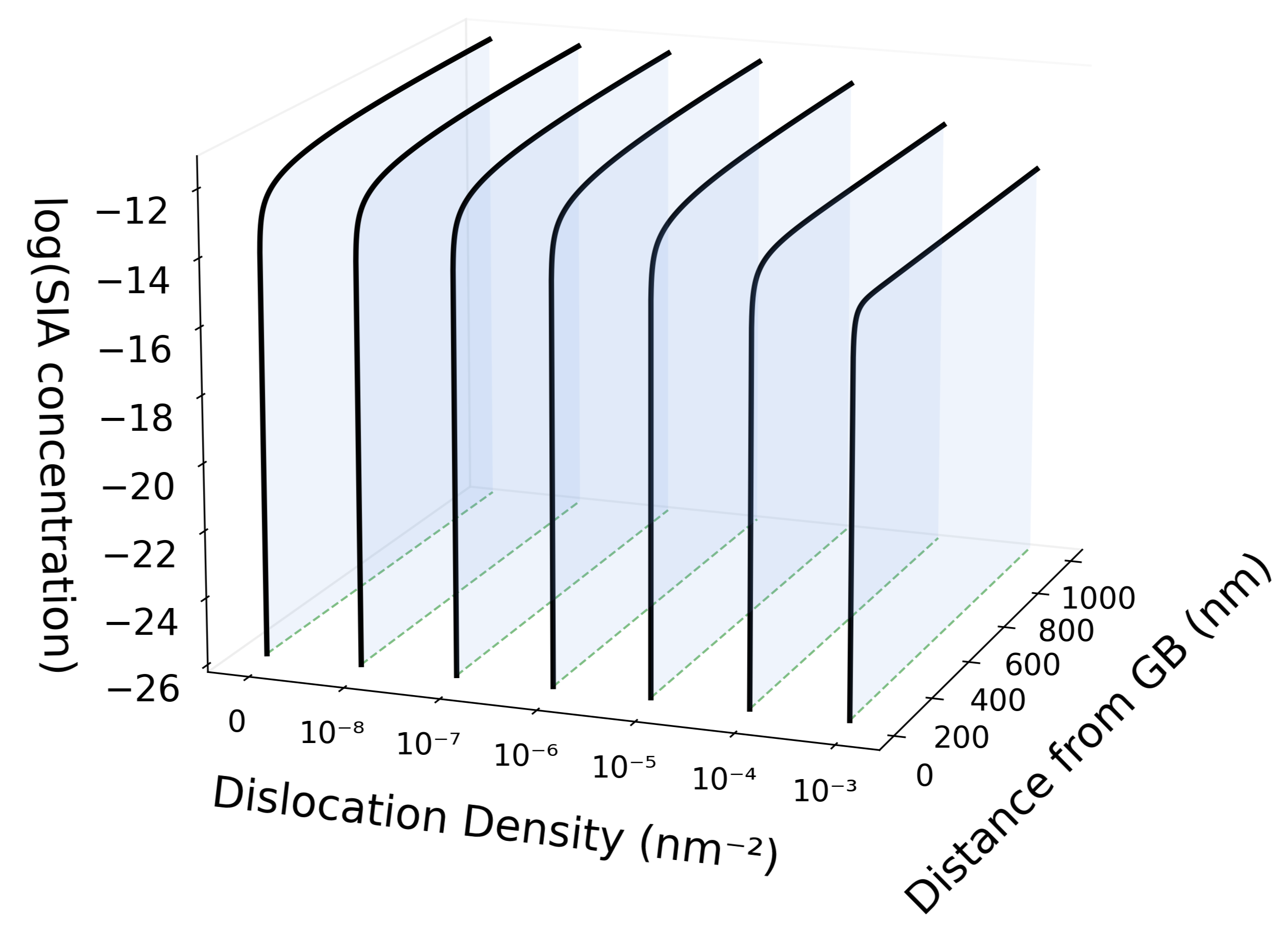}
    \caption{SIA concentration profiles}
    \label{fig:sink_sia_700}
\end{subfigure}
\begin{subfigure}{0.49\textwidth}
    \includegraphics[width=\textwidth]{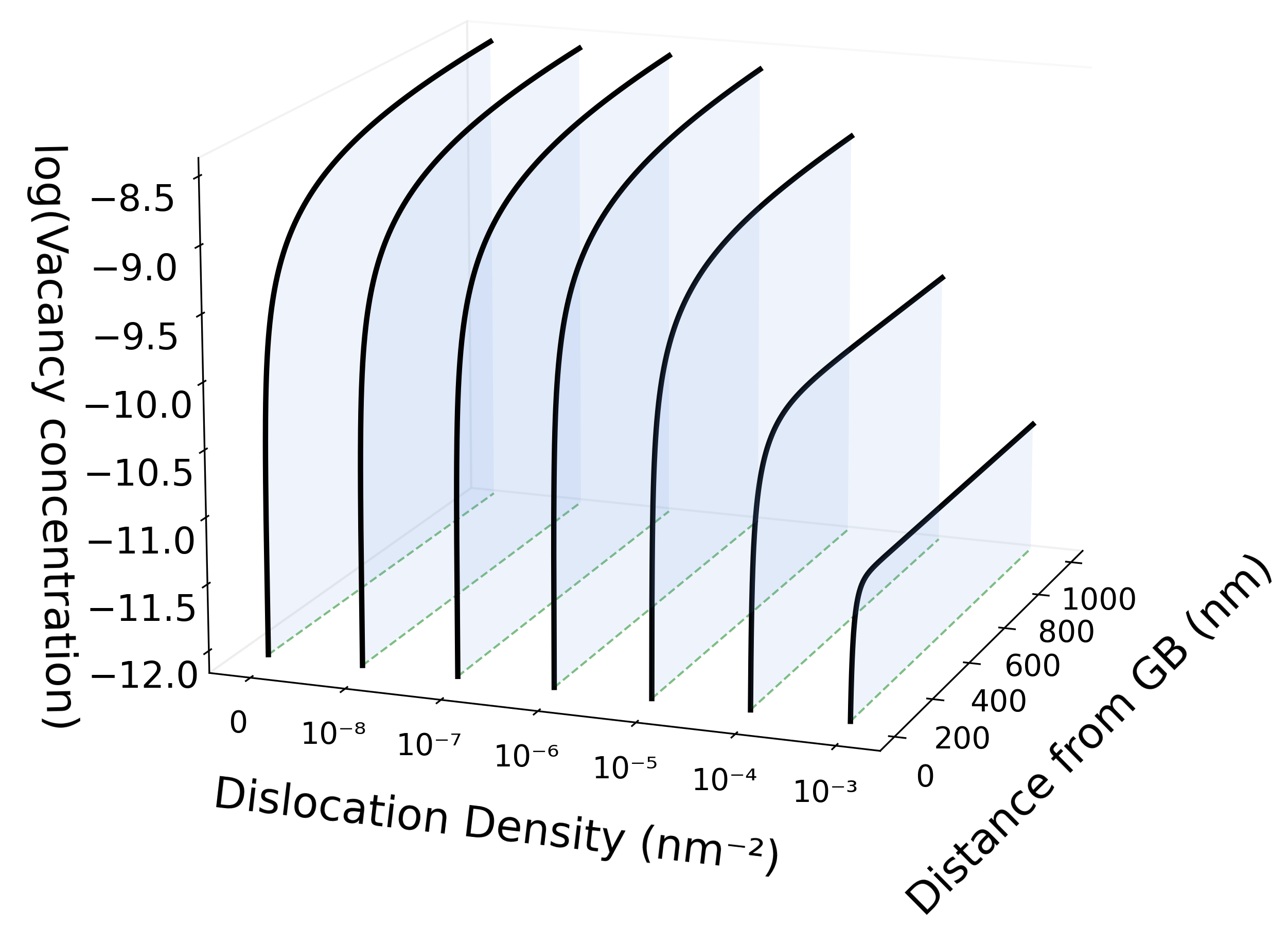}
    \caption{Vacancy concentration profiles}
    \label{fig:sink_vac_700}
\end{subfigure}
\caption{Point defect concentration profiles at 700~K corresponding to the Cr RIS profiles shown in Fig.~\ref{fig:sink_700} (depletion regime). }
\label{fig:sink_defect_700}
\end{figure}

Figs.~\ref{fig:sink_sia_500} and \ref{fig:sink_vac_500} present the point defect profiles at 500~K corresponding to the RIS profiles in Figs.~\ref{fig:sink_500}, while Figs.~\ref{fig:sink_sia_700} and \ref{fig:sink_vac_700} present the profiles at 700~K corresponding to Figs.~\ref{fig:sink_500}. 
At low dislocation densities, GBs serve as the dominant sinks, permitting substantial defect supersaturation to develop in the grain interior. As dislocation density increases, these bulk sinks capture point defects throughout the grain volume, reducing the bulk supersaturation. At the highest dislocation densities ($>10^{-3}$~nm$^{-2}$), the defect concentration gradients become highly localized near the GB, with minimal supersaturation in the grain interior.

\section{Production bias}

\begin{figure}[ht!]
\centering
\begin{subfigure}{0.49\textwidth}
    \includegraphics[width=\textwidth]{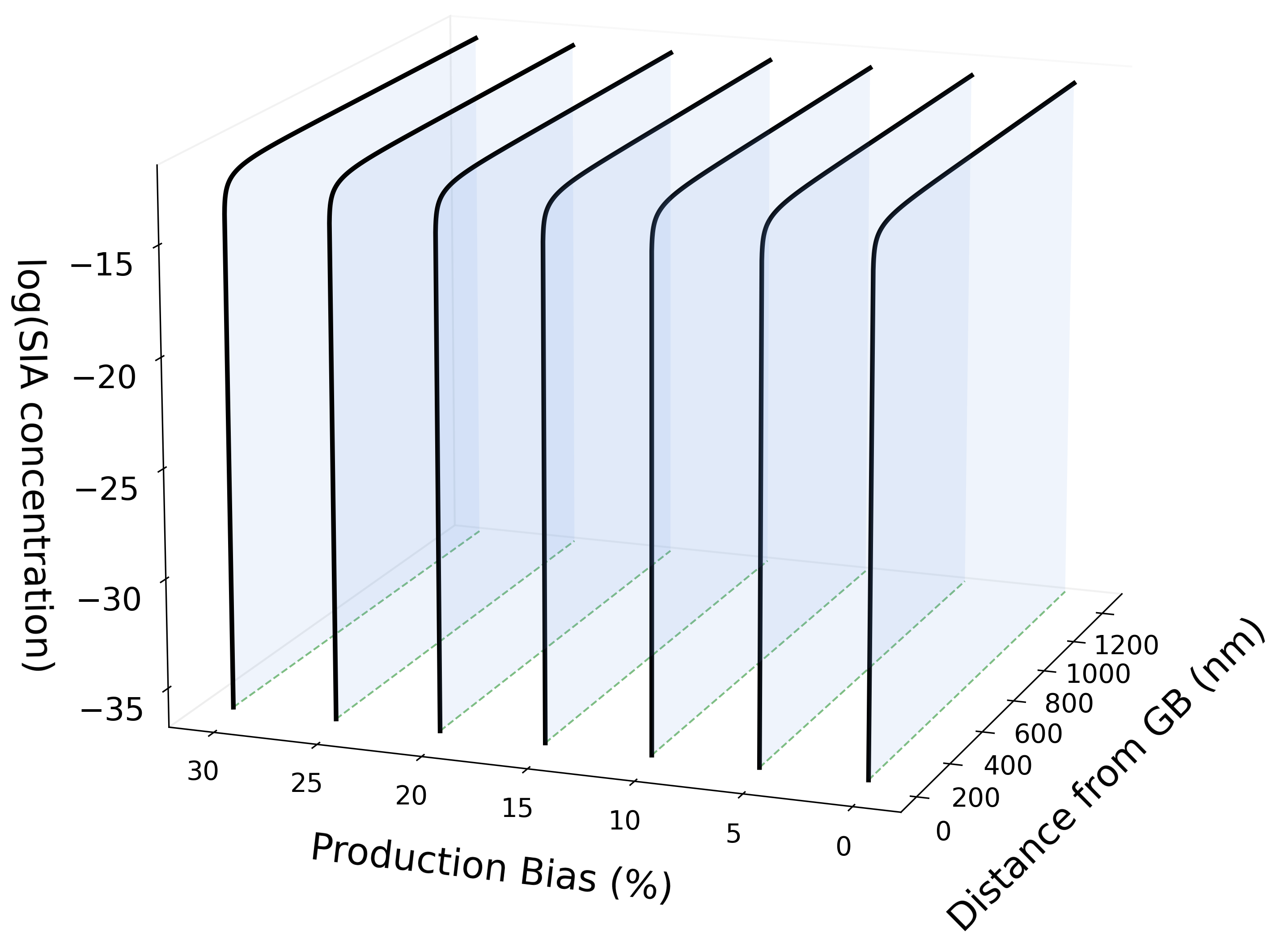}
    \caption{SIA concentration profiles}
    \label{fig:pb_sia_500_hs}
\end{subfigure}
\begin{subfigure}{0.49\textwidth}
    \includegraphics[width=\textwidth]{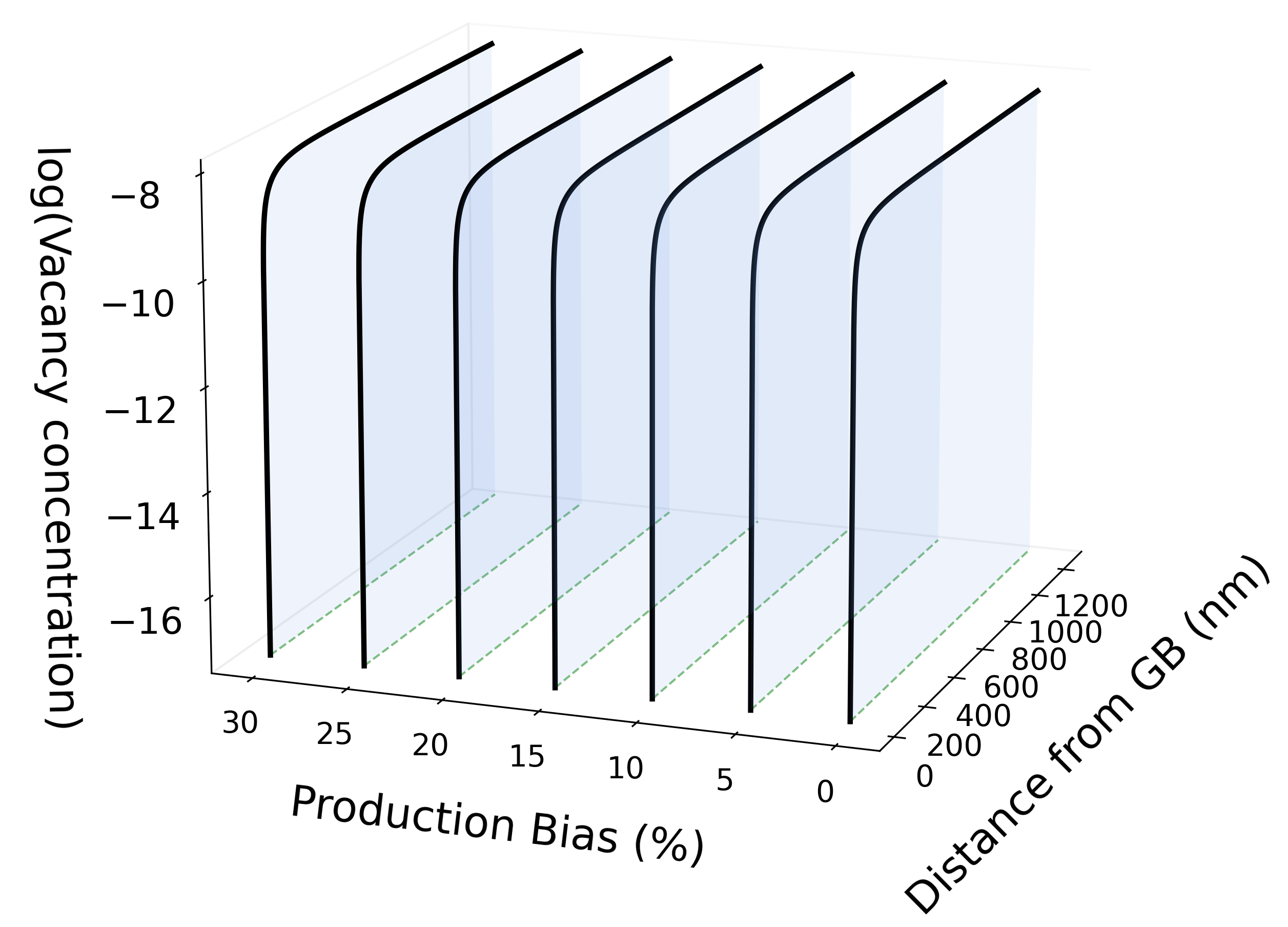}
    \caption{Vacancy concentration profiles}
    \label{fig:pb_vac_500_hs}
\end{subfigure}
\caption{Point defect concentration profiles at 500~K (high dislocation density) corresponding to the Cr RIS profiles shown in Fig.~\ref{fig:pb_500_hs}.}
\label{fig:pb_defect_500_hs}
\end{figure}

\begin{figure}[ht!]
    \centering
    \includegraphics[scale = 0.4]{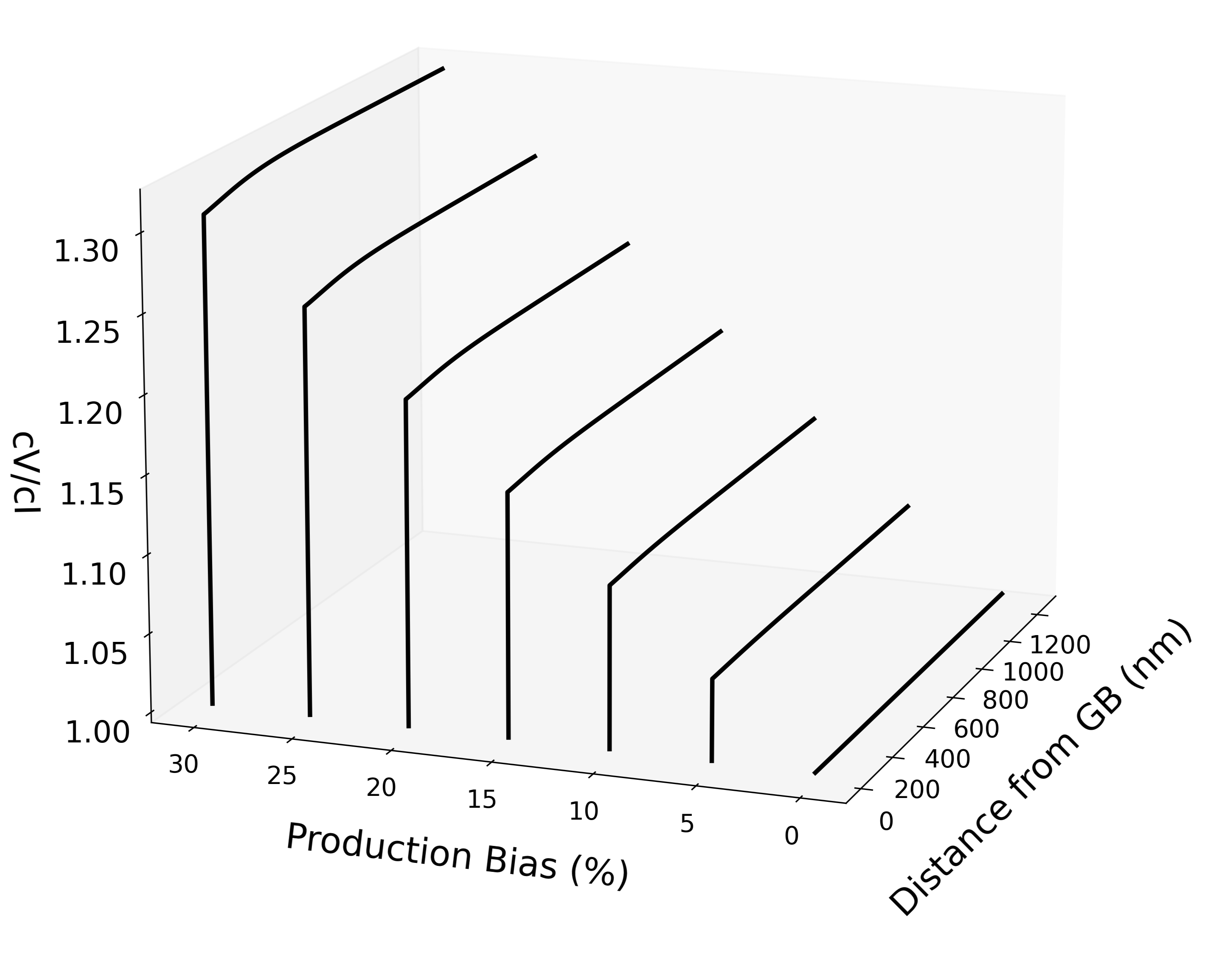}
    \caption{Ratio of the vacancy to SIA concentration for the biased case in Fig.~\ref{fig:pb_defect_500_hs} normalized with respect to the ratio ($D_V/D_I$) for the fully unbiased case.}
    \label{fig:cV_cI_500_pb_hs}
\end{figure}

\begin{figure}[ht!]
\centering
\begin{subfigure}{0.49\textwidth}
    \includegraphics[width=\textwidth]{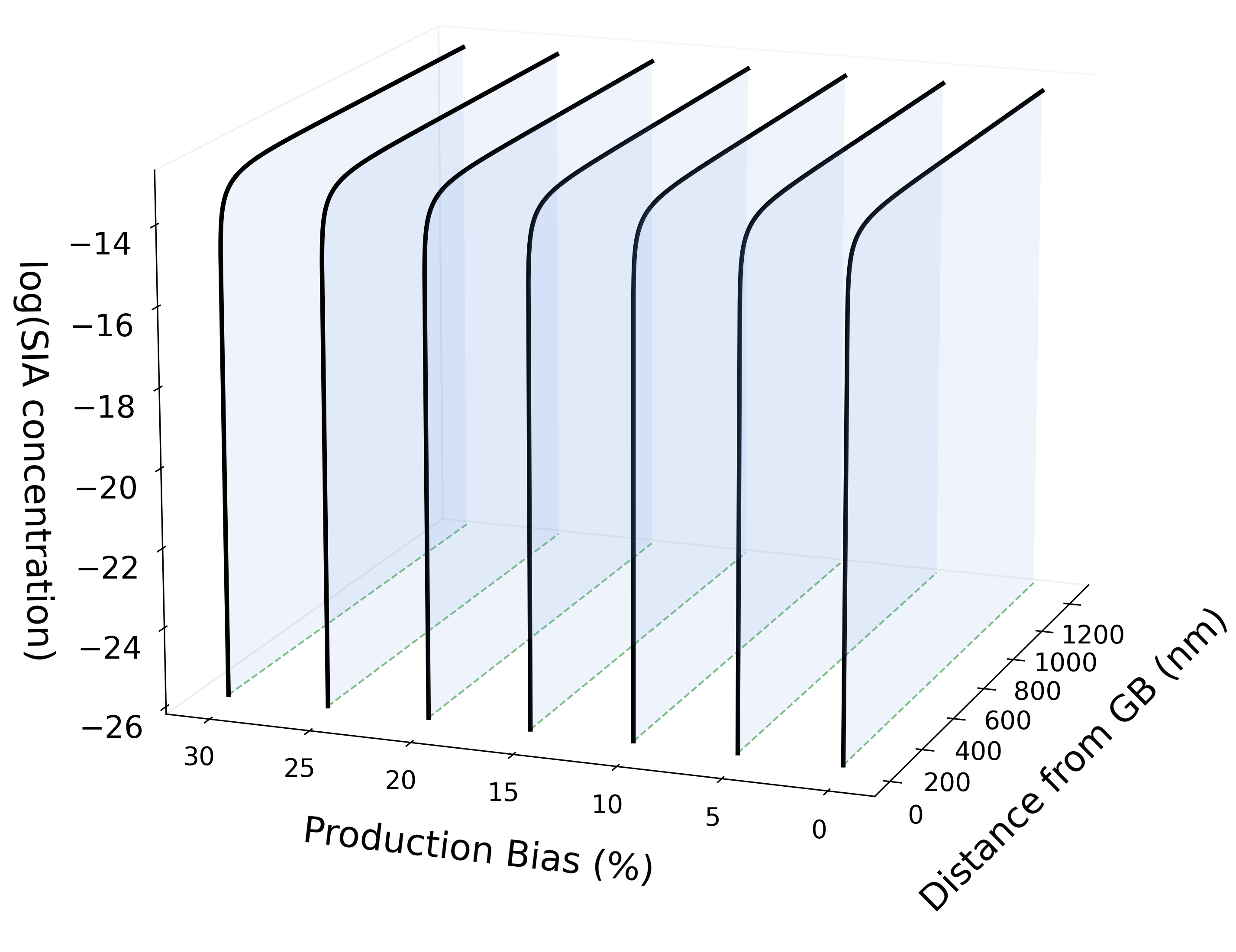}
    \caption{SIA concentration profiles}
    \label{fig:pb_sia_700_hs}
\end{subfigure}
\begin{subfigure}{0.49\textwidth}
    \includegraphics[width=\textwidth]{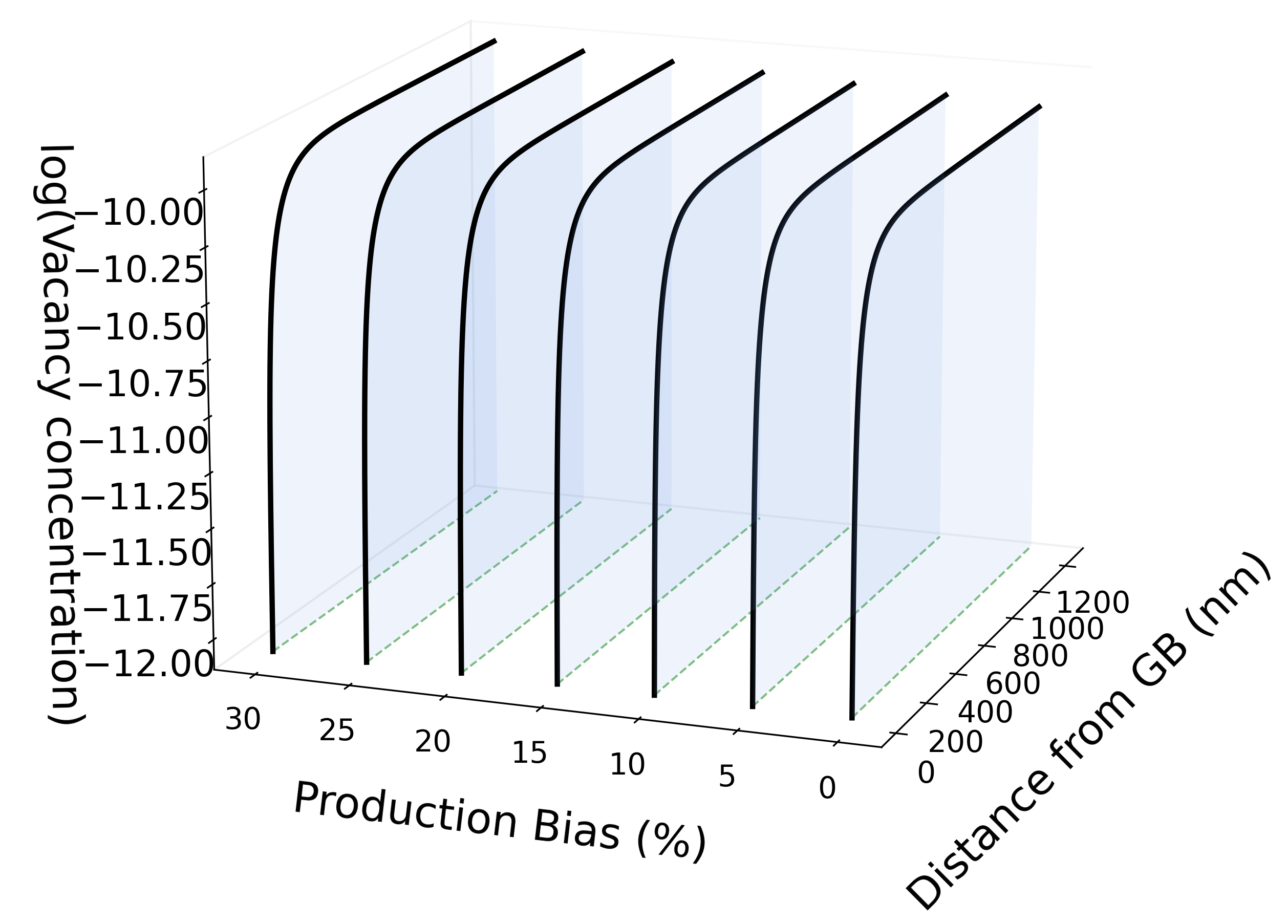}
    \caption{Vacancy concentration profiles}
    \label{fig:pb_vac_700_hs}
\end{subfigure}
\caption{Point defect concentration profiles at 700~K (high dislocation density) corresponding to the Cr RIS profiles shown in Fig.~\ref{fig:pb_700_hs}.}
\label{fig:pb_defect_700_hs}
\end{figure}

\begin{figure}[ht!]
    \centering
    \includegraphics[scale = 0.4]{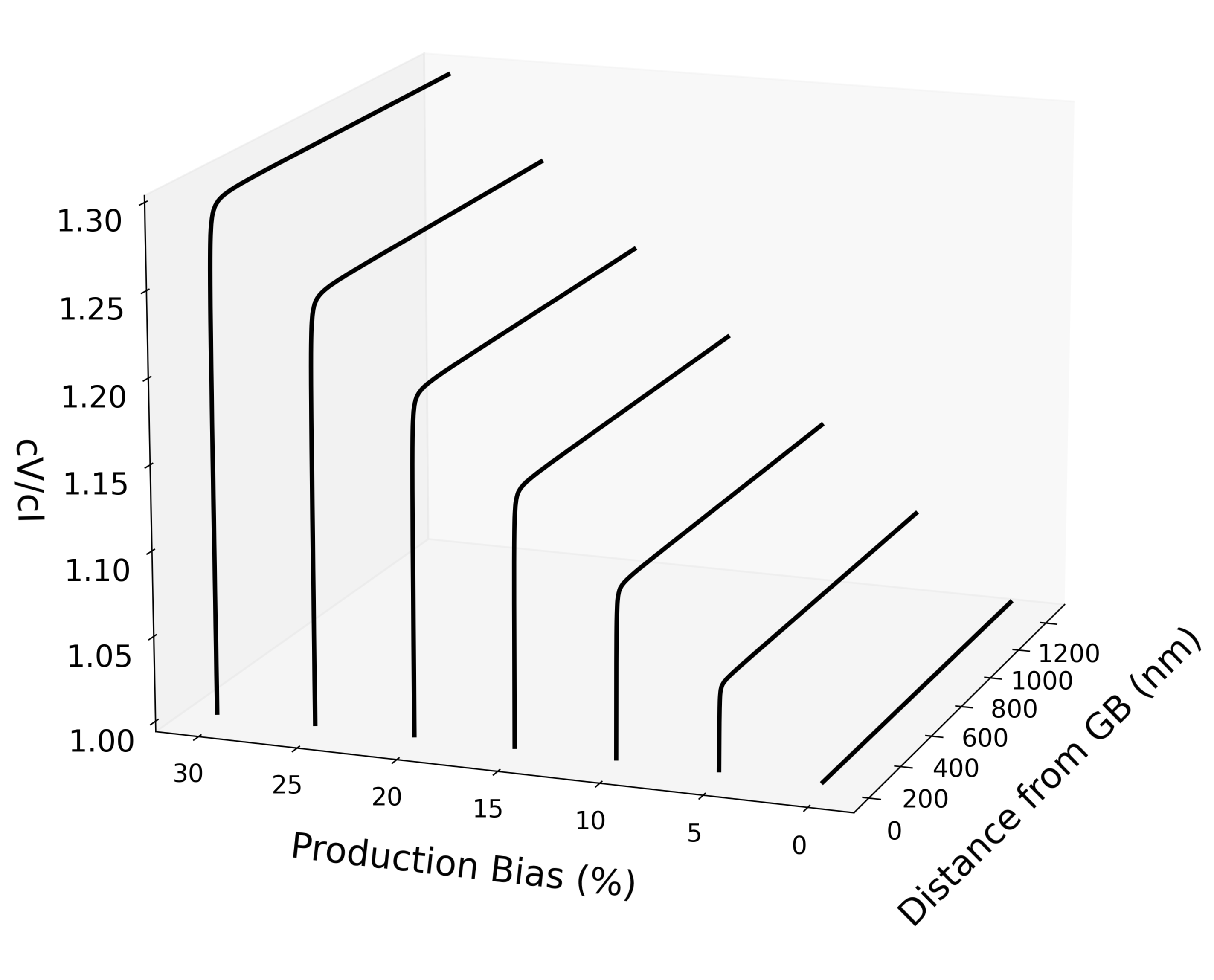}
    \caption{Ratio of the vacancy to SIA concentration for the biased case in Fig.~\ref{fig:pb_defect_700_hs} normalized with respect to the ratio ($D_V/D_I$) for the fully unbiased case.}
    \label{fig:cV_cI_700_pb_hs}
\end{figure}

Figs.~\ref{fig:pb_sia_500_hs} and \ref{fig:pb_vac_500_hs} present the defect profiles at 500~K corresponding to Fig.~\ref{fig:pb_500_hs}, while Figs.~\ref{fig:pb_sia_700_hs} and \ref{fig:pb_vac_700_hs} present the defect profiles at 700~K corresponding to Figs.~\ref{fig:pb_700_hs}; these results correspond to a dislocation density of $\rho=10^{-4}$ nm$^{-2}$.
Increasing the production bias to favor vacancies systematically elevates the bulk vacancy concentration, enhancing the vacancy flux to the GB, relative to SIA.
At $\rho=10^{-4}$ nm$^{-2}$, the system is in the dislocation sink-dominated regime, where production bias modifies the vacancy-to-SIA concentration ratio as $c_V/c_I = D_I k_I^2(1+\epsilon)/(D_V k_V^2) $. For the unbiased sink model ($k_I^2 \approx k_V^2$), this simplifies to $c_V/c_I \approx (D_I/D_V)(1+\epsilon)$. The $c_V/c_I$ ratio (normalized w.r.t the ratio $D_V/D_I$ for the fully unbiased case) for 500 \textdegree C and 700 \textdegree C scales linearly as $(1+\epsilon)$, as shown in Fig.~\ref{fig:cV_cI_500_pb_hs} and~\ref{fig:cV_cI_700_pb_hs}, respectively.

\begin{figure}[ht!]
\centering
\begin{subfigure}{0.49\textwidth}
    \includegraphics[width=\textwidth]{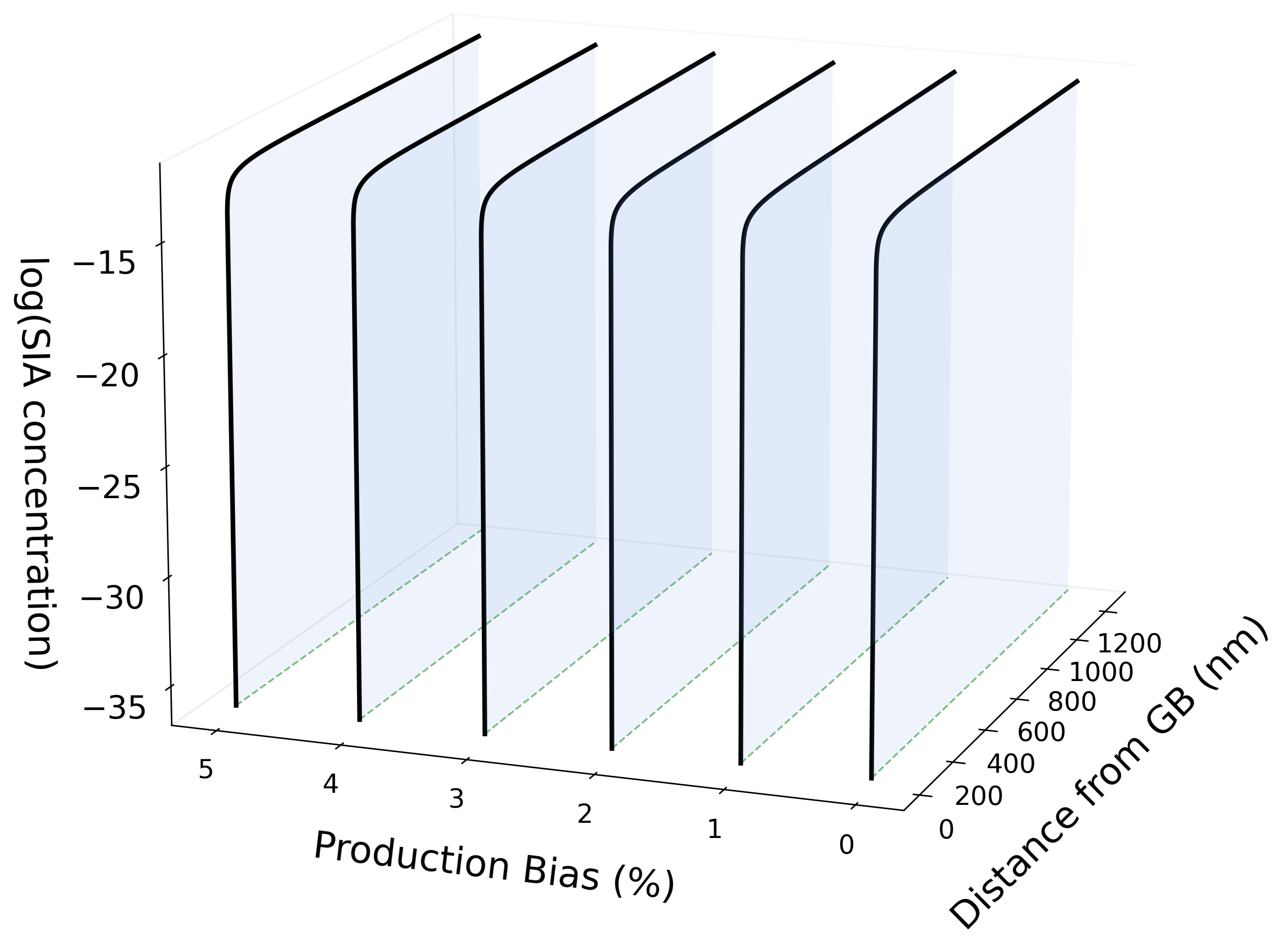}
    \caption{SIA concentration profiles}
    \label{fig:pb_sia_500_ls}
\end{subfigure}
\begin{subfigure}{0.49\textwidth}
    \includegraphics[width=\textwidth]{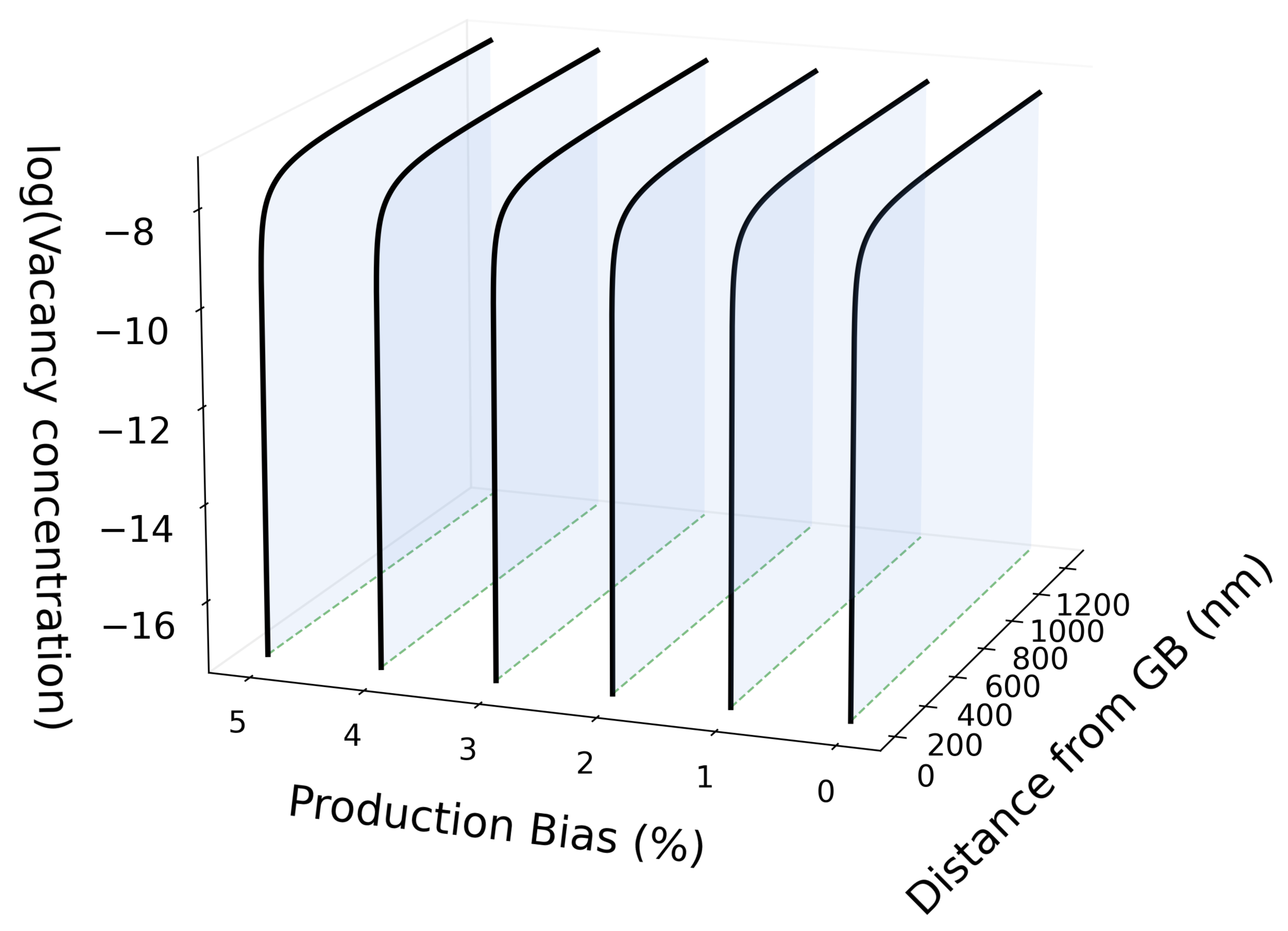}
    \caption{Vacancy concentration profiles}
    \label{fig:pb_vac_500_ls}
\end{subfigure}
\caption{Point defect concentration profiles at 500~K (low dislocation density, $\rho = 10^{-6}$~nm$^{-2}$) corresponding to the Cr RIS profiles shown in Fig.~\ref{fig:pb_500_ls}.}
\label{fig:pb_defect_500_ls}
\end{figure}

\begin{figure}[ht!]
    \centering
    \includegraphics[scale = 0.4]{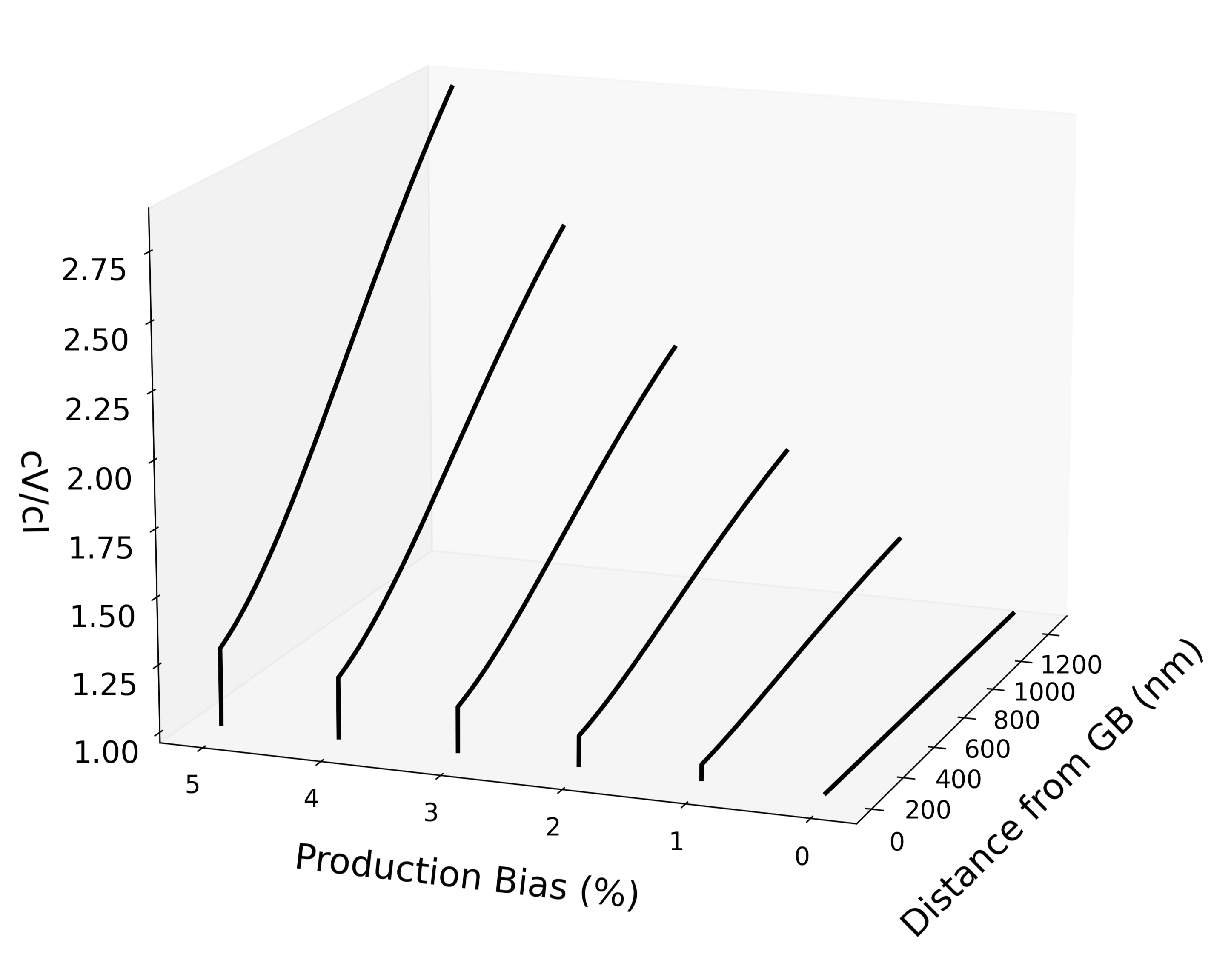}
    \caption{Ratio of the vacancy to SIA concentration for the biased case in Fig.~\ref{fig:pb_defect_500_ls} normalized with respect to the ratio ($D_V/D_I$) for the fully unbiased case.}
    \label{fig:cV_cI_500_pb_ls}
\end{figure}

For the case with reduced dislocation density ($\rho=10^{-6}$ nm$^{-2}$) in Fig.~\ref{fig:pb_500_ls}, Figs.~\ref{fig:pb_sia_500_ls} and \ref{fig:pb_vac_500_ls} present the corresponding defect profiles. 
Notably, these steady-state defect concentration profiles remain relatively smooth and monotonic under production bias. Non-monotonic defect profiles reported in previous production bias studies~\cite{ozturk2021_PB,gencturk2023_PB_RIS} are a result of numerical artifact rather than being of physical origin.
At the reduced dislocation density of $\rho=10^{-6}$, the simple analytic expression $c_V/c_I = (D_I/D_V)(1+\epsilon)$ derived for sink-dominated regime no longer applies. As shown in Fig.~\ref{fig:cV_cI_500_pb_ls}, even a modest 5\% production bias increases the normalized $c_V/c_I$ ratio to approximately 2.75, exceeding the ratio of 1.05 predicted by the sink-dominated assumption. This can be attributed to the shift in the system towards recombination-dominant kinetics. Moreover, the ratio exhibits spatial variation within the bulk.

\section{Dislocation absorption bias}

\begin{figure}[ht!]
\centering
\begin{subfigure}{0.49\textwidth}
    \includegraphics[width=\textwidth]{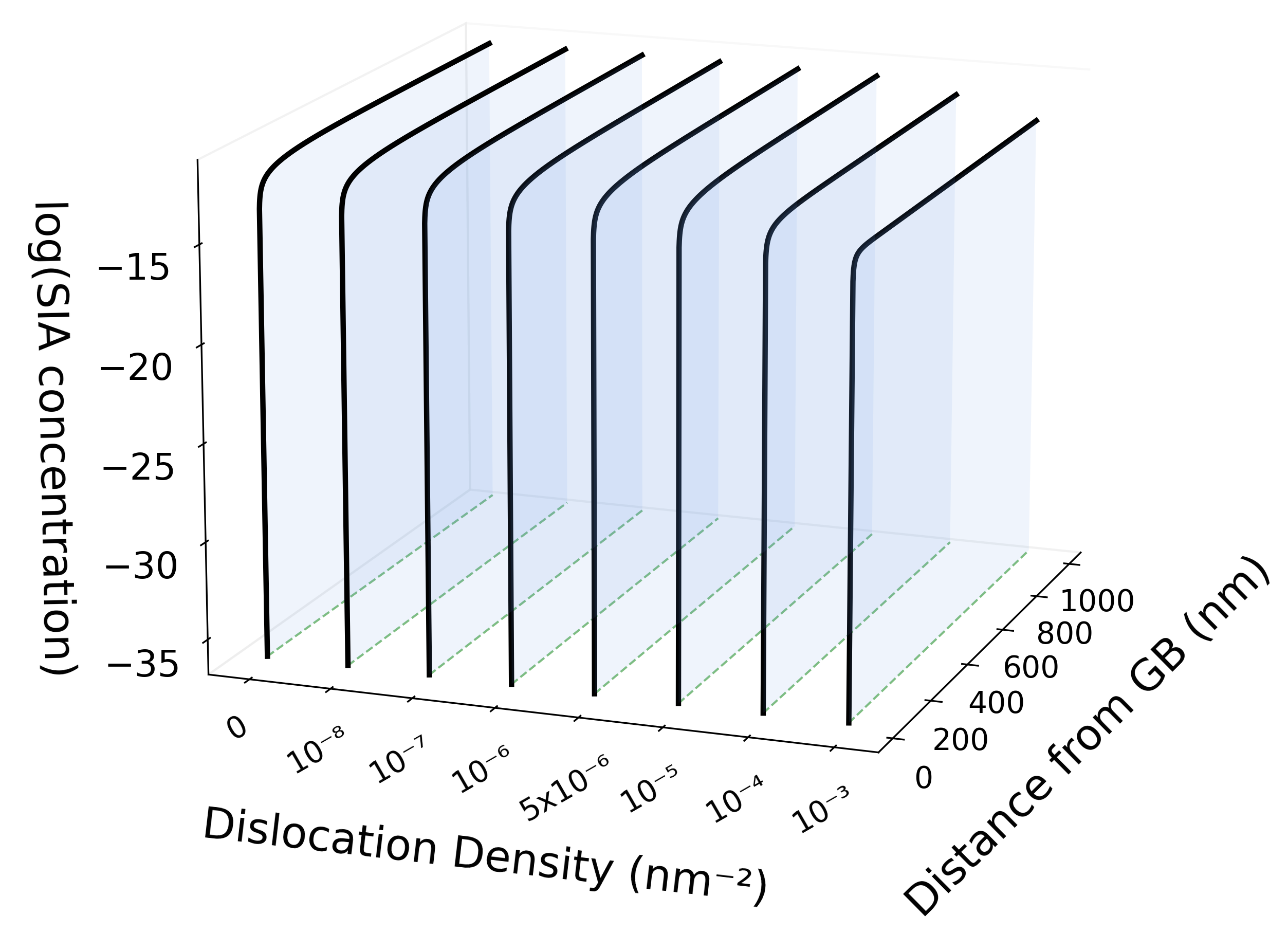}
    \caption{SIA concentration profiles}
    \label{fig:sb_sia_500}
\end{subfigure}
\begin{subfigure}{0.49\textwidth}
    \includegraphics[width=\textwidth]{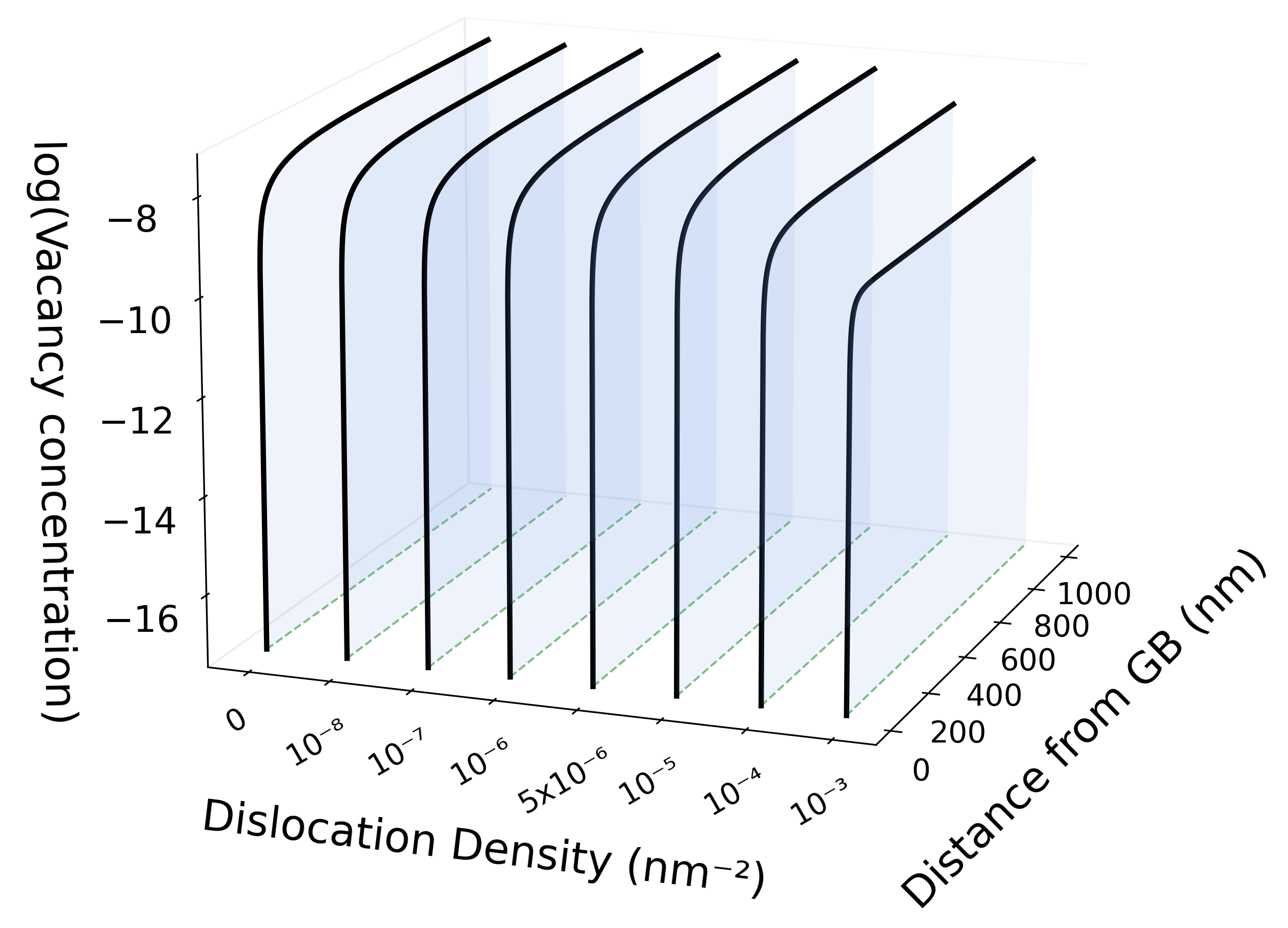}
    \caption{Vacancy concentration profiles}
    \label{fig:sb_vac_500}
\end{subfigure}
\caption{Point defect concentration profiles at 500~K with DDD-based absorption bias corresponding to the Cr RIS profiles shown in Fig.~\ref{fig:bias_sink_T500}.}
\label{fig:sb_defect_500}
\end{figure}

\begin{figure}[ht!]
\centering
\begin{subfigure}{0.49\textwidth}
    \includegraphics[width=\textwidth]{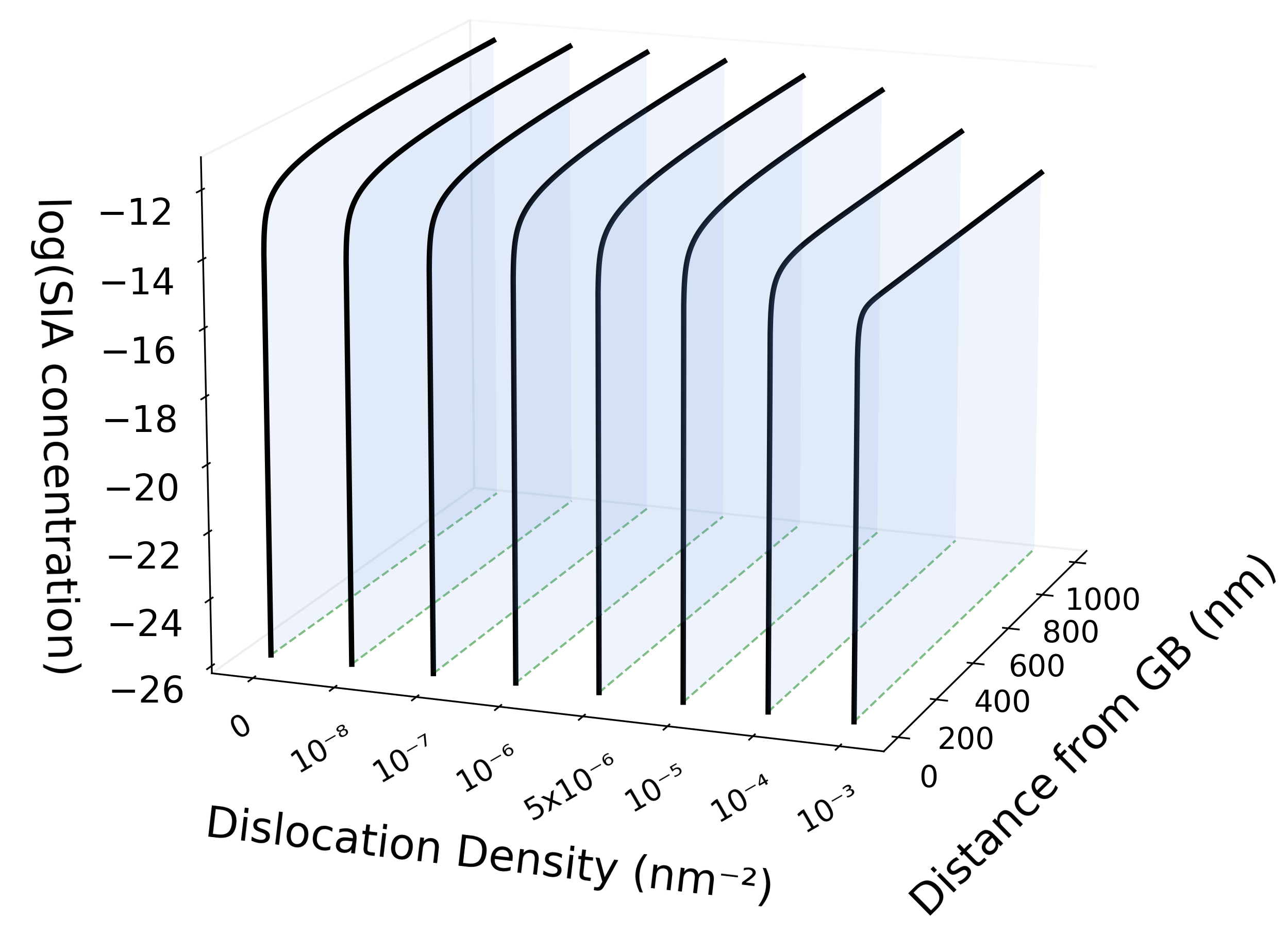}
    \caption{SIA concentration profiles}
    \label{fig:sb_sia_700}
\end{subfigure}
\begin{subfigure}{0.49\textwidth}
    \includegraphics[width=\textwidth]{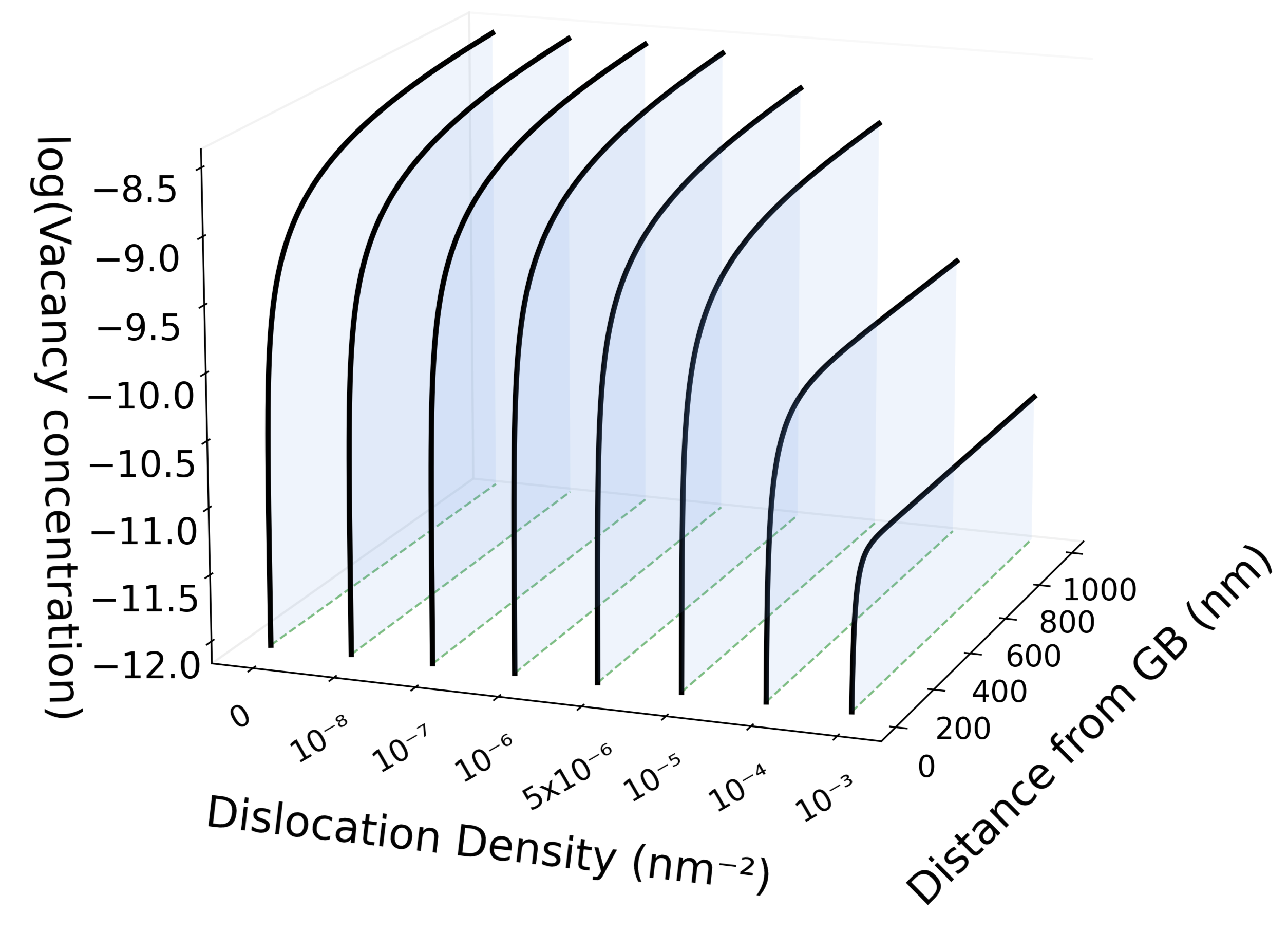}
    \caption{Vacancy concentration profiles}
    \label{fig:sb_vac_700}
\end{subfigure}
\caption{Point defect concentration profiles at 700~K with DDD-based absorption bias corresponding to the Cr RIS profiles shown in Fig.~\ref{fig:bias_sink_T700}.}
\label{fig:sb_defect_700}
\end{figure}

Figs.~\ref{fig:sb_sia_500} and \ref{fig:sb_vac_500} present the defect profiles at 500~K corresponding to Fig.~\ref{fig:bias_sink_T500}, while Figs.~\ref{fig:sb_sia_700} and \ref{fig:sb_vac_700} present the defect profiles at 700~K corresponding to Figs.~\ref{fig:bias_sink_T700}. 
At steady state, the vacancy-to-SIA concentration ratio 
is given by $c_V/c_I = (D_I k_I^2)/(D_V k_V^2)$, where $k_V^2$ and $k_I^2$ are the total sink strengths for each defect species \cite{wiedersich1979theory,wasfundradmatsci2017}. At low dislocation densities where unbiased GB sink or recombination reaction dominate, $k_I^2 \approx k_V^2$, and consequently 
Figs.~\ref{fig:sink_defect_500} and \ref{fig:sb_defect_500} for 500~K and Figs.~\ref{fig:sink_defect_700} and \ref{fig:sb_defect_700} for 700~K exhibit nearly identical concentration profiles. At high 
dislocation densities where $k_{disl}^2 \gg k_{GB}^2$, the absorption bias 
becomes significant with
$k_I^2 \approx (1+B)k_{disl}^2$ while $k_V^2 \approx k_{disl}^2$. With bias factors of 65--75\% at $\rho \geq 10^{-3}$~nm$^{-2}$
(Fig.~\ref{fig:sink_bias}), this yields $k_I^2/k_V^2 \approx 1.7$, increasing the bulk $c_V/c_I$ ratio by a corresponding factor relative to the unbiased case.

\begin{figure}[ht!]
    \centering
    \includegraphics[scale = 0.5]{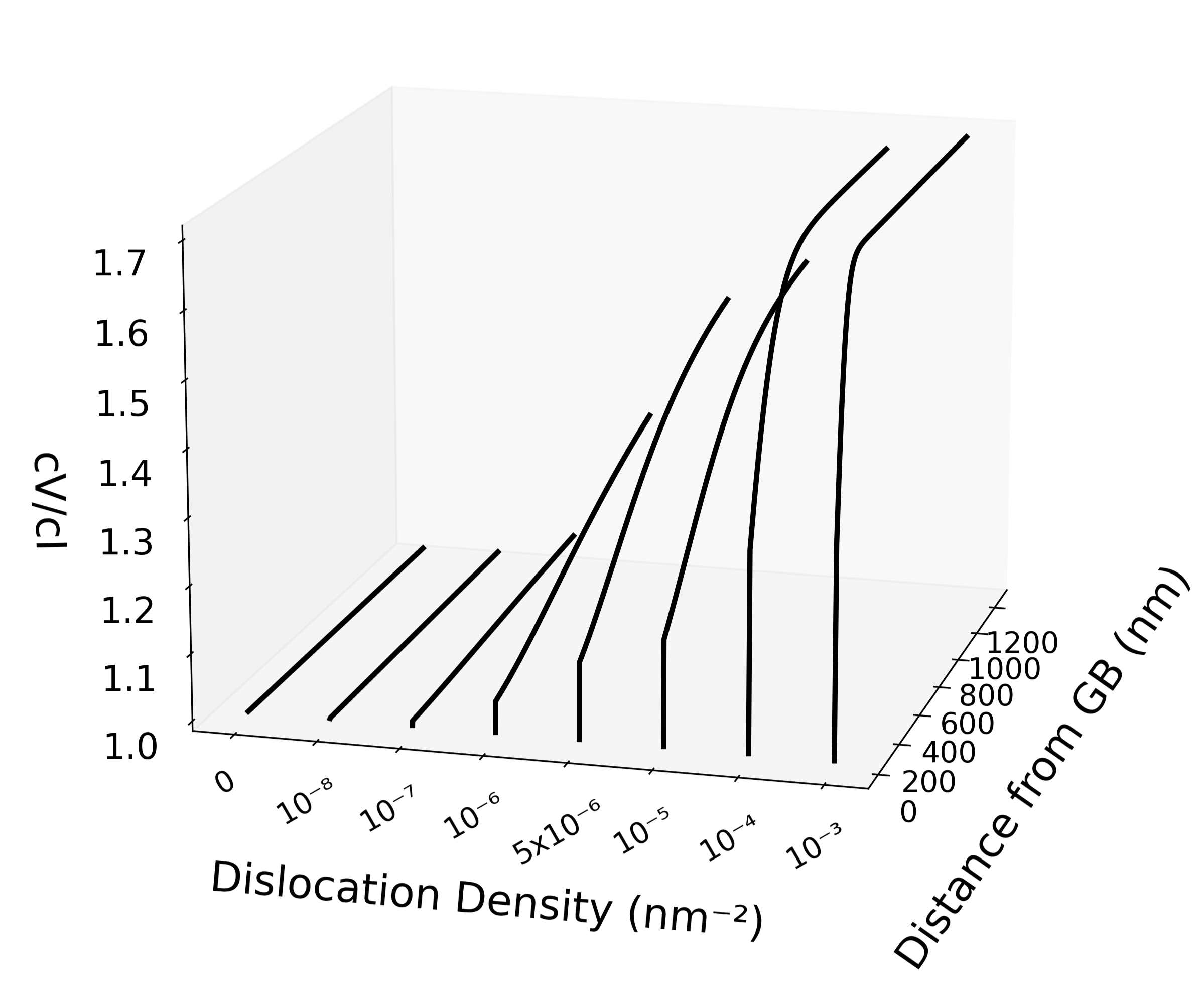}
    \caption{Ratio of the vacancy to SIA concentration for the biased case in Fig.~\ref{fig:sb_defect_500} normalized with respect to the ratio ($D_V/D_I$) for the fully unbiased case.}
    \label{fig:cV_cI_500_sink}
\end{figure}

\begin{figure}[htp!]
    \centering
    \includegraphics[scale = 0.5]{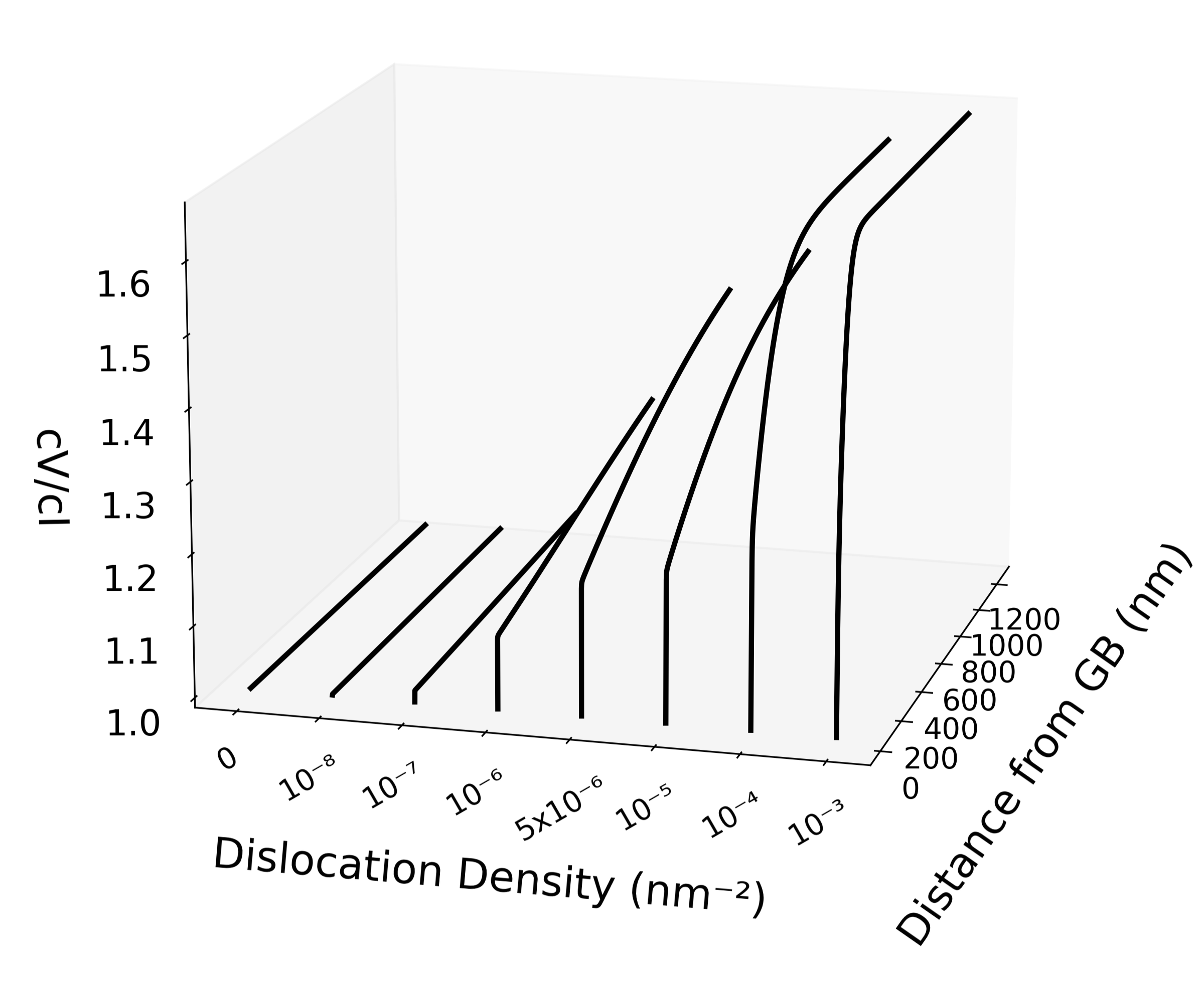}
    \caption{Ratio of the vacancy to SIA concentration for the biased case in Fig.~\ref{fig:sb_defect_700} normalized with respect to the ratio ($D_V/D_I$) for the fully unbiased case.}
    \label{fig:cV_cI_700_sink}
\end{figure}

\end{document}